\documentclass[numberedappendix]{emulateapj}
\usepackage{enumitem}

\begin{document}

\newcommand{\Msun}{\hbox{M}_{\sun}}
\newcommand{\Lsun}{\hbox{L}_{\sun}}
\newcommand{\Ms}{M_{\rm s}}
\newcommand{\Mhalo}{M_{\rm halo}}

\title{Simulating Deep Hubble Images With Semi-empirical Models of Galaxy Formation}
\shorttitle{Simulating Hubble Images With Galaxy Formation Models}
\shortauthors{Taghizadeh-Popp et al.}
\author{Manuchehr Taghizadeh-Popp\altaffilmark{1,2}}
\author{S. Michael Fall\altaffilmark{1}}
\author{Richard L. White\altaffilmark{1}}
\author{Alexander S. Szalay\altaffilmark{2}}
\altaffiltext{1}{Space Telescope Science Institute, 3700 San Martin Drive, Baltimore MD 21218, USA; mtaghiza@pha.jhu.edu}
\altaffiltext{2}{Department of Physics and Astronomy, Johns Hopkins University, 3400 North Charles Street, Baltimore, MD 21218, USA}


\begin{abstract}
We simulate deep images from the \textit{Hubble Space Telescope} (\textit{HST\/}) using semi-empirical models
of galaxy formation with only a few basic assumptions and parameters.
We project our simulations all the way to the observational domain,
adding cosmological and instrumental effects to the images, and
analyze them in the same way as real \textit{HST\/} images (``forward modeling'').
This is a powerful tool for testing and comparing
galaxy evolution models, since it allows us to make unbiased
comparisons between the predicted and observed distributions of
galaxy properties, while automatically taking into account all
relevant selection effects.

Our semi-empirical models populate each dark matter halo
with a galaxy of determined stellar mass and scale radius.  We
compute the luminosity and spectrum of each simulated galaxy from
its evolving stellar mass using stellar population synthesis models.
We calculate the intrinsic scatter in the stellar mass-halo mass
relation that naturally results from enforcing a monotonically
increasing stellar mass along the merger history of each halo.  The
simulated galaxy images are drawn from cutouts of real galaxies
from the Sloan Digital Sky Survey, with sizes and fluxes rescaled
to match those of the model galaxies. 

The distributions of galaxy luminosities, sizes, and surface
brightnesses depend on the adjustable parameters in the models, and
they agree well with observations for reasonable values of those
parameters.  Measured galaxy magnitudes and sizes have significant
magnitude-dependent biases, with both being underestimated near the
magnitude detection limit.  The fraction of galaxies detected and
fraction of light detected also depend sensitively on the details
of the model.

\end{abstract}

\section{Introduction}\label{Introduction}

The \textit{Hubble Space Telescope} (\textit{HST\/}) has spent much of its operational
lifetime staring into deep space, surveying galaxies in their infancy
and youth. The motivation for these deep surveys (the Hubble Deep
Field and its successors) was to obtain time-lapse images that would
reveal how galaxies formed and evolved. While we have made great
progress in interpreting the deep \textit{HST\/} surveys, this program remains
challenging and far from complete \citep[ and references
therein]{galametz2013,guo2013b,bouwens2014}. There are two major
reasons for this. First, the samples of high-redshift galaxies
have been severely edited by selection effects, primarily limits
on flux and surface brightness, effectively biasing the observable 
universe toward
bright, compact galaxies.  Second, on the theoretical side, there
are still many significant gaps in our understanding of the physical
processes that affect the baryonic components of galaxies (stars,
gas, and dust) and the radiation they emit. These uncertainties are
reflected in the many free parameters of the semi-analytical models
and in the analogous sub-grid physics of the hydrodynamical models.
In this paper, we present a new approach to the analysis and interpretation
of deep galaxy surveys that addresses both of these issues.

To account for selection effects, we create
simulated \textit{HST\/} images of model galaxy populations, and we then analyze
these images in the same way as real \textit{HST\/} images, extracting catalogs
from the simulated images to detect and measure the fluxes, sizes,
and other properties of the galaxies.  Our simulated images include
both cosmological effects (projection along pencil beams, redshifting
of passbands, dimming of flux and surface brightness) and instrumental
effects (point-spread function [PSF], pixelation, noise, sky
backgrounds) for any given \textit{HST\/} camera, filter, and exposure time.
We then extract catalogs of objects in the images with the widely
used {\tt SExtractor} software \citep{bertin1996}.  Thus, our
procedure automatically takes into account all relevant selection
effects, allowing us to make unbiased comparisons between the
predicted and observed distributions of luminosities, sizes, and
other properties of galaxies.  This is the approach recommended by
textbooks on statistical inference: map the predictions all the way
into the observational domain and make the comparisons there,
often called ``forward modeling.''

The forward modeling method in not sensitive to small errors in the
luminosities, sizes, and other properties of galaxies, or even to the
exact definitions of these quantities, because such errors affect
measurements of both the simulated and real images in the same way.
In other words, these errors cancel out of the comparisons of the
simulated and real distributions of luminosity, size, and so forth.
This is one of the main advantages of the forward modeling approach.

In contrast, nearly all work in this field is based on the opposite,
but simpler, approach of comparing predictions with observations
in the theoretical domain (``backward modeling''). 
Some exceptions are the
reconstruction of mock images or data starting from semi-analytical
models, \citep[e.g.][]{blaizot2005,overzier2013} or hydrodynamical
simulations \citep[e.g][]{lotz2008,devriendt2010,mozena2013},
although the latter suffer from unrealistic star formation histories
and too rapid growth of stellar masses at early times \citep{bouche2010}.
In this paper, we go beyond the creation of mock galaxy images
by deriving simulated distributions of galaxy properties and comparing
them with the corresponding observated distributions.

To limit the number of assumptions and parameters in our models,
we adopt a semi-empirical description of galaxy evolution.
This description is based on the evolution of dark matter halos 
in cosmological $N$-body simulations, which is now well understood,
in contrast to the evolution of the baryonic parts of galaxies.
The main assumption of the semi-empirical description is that
most of the information needed for simulating a population of 
galaxies is already encoded in the merger trees of their dark halos.
Each halo is assumed to host one model galaxy, and the properties 
of that galaxy are then uniquely determined by those of its halo, 
including its mass and size. 
The advantage of this method is that it sidesteps much of the complex
and uncertain baryonic processes in galaxy formation; the disadvantage
is that it likely oversimplifies some aspects of these processes.
This semi-empirical description has been developed in numerous studies
over the past decade
\citep[e.g.,][]{vale2004,conroy2006,conroy2009,guo2010,moster2010,behroozi2010,behroozi2013,guo2014,moster2013,kravtsov2013,reddick2013}.

We derive the radiative spectrum of each simulated galaxy from its 
star formation history using stellar population synthesis models. The star
formation history in turn follows from the growth of its halo mass, including 
both smooth accretion and discrete mergers with other halos. 
The radiative spectrum also depends on the metallicity of the stellar population and the 
absorption by gas and dust in the galaxy and by gas in the intergalactic medium.

In our implementation of this method, we use cutout images of real 
galaxies in the Sloan Digital Sky Survey (SDSS) as templates for the 
visual appearance of our model galaxies, with their fluxes
and sizes rescaled to reflect galaxy evolution according to our
models. This generates much more realistic morphologies than the
smooth S\'ersic bulge+disk light profiles commonly used in previous
semi-empirical or semi-analytical models
and improves modeling of the detection incompleteness
effects that arise from the internal clumpiness of real
galaxies.

Our approach gives us a valuable science tool for comparing models
of galaxy evolution. In this paper we build a reference 
model using plausible choices of parameters, and
explore other models by changing one parameter at a time.
Thus, we can test the sensitivity of the simulated universe to each
of these parameters by comparing the statistics derived from their
respective simulated images to those from reference model
and real \textit{HST\/} images.
Moreover, since the \textit{input} properties of each model
galaxy on the simulated images are known, we are able to quantify
the model-dependent \textit{output} galaxy detection efficiency by
comparing the input and output distributions of galaxy properties
(such as luminosity or size).  
Our approach can be used to  
inform the design of future surveys (e.g., choice of filters and
exposure times) by addressing directly the question of which
data are most useful to discriminate between different theoretical
models.
This will be especially valuable in planning the deepest surveys with
the \textit{James Webb Space Telescope} (\textit{JWST\/}).

We emphasize that the main purpose of this paper is to demonstrate the
validity and utility of a general method for analyzing and interpreting deep 
galaxy surveys.
This is an initial, exploratory study.
We regard our specific implementation of the method and the first results 
obtained from it and presented here as being illustrative rather than definitive.
There is scope for further development of the method and refinement of the 
results. 
Nevertheless, the overall agreement we find between our simulations and 
observations---with no fine-tuning of parameters---is remarkable and
encouraging.

This paper is arranged as follows. \S\ref{Sec:MakingSimUniverse} explains the method
for modeling and building simulated universes, from the
selection of semi-empirical models and the dark matter simulation
to the building of simulated \textit{HST\/} images. The main results, presented in
\S\ref{Sec:Results}, include the implementation of stellar
mass-halo mass relations (\S\ref{Sec:Halo-stellMassRESULTS}), 
present-day mass and luminosity distributions derived from the
semi-empirical models of a simulated universe
(\S\ref{Sec:ComparingZ0}),
and a comparison of the luminosity, size and surface brightness
distributions extracted from the simulated images with measurements
from real \textit{HST\/} images (\S\ref{Sec:SimImageAndStatistics}). 
We devote \S4 to an analysis of the cosmic star formation rate density.
Lastly,
\S\ref{Sec:Conclusion} discusses the results and summarizes our
conclusions.

\section{Building a Simulated Universe }\label{Sec:MakingSimUniverse}

In this section we describe in detail the steps required to
build a self-consistent simulated universe and associated \textit{HST\/} images.  We also
describe the parameters chosen for our \textit{reference model} for
the universe, as well as variations on those parameters explored
in the other models.

\subsection{The Dark Matter Simulation}\label{Sec:Thedarkmattersim}

Following standard practice, we use a $\Lambda$CDM simulation as the
three-dimensional (3-D) skeleton of our simulated universes, placing a \textit{model
galaxy} at the center of each dark (sub)halo\footnote[1]{We use ``halo" and
``sub-halo" interchangeably, following \cite{guo2010}.}. That defines
the spatial distribution and number density of galaxies as a
function of redshift. Most of the information needed in our method
is in fact provided by the halo mass and size evolution along halo
merger trees, which does not involve any fitting of free parameters
(see \S\ref{Sec:Halo-StellMassRelation} and
\S\ref{Sec:GalSizeFromHaloSize}).

For the merger trees, we use the milli-Millennium cosmological dark matter
simulation (mMS) \citep{springel2005,lemson2006}.
Its smaller size compared to the full
Millennium or Millennium II simulations \citep{springel2005,boylan-kolchin2009} makes it easier to use for this exploratory
work; we show below that the coarser mass resolution
in the mMS is adequate for our simulations (\S\ref{Sec:ResultsFromStandardModel}).
Halos are defined using friends-of-friends groups as explained in \cite{guo2010}.
Bound dark matter structures or (sub)halos are 
composed of the most massive main (or central) sub-halo surrounded
by satellite sub-halos.

The mMS has the same cosmology and particle mass ($1.18\times
10^9\Msun$) as the much larger Millennium simulation, but with
both a smaller box size (85.62 Mpc) and a reduced number of particles ($270^3$).
The cosmological parameters used are the ones obtained by \textit{WMAP\/}1
\citep{spergel2003}, i.e., $\Omega_{M}=0.25$, $\Omega_{\Lambda}=0.75$,
$h_{0}=0.73$ and $\sigma_{8}=0.9$. As explained by \cite{guo2013a},
the difference between the \textit{WMAP\/}1 and the standard \textit{WMAP\/}7 \citep{komatsu2011}
cosmologies does not affect significantly the relevant aspects of
dark matter structure. In fact, a smaller \textit{WMAP\/}7 $\sigma_{8}=0.807$
is counterbalanced by a greater $\Omega_{M}=0.272$, which results,
for example, in the \textit{WMAP\/}1 and \textit{WMAP\/}7-derived halo mass functions
being very similar at $z=0$.

\subsection{Constructing Stellar Mass-Halo Mass Relations with an Intrinsic Scatter}\label{Sec:Halo-StellMassRelation}

We obtain the stellar mass of model galaxies in the
simulation using semi-empirical modeling. This approach defines
the stellar mass $\Ms$ of the galaxy as a function of the dark matter mass of the
halo, $\Ms = \Ms(\Mhalo)$, with the function $\Ms(\Mhalo)$ defined to be
the stellar mass-halo mass (SMHM) relation.
Although this is a simple one-to-one relation, it can
be readily modified to
include statistical scatter or redshift dependence
\citep[e.g.,][]{behroozi2013}.  In this paper
we adopt several SMHM relations from the literature, and use them 
for building our simulated images.
We introduce a novel,
self-consistent approach that naturally adds scatter to the SMHM
mass relation.

As a measure of the halo mass, we use the virial mass
$M_{\rm vir}$ (mass enclosed inside the maximum radius within which
the mean density is 200 times the critical value), which is obtained
from the value-added catalog of \cite{guo2010} based on the
milli-Millennium simulation.
In this catalog, the mass of a central halo is given by $M_{\rm vir}$, whereas for a satellite
halo it is the maximum $M_{\rm vir}$ ever attained before becoming a satellite.
In the semi-empirical approach, this preserves the stellar mass of
satellite galaxies even while the outer parts of their dark halos
are being tidally stripped as they orbit within a central halo.

\begin{figure}
\begin{center}
\epsscale{1.0}
\plotone{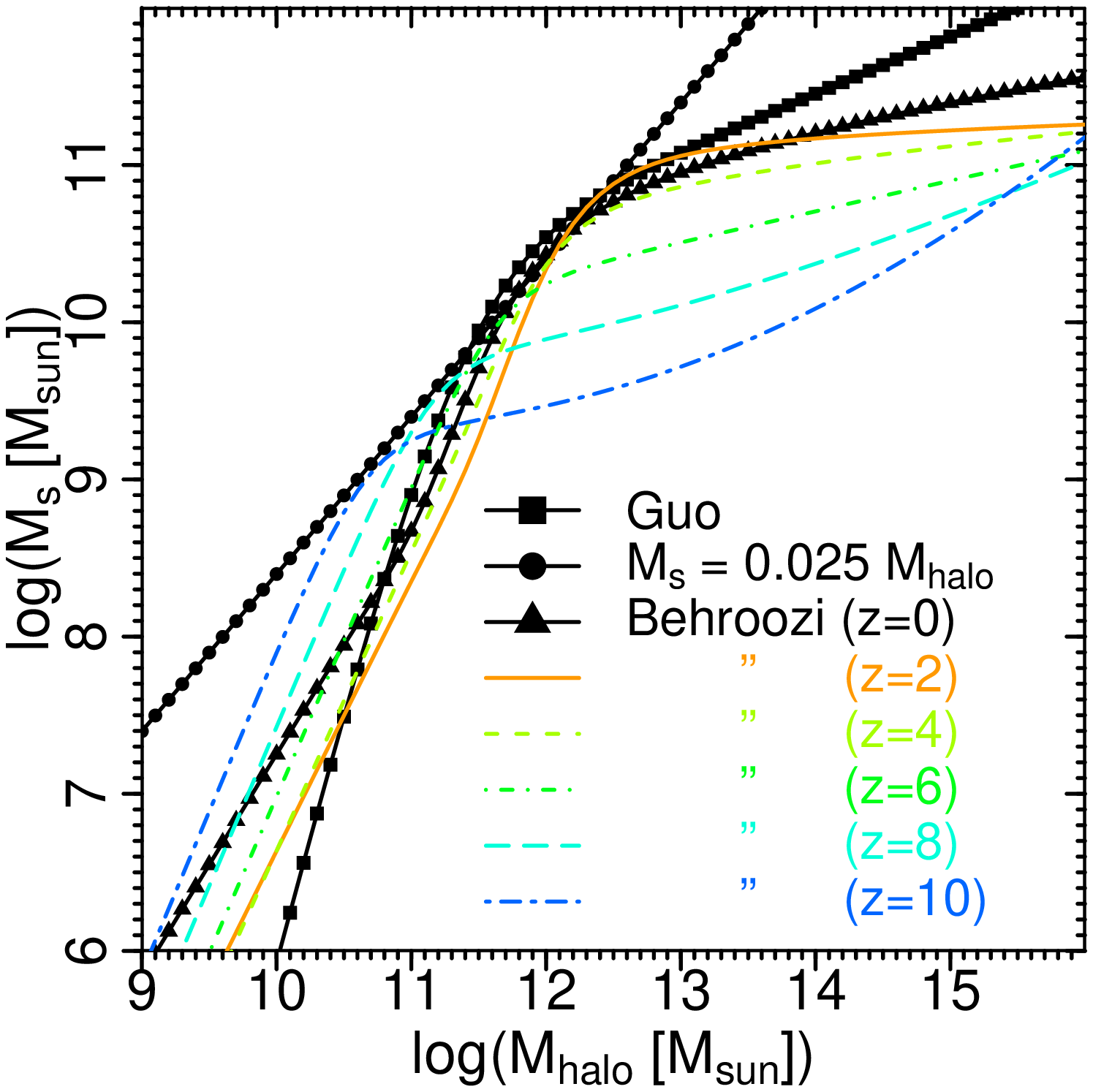}
\caption{
Stellar mass $\Ms$ versus halo mass $\Mhalo$ relations
tested in this paper. Included are the models of \cite{guo2010} and
\cite{behroozi2013}, as well as a linear SMHM relation $M_{s}=0.025
\Mhalo$. The low and high mass tails are necessarily
extrapolations of the fits due to the limited resolution of the DM
simulations ($\log\Mhalo<10.8$ for the \cite{guo2010} model and
$\log\Mhalo<10.3$ for the \cite{behroozi2013} model).
Note that this last model evolves with redshift, with
$M^{*}$ (characterizing the
knee in the SMHM relation) becoming smaller at higher redshift.
}
\label{Fig:HaloMassVsStellarMass}
\end{center}
\end{figure}

Figure~\ref{Fig:HaloMassVsStellarMass} shows some of the SMHM
relations found in the literature. Our reference model (or
Model 1) for simulating the universe uses the 
redshift-independent SMHM relation
from \cite{guo2010}. They computed this relation
using a halo abundance matching
technique that equates
quantiles of the low redshift
stellar mass function from SDSS \citep{li2009} to those of the
present-day halo virial mass function from the Millennium simulations. 
The quantile matching is made over a range of masses 
around the typical value $M^{*}$ (i.e., the ``knee"
in the SMHM relation as well as in  the mass function), with extrapolations
to the very low and high mass regimes, where both the stellar and halo
mass distributions are not well constrained. 
Another option is the redshift-evolving SMHM relation of \cite{behroozi2013}.
This relation is based on stellar
mass functions and cosmic star formation rates up to $z=8$. 
For our non-evolving Model 2, we adopt the \cite{behroozi2013} SMHM relation evaluated at $z=0$,
while for our evolving Model 3, we adopt the full redshift dependence of the \citeauthor{behroozi2013} SMHM relation.
Our Model 4 does not involve halo abundance matching, but
is a simple linear relation given by $\Ms=0.025 \Mhalo$, where
the slope has been chosen by eye to coincide with the other relations
in the vicinity of $M^{*}$.
We include this last model not because it is realistic but to study the sensitivity of our 
results to a SMHM relation that is very different from those of our first three models
(based on the results of \citealp{guo2010} and \citealp{behroozi2013})

In this
paper, we do not explicitly impose random scatter in the SMHM relation. Instead we
explore the scatter that emerges naturally from the dark
matter simulation as a consequence of one basic assumption:
\textit{we assume that the stellar mass along merger trees is a monotonically 
increasing function of time}.  This is physically plausible because
the stellar mass is concentrated in the center of halos due
to dissipation in the baryons, and as a result it
tends to be retained during mergers.  This is the
simplest physically motivated rule we have found for creating
consistent galaxy stellar masses from dark matter simulations.
The alternative is to assume that galaxies lose and gain
mass willy-nilly as halo masses decrease and increase and as
sub-halos merge; that appears much less plausible based on our
current understanding of galaxy formation and dynamics.
Another alternative is to impose scatter on the SMHM relation in a predetermined manner 
(as in the approach adopted by \cite{behroozi2013}) 

Our assumption of monotonically increasing stellar mass naturally leads to scatter in
the SMHM relation as a consequence of the following three related
effects:

\begin{enumerate}

\item In dark matter simulations, individual halos can
decrease in mass
from one time step to the next, for example in events of tidal
stripping. In that case, we follow
the approach of \cite{guo2010} and do \textit{not} reduce the stellar
mass accordingly, but retain the stellar mass present
before the decrease in $\Mhalo$.

\item In halo mergers, the dark matter mass of a descendant $M_{\rm
halo,desc}$ can be smaller than the sum of the progenitor dark
matter masses $\Sigma M_{\rm halo,prog}$ as some particles become unbound during the collision.
This is not a rare
occurrence in the simulations.
In order to conserve the stellar mass content,
we must break with the one-to-one fixed SMHM relation.

\item Observed SMHM relations are intrinsically non-linear (see
Figure~\ref{Fig:HaloMassVsStellarMass}).  That leads to
a conflict
with the assumed monotonic growth of $\Ms$ in halo
mergers. Consider the case when two halos 
with $M_{\rm halo,prog} \sim \Mhalo^{*}$ merge to create 
a new halo with $M_{\rm halo,desc} > \Mhalo^{*}$.  The 
SMHM relation increases
more slowly than linearly above $\Mhalo^{*}$,
which means that $\Ms$ for the new halo is supposed to
be smaller than the sum of the stellar masses for the
merging halos.  That conflicts with the assumption
that the stellar mass along the merger tree must increase
monotonically.

\end{enumerate}

To eliminate these conflicts we adjust stellar masses retroactively as follows. 
If a halo is
found to have a stellar mass (according to the imposed SMHM relation)
that decreases with time, we decrease
the stellar masses of its immediate progenitors to enforce a monotonic
increase in the stellar mass content of dark matter halos across
their merger trees.
This adjustment is applied recursively to all the progenitors
of halos with modified stellar masses.  Note that we are not
assuming reverse causality with this scheme!  Our assumption is
that lower mass halos in denser environments (which are going to
merge in the future) have their star formation rates suppressed
by these environments. 
This procedure is described in more detail in Appendix \ref{Sec:Appendix1}. 

One natural consequence of this procedure
is that, at a given halo mass, there is a
scatter in the stellar masses that 
tends to fall slightly below
the one-to-one SMHM relation at some points in the merger history.
The results with the modified SMHM relations will be shown in
\S\ref{Sec:Halo-stellMassRESULTS}.

\subsection{Illuminating Galaxies in Dark Matter Halos}\label{Sec:IlluminatingGalaxies}

The luminosity and spectrum of a model galaxy
at any time are determined by its star formation history
computed along the past merger history of its host
dark matter halo. Star formation is implied when the stellar
mass of a descendant halo is greater than the total stellar mass
of its progenitors in consecutive simulation time steps, or when
a single halo increases its dark
matter mass (and hence its stellar mass) due
to accretion of surrounding dark matter particles. The stellar
mass increase between simulation time steps implies a star
formation rate, which is used in stellar population synthesis
models to compute the emitted spectrum of the galaxy.
We assume that the star formation rate between time steps is uniform,
so that the star formation history is completely determined by the stellar
mass history of a halo.  Note that since we have forced the stellar
mass to increase monotonically with time, negative star formation rates are automatically excluded.

We thus simply reconstruct the spectrum of the model galaxy (and derived photometry) 
as the sum of a series of uniform
starbursts between each of the time steps in the simulation.
When a galaxy is placed in a simulated image at a particular
redshift $z_g$ (implying an age $t_g$), its rest-frame luminosity as a function of wavelength
is computed using the star-formation history up to time $t_g$
using the evolutionary stellar synthesis code {\tt GALAXEV} by
\cite{bruzual2003}. Note that the redshift $z_g$ is not required to be at one of the
discrete simulation time steps, since we can integrate the star formation rate 
from the previous time step to the actual time $t_g$.
\citep[The need of similar interpolation schemes has also been noticed by][]{yip2010}. In our reference model, we adopt the
Chabrier initial mass function, with a fixed solar metallicity
($Z=0.02$) and the standard dust model from \cite{charlot2000}
($\tau_{\nu}=1$ and $\mu=0.3$). Our modeling of galaxy spectra is flexible, as we can
in principle elaborate this model with additional variables, such as a
redshift-dependent metallicity.

\begin{figure}
\begin{center}
\epsscale{1.0}
\plotone{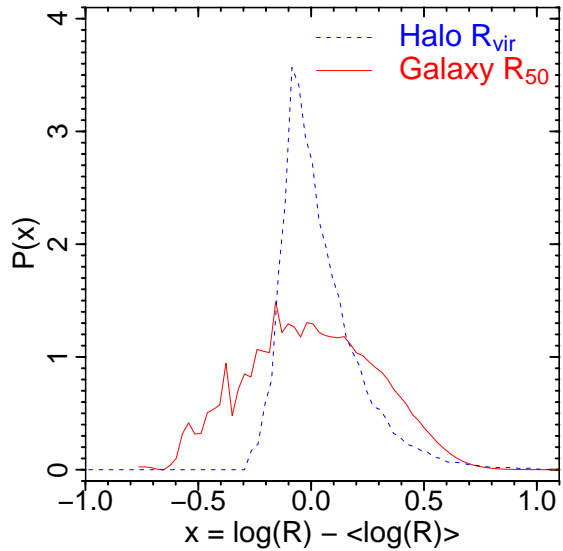}
\caption{
Probability density distributions of the (median subtracted)
logarithms of $R_{\rm vir}$ (dark matter halo virial radius) and
$R_{50}$ (SDSS galaxy $r$-band half-Petrosian-flux radius). Note that
the radius distributions resemble Gaussian curves, as
expected for approximately log-normal random variables. The median values
are $\langle\log R_{\rm vir}[{\rm Mpc}]\rangle$=-1.159 and
$\langle\log R_{50}[{\rm Mpc}]\rangle$=-2.824, with standard deviations
$\sigma(\log R_{\rm vir}[{\rm Mpc}])= 0.176$ and
$\sigma(\log R_{50}[{\rm Mpc}])=0.285$.
}
\label{Fig:HaloRadiusVsGalRadius} 
\end{center}
\end{figure}

\subsection{Deriving Galaxy Sizes from Dark Matter Halo Sizes}\label{Sec:GalSizeFromHaloSize}

We adjust the size of a model galaxy so that it is always a fixed fraction of the evolving size of its 
dark matter halo \citep[e.g.,][]{fall1980,kravtsov2013}. We determine the 
constant of proportionality between the
galaxy size and the halo size by comparing their
respective $z=0$ distributions, as in the halo abundance matching
method. However, we match only their medians instead of the full
distributions, which suffer from incompleteness in the tails.
It is well
known that size distributions for halos and galaxies are close to
being log-normal \citep[e.g.,][]{shen2003}. Since we choose to make
galaxy size linearly proportional to halo size, the scale
parameter and the median subtracted logarithmic distributions of
both distributions should be the same.
Figure~\ref{Fig:HaloRadiusVsGalRadius} shows the distributions
of halo virial radii from Millennium and $r$-band half Petrosian flux
radii $R_{50}$ for our main sample of SDSS galaxies from
\S\ref{Sec:GalaxiesFromSDSS}. The latter has been corrected for
incompleteness using the $V_{\rm MAX}$ method \citep{schmidt1968},
as shown with a similar galaxy sample by \cite{Taghizadeh-Popp2012}.
We find the relation $R_{50} = 0.022 R_{\rm vir}$.  The dispersions are
different (indicating that our assumption is not completely
accurate) but are similar enough for our purposes.
As in the case of
the spectra, we interpolate the galaxy size between discrete time steps
to the redshift $z_{g}$ of the model galaxy.

\begin{deluxetable*}{lp{0.7\columnwidth}} 
\tablecolumns{5}
\tablewidth{0pt}
\tablecaption{Galaxy Evolution Models\label{Tab:Models}}
\tablehead{
\colhead{Model (short title)} & \colhead{Details}
}
\startdata
\begin{minipage}[t]{0.2\columnwidth}%
Model 1 \\
(\textit{Reference Model\/})
\end{minipage} &
\begin{minipage}[t]{0.7\columnwidth}%
\begin{itemize}[noitemsep,topsep=0pt,parsep=0pt,partopsep=0pt,leftmargin=*]
\item Stellar mass-halo mass relation from \cite{guo2010}
\item Dust model from \cite{charlot2000}
\item Fixed solar metallicity ($Z=0.02$) at all redshifts
\item Apparent sizes of SDSS galaxy cutouts on image are scaled to the theoretical size predicted by the dark matter halo size
\item SDSS galaxy cutouts are chosen to be the closest match to the theoretically predicted model galaxy $u-r$ color and stellar mass.
\end{itemize}
\end{minipage} \\
 & \\
Model 2 & Same as Model 1, but using the \cite{behroozi2013} SMHM relation evaluated at $z=0$. \\
 & \\
Model 3 & Same as Model 1, but using the \cite{behroozi2013} redshift dependent SMHM relation.  \\
 & \\
Model 4 & Same as Model 1, but using the a linear SMHM relation $\Ms=0.025 \Mhalo$. \\
 & \\
Model 5 & Same as Model 1, but using no dust model for galaxies.\\
 & \\
Model 6 & Same as Model 1, but using a fixed very low metallicity ($Z=0.0001$) at all redshift.\\
 & \\
Model 7 & Same as Model 1, but without rescaling the intrinsic size of SDSS galaxy cutouts (except for the angular diameter distance scaling, also applied in the reference model). \\
 & \\
Model 8 & Same as Model 1, but the PetroR50 radius as well as $u-r$ and $\Ms$ are used for matching SDSS to model galaxies.\\
\enddata
\tablecomments{Description of eight different models used for building a
simulated universe. Our reference model (Model 1) contains
the most plausible parameters and sub-models. Other models are defined
by changing one of these parameters at a time.}
\end{deluxetable*} 

\subsection{Galaxy Image Cutouts from SDSS}\label{Sec:GalaxiesFromSDSS}

Our method places cutouts of real galaxy observations onto our
simulated image, which has advantages over using smooth analytic galaxy
light profiles. Galaxies often have a clumpy internal structure,
which affects source detections. A real galaxy could be
detected as two or more separate objects, especially if its surface
brightness approaches the image noise level. Real galaxy cutouts
recreate this effect and are more realistic than smooth bulge and disk
light profiles adopted in previous semi-empirical and semi-analytical 
models. Of course, this effect is not as important when the
apparent galaxy sizes are comparable to the width of the point spread function,
as may happen in the high redshift limit.

A database of SDSS galaxy images is used for the cutouts.  We select
from the SDSS image database the real galaxy that is the closest
neighbor to the model galaxy in a multi-dimensional space of
galaxy properties.  Once the closest
matching SDSS galaxy is found, we rescale its flux and size in order 
to make them equal to those of the model galaxy. No
free parameters are fitted or required in this step.  The full
details of the SDSS data and the selection procedure is described
in Appendix~\ref{Sec:Appendix2}.

One might wonder whether SDSS galaxies are clumpy enough to be 
accurate models of high-redshift galaxies, since
galaxies at higher redshifts tend to be more irregular than local galaxies.
The changes in morphology are due both to the shift of optical band
filters into the rest-frame ultraviolet for distant objects and also
to a higher merger rate and dynamically less-relaxed structures
in the early universe.

While these effects are worthy of further exploration in the
future, for this paper the SDSS images are a good basis
for simulations.
Our analysis of galaxy counts and detection efficiencies relies
on observations and simulations in the \textit{HST\/} WFC3/IR F160W filter
($\lambda = 1.6\,\mu$m); this implies that objects with redshifts
less than 3 have an SDSS filter (from $griz$) that is at the
appropriate rest-frame wavelength, and the different morphologies in the
ultraviolet are irrelevant.  Moreover, galaxies
at redshifts beyond 3 tend to be sufficiently compact that their internal clumpiness has
little impact on their detectability.
The median FWHM size of detected galaxies with $z>3$ in our simulations
is 0.3 arcsec, which is only twice the FWHM of the $1.6\,\mu$m
WFC3 PSF.
While there is room for improvement in
modeling the internal structure of galaxies, particularly for bluer filters in the $1<z<3$
redshift range, the SDSS cutout images are certainly an improvement over models
that use smooth analytical profiles.

\subsection{Assembling the Simulated Image}\label{Sec:CreatingSimImage}

Once model galaxy properties are calculated, we generate 
pencil-beams through our simulated volume and project them onto the plane of the sky. We then 
simulate ACS/WFC camera images, with their visible filters, as well as corresponding WFC3/IR images (sampled to the ACS/WFC pixel size), with their 
infrared filters. We include realistic PSFs
for both cameras.

To determine the 3-D structure within these pencil beams, we use a 
Monte Carlo method. This approach is much simpler and faster than other alternatives, such as the replication of a simulation box much smaller than the depth spanned by the 
simulated image. We first sample a random redshifts $z_{g}$
from a distribution that gives constant comoving volume per redshift
interval. Then we randomly select a dark matter halo
at $z=0$, choose all of its progenitors found at $z_{g}$,
and place them in the simulated image (at redshift $z_{g}$)
according to their relative 3-D spatial
positions in the simulation box (interpolating between time steps).
Although the large-scale correlations are discarded, we still preserve the short-range correlations between progenitors, while reducing considerably the computation time.

The visual appearance of the model galaxy on the simulated image
is given by the best matching SDSS galaxy cutout
(Appendix \ref{Sec:Appendix2}).  We select the SDSS band whose redshifted central
wavelength $\lambda_{\rm c}$ (to redshift $z_{g}$) is the closest
to the band of the simulated image. Due to the redshift of wavelengths, we end up using
the bluest visible SDSS bands to represent most high-redshift simulated
galaxies in the visible and infrared \textit{HST\/} filters, since SDSS lacks the UV and far UV bands 
that would be more suitable for this redshift regime\footnote{A cross-match between 
SDSS and \textit{GALEX\/} might be useful in the future for adding ultraviolet bands to our suite of 
galaxy cutouts, but that is out of the scope of this exploratory study and analysis.}.
Since the $u$ band is noisy in SDSS, we use the $g$ band as the bluest bandpass.
In fact, all model galaxies placed on the simulated image
are represented by a $g$-band cutout
at $z>2$ for ACS/WFC images or $z>3$ for WFC3/IR images.

We rescale the flux of the image
cutout to match that of the model galaxy. We apply to the model
spectrum the effects of redshift, cosmological distance dimming and
intergalactic absorption \citep[as in][]{madau1995}, before applying
the photometric filter response.

We rescale the sizes of the galaxy cutouts placed on the simulated
image according to one of the following rules:

\begin{enumerate}
\item \textit{No size scaling:\/} The proper size of the original SDSS physical galaxy is left intact,
irrespective of the model galaxy size and hence the size of its halo. The physical size is scaled
to the apparent size on the image using the angular diameter distance $D_{A}(z_{g})$.

\item \textit{Size scaling:\/} The proper size of the SDSS galaxy is scaled to match
the physical size of the model galaxy, which in turn is a fixed fraction of the halo size (as described above).
The apparent size is then computed from $D_{A}(z_{g})$.

\end{enumerate}

In the first model, the $z{=}0$ size-mass relation of galaxies is
preserved at all redshifts, whereas in the second model, the size-mass
relation evolves with redshift, driven by the growth of the dark
halos.  Galaxy sizes tend to be smaller at higher redshifts in the
second model due to evolution in the halo sizes.

To add instrumental effects, we convolve with the \textit{HST\/} point
spread function, apply the \textit{HST\/} instrument efficiency and pixelation,
and add noise. The \textit{HST\/} instrument modeling uses the same software
({\tt pysynphot, Tiny Tim}), instrumental parameters and sky
background as used in the standard \textit{HST\/} Exposure Time Calculators,
providing a high fidelity model of the real \textit{HST\/} performance.
	
\subsection{Source Detection and Photometry of Simulated Images}

The detection, extraction, and photometry of galaxies are performed by
running {\tt SExtractor} \citep{bertin1996} on the simulated
images. Here we directly follow the method and parameters described
in \cite{galametz2013} and designed for the analysis of real \textit{HST\/} images. The complete output catalog merges {\tt SExtractor} runs using
two different detection modes.
The Cold mode is used for detecting bright and
extended sources, while the Hot
mode is optimized for extracting faint and small objects. After
extraction, detected or \textit{output} galaxies on the final
simulated image are matched (using the detected position and luminosities)
to the \textit{input} galaxies on the same image before adding the
instrumental effects. This provides a direct measure of the detection
completeness by comparing what {\tt SExtractor} detects
to what was originally placed on the image.

To compare our simulated galaxies to
those from real galaxy surveys we use {\tt AUTO} magnitudes and
Petrosian radii $R_{\rm P}$, as calculated by {\tt SExtractor}.
The Petrosian radii of our input galaxy cutouts, as given by the
SDSS pipeline, differ from those returned by {\tt SExtractor}
on the already simulated images, probably due to different
definitions for the radius in the SDSS pipeline and in {\tt SExtractor}.
Since we observe a
linear offset between the distributions of these radii, we change the input
SDSS values to match those from {\tt SExtractor} by using
$\log R_{\rm P}({\tt SExtractor})$=$\log R_{\rm P}(\hbox{\rm SDSS})+0.41$.

\begin{figure*}[]
  \centering
  \begin{tabular}{cc}
    \includegraphics[width=79mm]{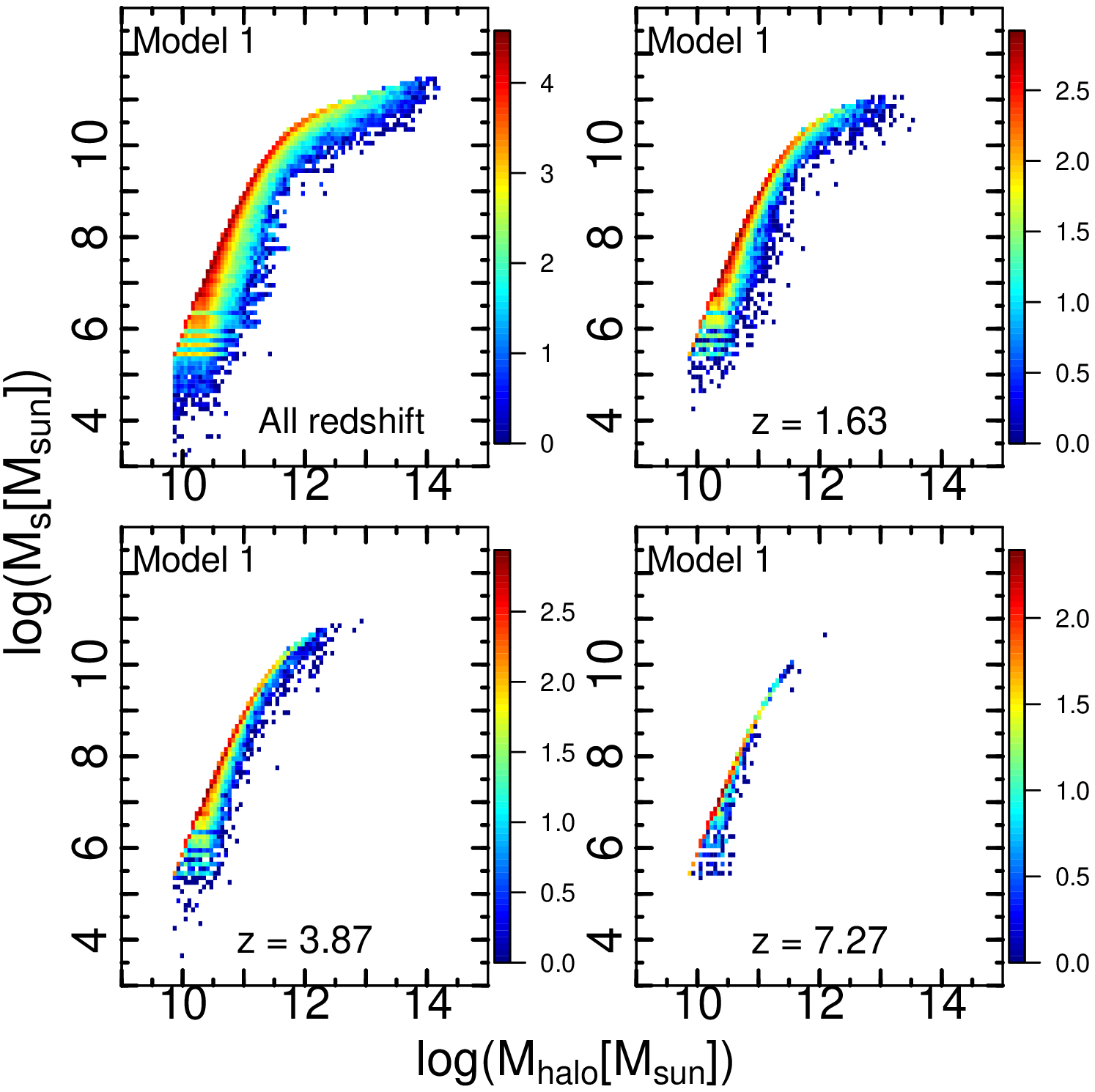}&
    \includegraphics[width=79mm]{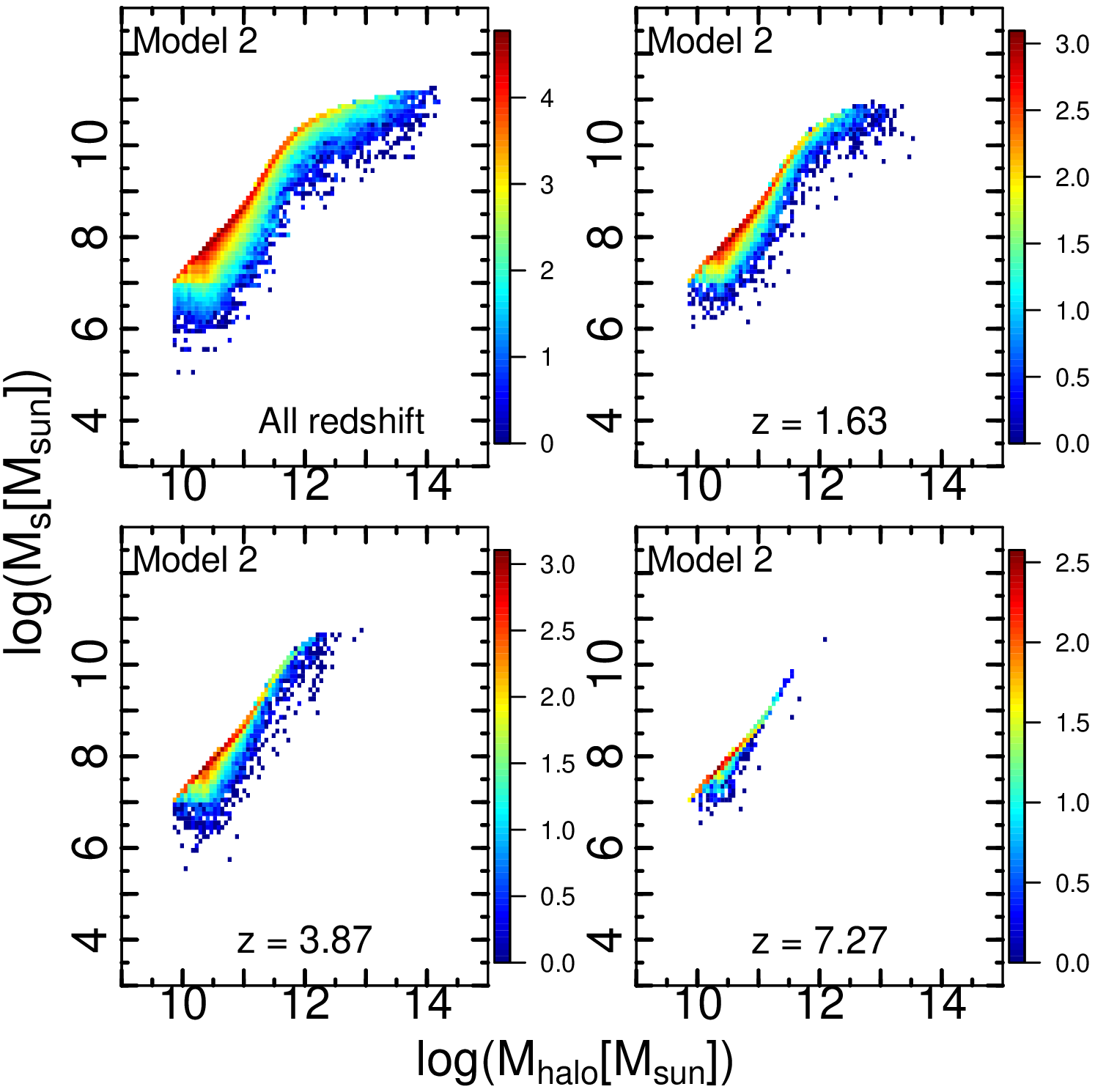}\\[1mm]
    \includegraphics[width=79mm]{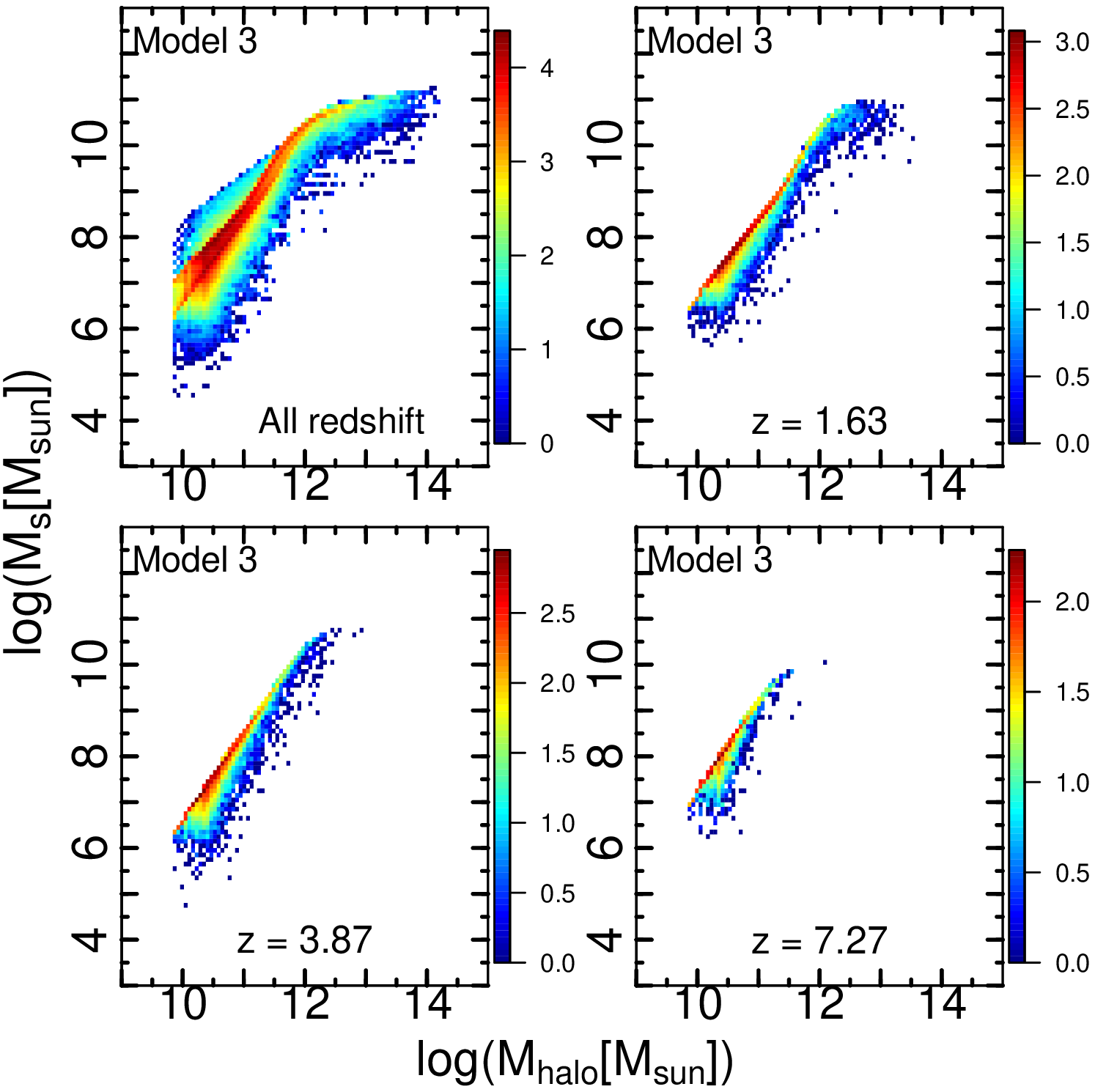}&
    \includegraphics[width=79mm]{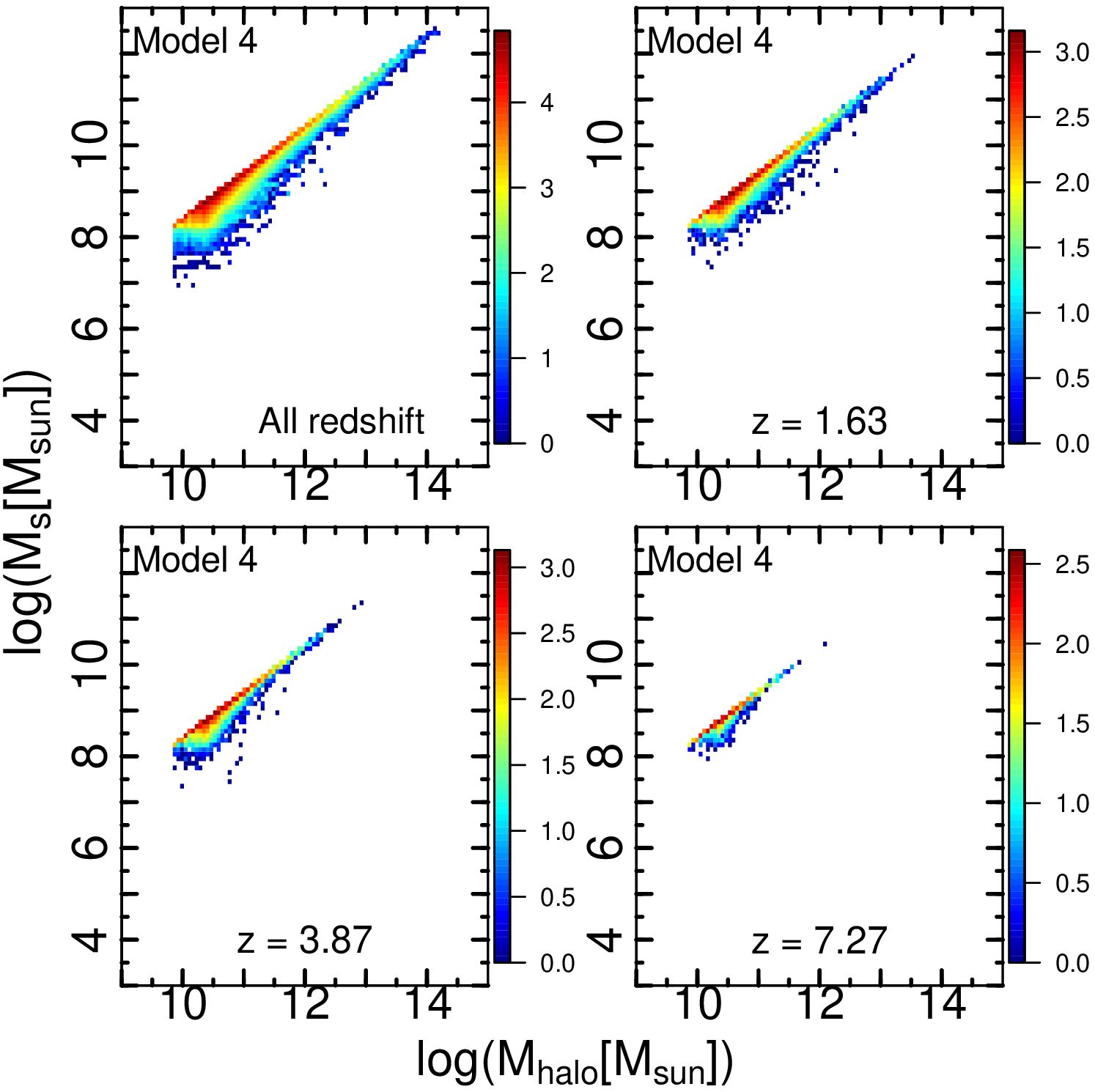}\\[1mm]
	\end{tabular}
\caption{
Modified stellar mass-halo mass relations
for the four models tested in this paper, obtained after retroactively
reducing the $\Ms$ values predicted by the one-to-one SMHM
relations in Fig.~\ref{Fig:HaloMassVsStellarMass} to enforce a
monotonically increasing $\Ms$ as a function of time
along merger trees. Data from three redshift time steps as well as
the combination from all
time steps is shown. The colors show the log-scaled number counts
in bins of size $\Delta \log \Mhalo=0.06$ by $\Delta \log \Ms=0.01$.
}
\label{Fig:StellMassHaloMassDist}
\end{figure*}

\begin{figure*}[]
  \centering
  \begin{tabular}{cccc}
    \includegraphics[width=40mm]{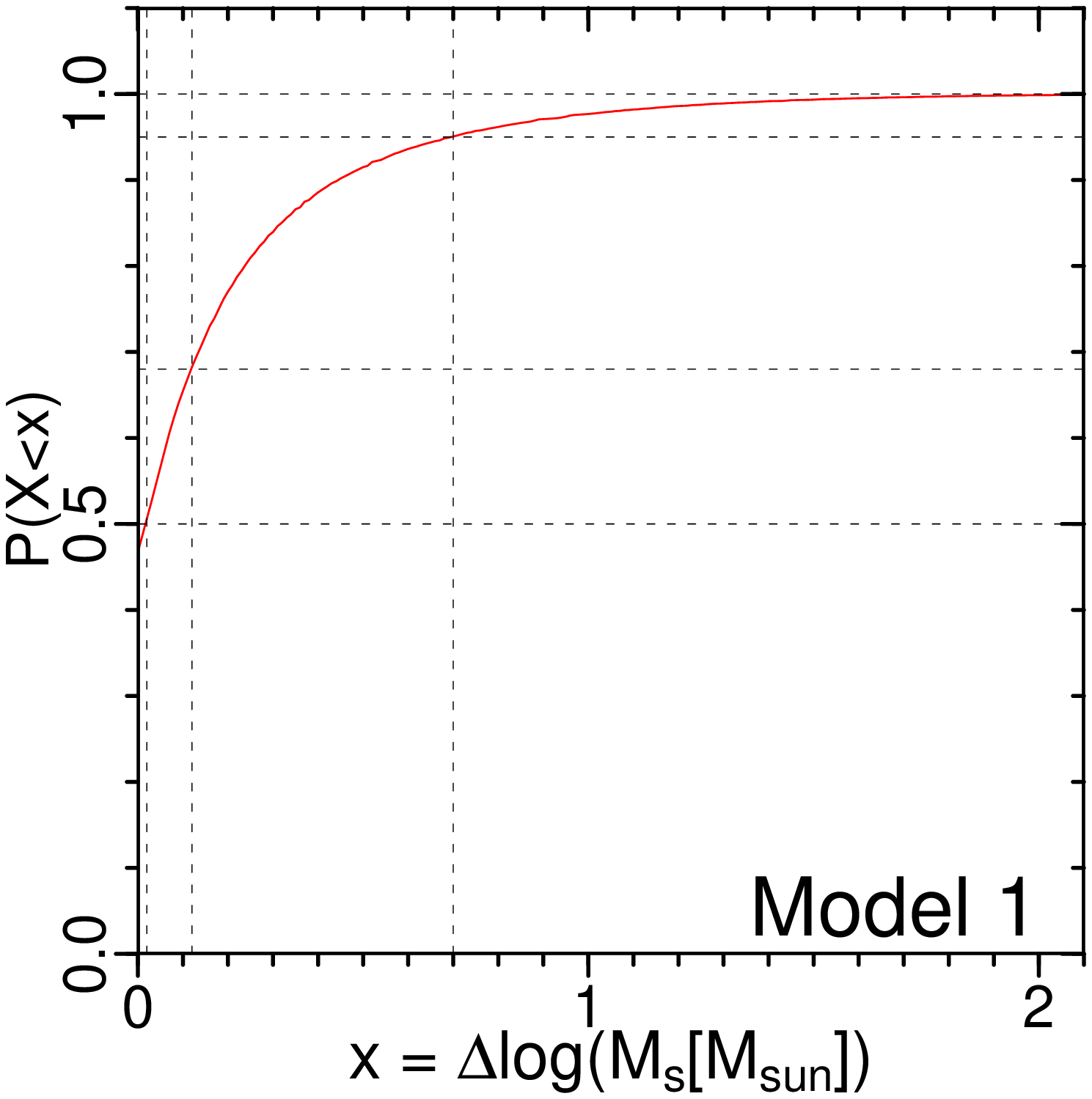}&
    \includegraphics[width=40mm]{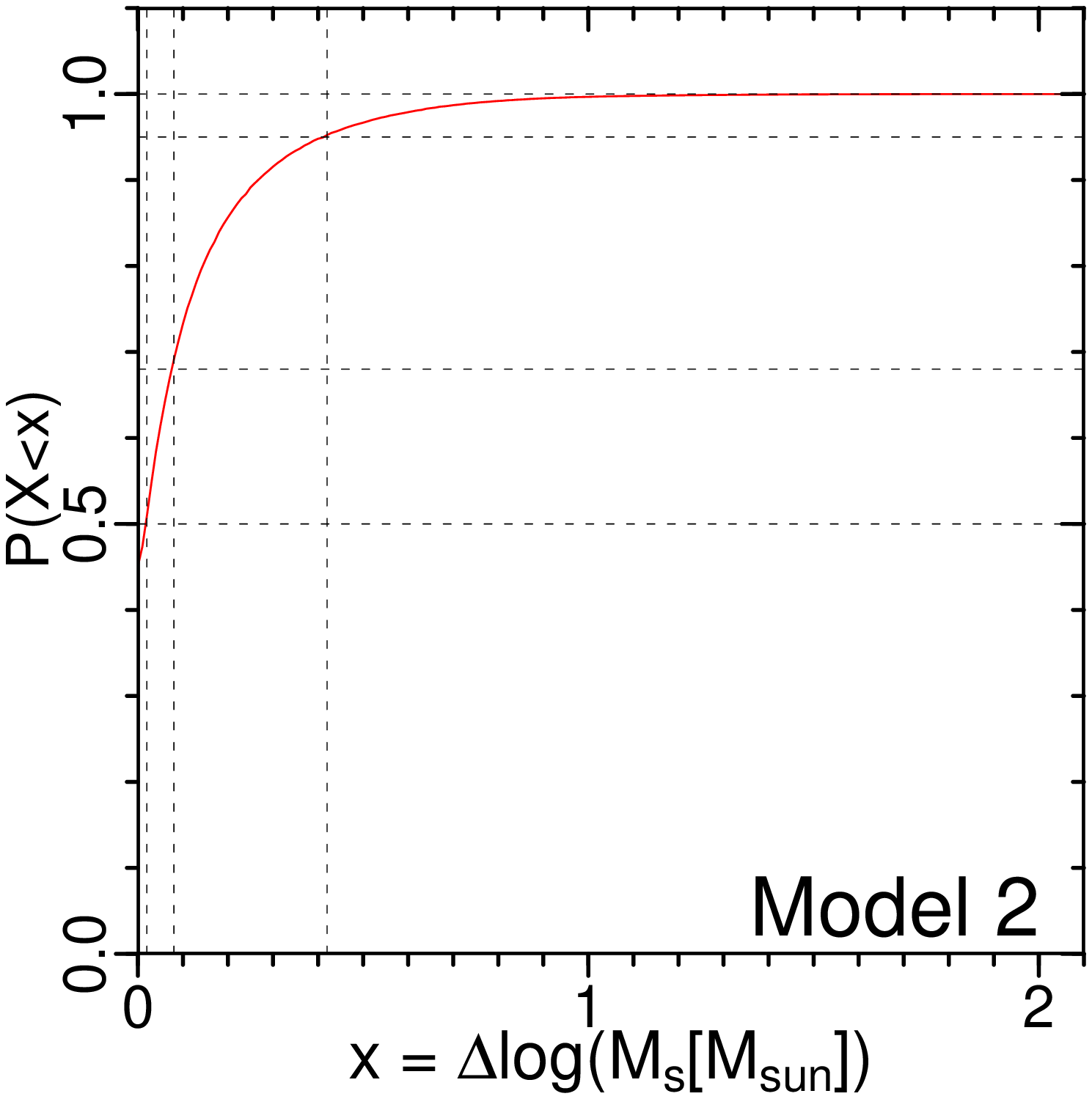}&
    \includegraphics[width=40mm]{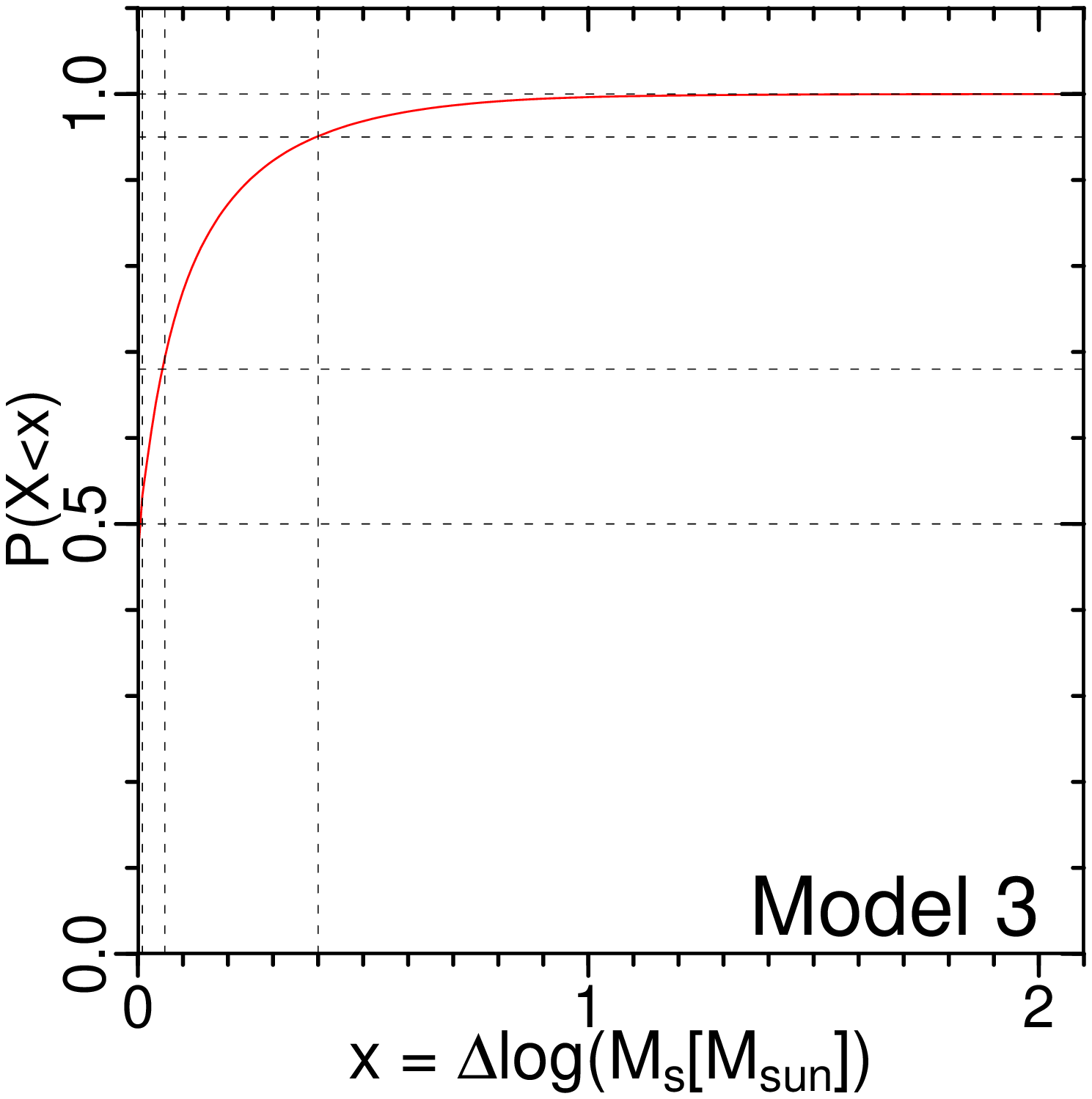}&
    \includegraphics[width=40mm]{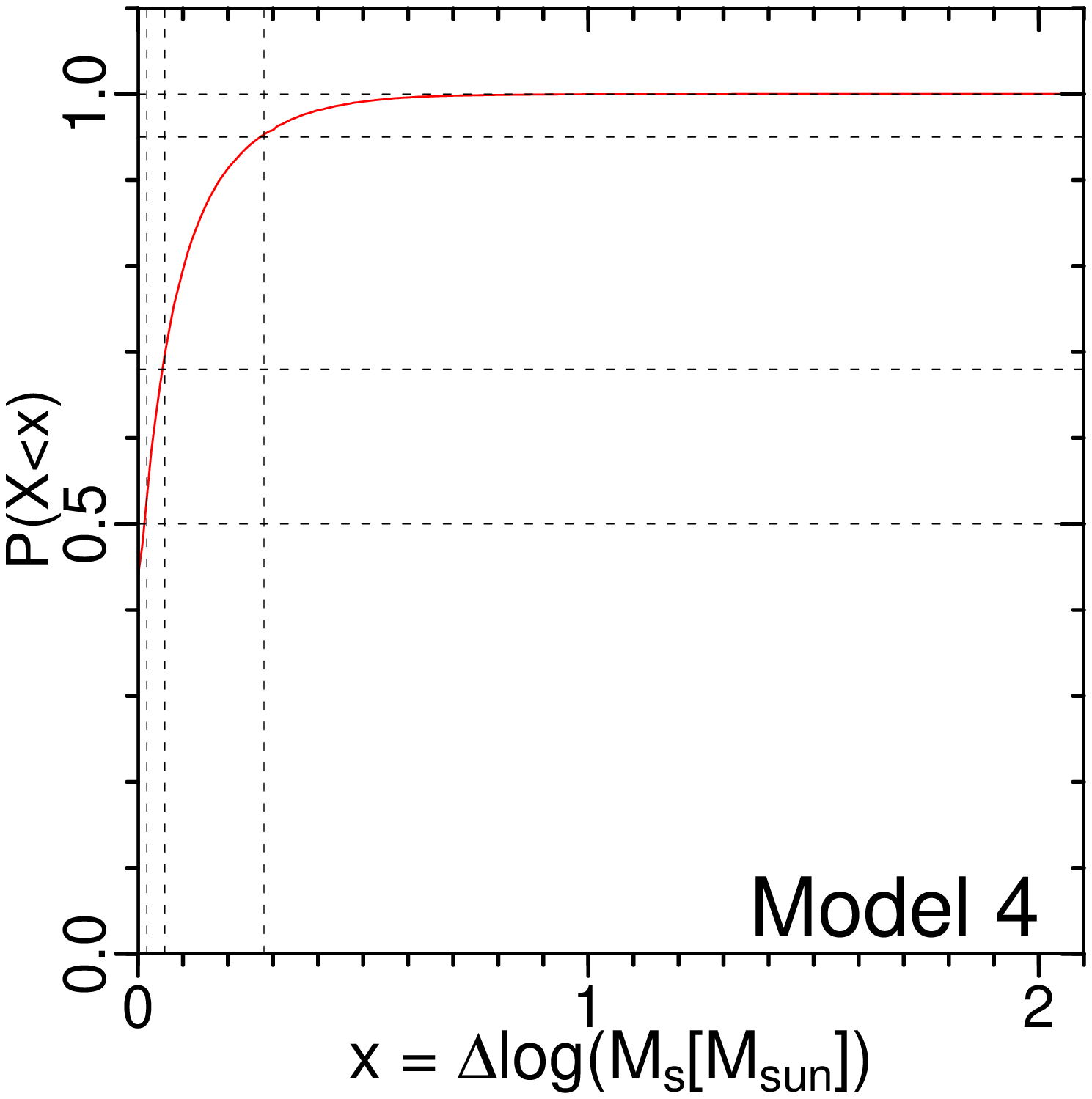}\\[1mm]
  \end{tabular}
\caption{
Cumulative distributions
of the difference between the modified SMHM relation shown in
Fig.~\ref{Fig:StellMassHaloMassDist}
and the original one-to-one relation from
Fig.~\ref{Fig:HaloMassVsStellarMass} (values from all redshift time steps
are included). The 50\%, 67\% and 95\% quantiles are indicated by
dashed vertical lines. About half of the halos in each
model have their masses unmodified ($x=0$).
}
\label{Fig:StellMassHaloMassErrCumDist}
\end{figure*}

\section{Results}\label{Sec:Results}

The following subsections present our results for eight different models of 
simulated universes. These models
are detailed in Table \ref{Tab:Models}, where the reference model
involves the most common choices of parameters 
as explained in the previous sections, and the seven remaining
models vary one parameter of the reference model 
at a time. In models 2--4 we change the SMHM
relation; in models 5--8 we explore extreme 
values of other models parameter (often deliberately unrealistic values) to test
the sensitivity of the observations to
these parameters.

\eject 
\subsection{Modified SMHM Relations with Natural Intrinsic Scatter}\label{Sec:Halo-stellMassRESULTS}

As described in \S\ref{Sec:Halo-StellMassRelation} and Appendix~\ref{Sec:Appendix1},
we modified the fixed SMHM relation, selectively reducing the values
of $\Ms$ assigned to halos to enforce a monotonically
increasing $\Ms$ along merger trees. This
leads to the natural dispersion of $\Ms$ values
shown in Figures~\ref{Fig:StellMassHaloMassDist} and \ref{Fig:StellMassHaloMassErrCumDist}
for
the various SMHM relations explored in this paper. For all models, about
half of halos do not have their stellar masses modified (meaning that they
lie on the imposed SMHM relation).
The scatter in the $\Ms$-$\Mhalo$ distributions reaches
about 0.07--0.12~dex in our simulations, similar to the scatter inferred indirectly from theory and observations
\citep[e.g.,][]{reddick2013,behroozi2013}.
The linear SMHM relation (Model 4) is the least affected by our
modifications and shows the smallest dispersion because it lacks the
SMHM non-linearity present in the other models.
For the non-linear Models 1--3, the scatter is smallest near $\Mhalo \sim 10^{12}\Msun$
because the stellar mass corrections are minimized near the peak of the SMHM relation.

\subsection{Simulated Universe Models at Low Redshift}\label{Sec:ComparingZ0}

We evolve all models of simulated universes to the present day.  In this
section we compare the model properties with low-redshift observations to test the
overall accuracy of our method.

\begin{figure}
  \centering
\includegraphics[width=83mm]{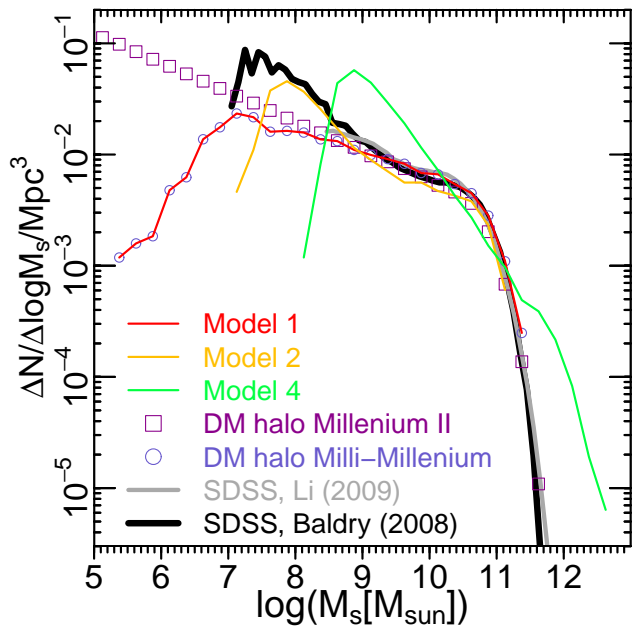}
\caption{
Present day stellar mass function for a simulated universe according
to the different models of stellar mass-halo mass relations tested
in this paper: Model 1 \citep{guo2010}, Model 2 \citep{behroozi2013} and Model 4 (a linear
SMHM relation). The open symbols show the stellar mass functions
at $z=0$ from the milli-Millennium and Millennium II
simulations, computed by converting dark matter halo masses
into stellar masses using the \cite{guo2010} relation. Observed stellar
mass functions measured from local SDSS galaxies \citep{baldry2008,li2009} are also shown.
}
\label{Fig:HaloMassVsStellarMassAndMassFunAtZ0}
\end{figure}

\begin{figure}
  \centering
\includegraphics[width=83mm]{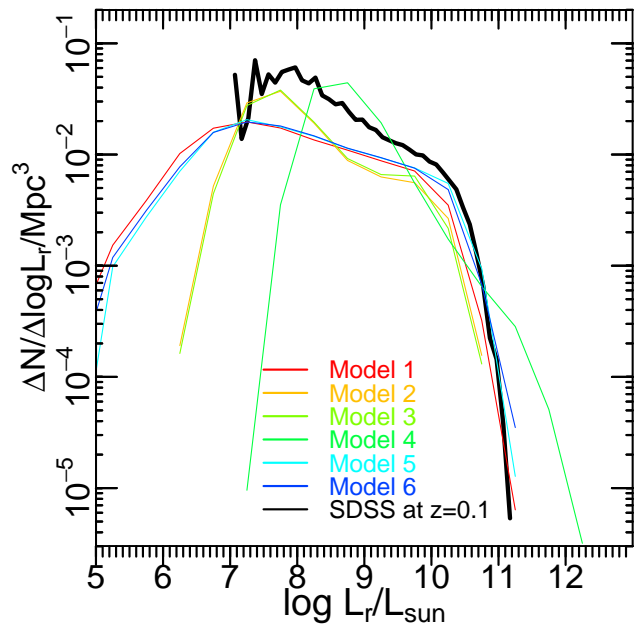}
\caption{
Global luminosity functions ($r$-band) calculated after evolving our simulated
universe to $z=0.1$ under six different models. Also shown is the
luminosity function of local SDSS galaxies measured by \cite{blanton2005}.
Note the similarity with
Fig.~\ref{Fig:HaloMassVsStellarMassAndMassFunAtZ0}. The downturn
in the computed luminosity distributions at low
luminosities is due to the finite particle mass resolution in
the milli-Millennium simulation.
}
\label{Fig:LumFunAtZ0}
\end{figure}

\subsubsection{Stellar Mass Functions}

Figure~\ref{Fig:HaloMassVsStellarMassAndMassFunAtZ0} compares the
stellar mass function of model galaxies at $z=0$ to those derived
from SDSS.  The stellar mass function of the reference 
model, with its \cite{guo2010} SMHM relation, agrees with
the observed stellar mass function from \cite{li2009} in the
range $\Ms> 10^{9}\Msun$. That is expected since this SMHM
relation was derived from halo abundance matching on
mostly the same data.
We also confirm that the stellar mass function predicted by the reference model fits perfectly the stellar mass function derived by combining
the Guo et al.\ 
SMHM relation with the halo mass function from the milli-Millennium simulation, as required by our semi-empirical modeling. 
However, the stellar mass function of the reference model has an artificial downturn at $\Ms<10^7
\Msun$, which corresponds to the mass resolution of the milli-Millennium simulation in the
identification of friends-of-friends groups (composed of a minimum of 20 dark matter particles).
Similar downturns are present in the stellar mass functions of
the other SMHM relations tested. In contrast, the Millennium II simulation \citep{boylan-kolchin2009}, which has a dark matter particle
mass 125 times smaller than the mMS, shows a power-law tail at the low-mass end.

The halo abundance matching in
the Guo model covered the range $10^{8.3}<M_{\rm
s}/\Msun<10^{11.8}$, which misses the upturn
at $\Ms<10^{9}\Msun$ that is seen in
the more complete SDSS stellar mass function from \cite{baldry2008}
(shown by the black line in Fig.~\ref{Fig:HaloMassVsStellarMassAndMassFunAtZ0}).
The SMHM relation from \cite{behroozi2013}
(Model 2, yellow line) fits this upturn much better since it was built using the
\cite{baldry2008} stellar mass function.


The linear SMHM relation (Model 4) tracks the
approximately $M^{-2}$ power-law mass distribution of dark matter halos.
The proportionality constant for Model 4 was chosen to roughly match
the observed
stellar mass function around its knee at $\Ms\sim M^{*}$.
Again, a clear mass resolution cutoff is present at lower stellar
masses, also shown in the non-linear relations.
This linear SMHM model is obviously in strong conflict with the
observations at both higher and lower masses.

\subsubsection{Luminosity Functions}

Figure~\ref{Fig:LumFunAtZ0} compares the simulated and observed luminosity functions
at $z=0.1$. 
To first order, the shapes are Schechter functions.
The present-day 
luminosity functions from our models are similar in shape to their respective stellar mass functions
in Figure~\ref{Fig:HaloMassVsStellarMassAndMassFunAtZ0}. This is
expected since the $r$-band luminosity roughly traces the old stellar
population that constitutes most of the stellar mass in present-day galaxies
(e.g., \cite{bell2003}). Models 5 (no dust) and 6 (low-metallicity) show good agreement with
the \cite{blanton2005} observed luminosity function in the high-luminosity tail.
Since our reference model contains dust and metals, its high-luminosity tail is shifted toward fainter 
luminosities with respect to those of Models 5 and 6.
(Note that the luminosity in the $r$-band is reduced by dust absorption even though the
bolometric luminosity is conserved.)
On the other hand, the flatter slope at low luminosities for these three models 
mimics the flat tail present in their corresponding stellar mass
functions, which is the result of applying halo abundance matching to the stellar
mass function of \cite{li2009} as explained earlier. A better fit
is obtained for Models 2 and 3 with the SMHM relation from \cite{behroozi2013},
which includes the low-luminosity upturn.
The luminosity function derived from the linear SMHM relation for model galaxies tracks the
approximately power-law mass distribution of dark matter halos.

\begin{figure*}
  \centering
  \begin{tabular}{cc}
    \includegraphics[width=84mm]{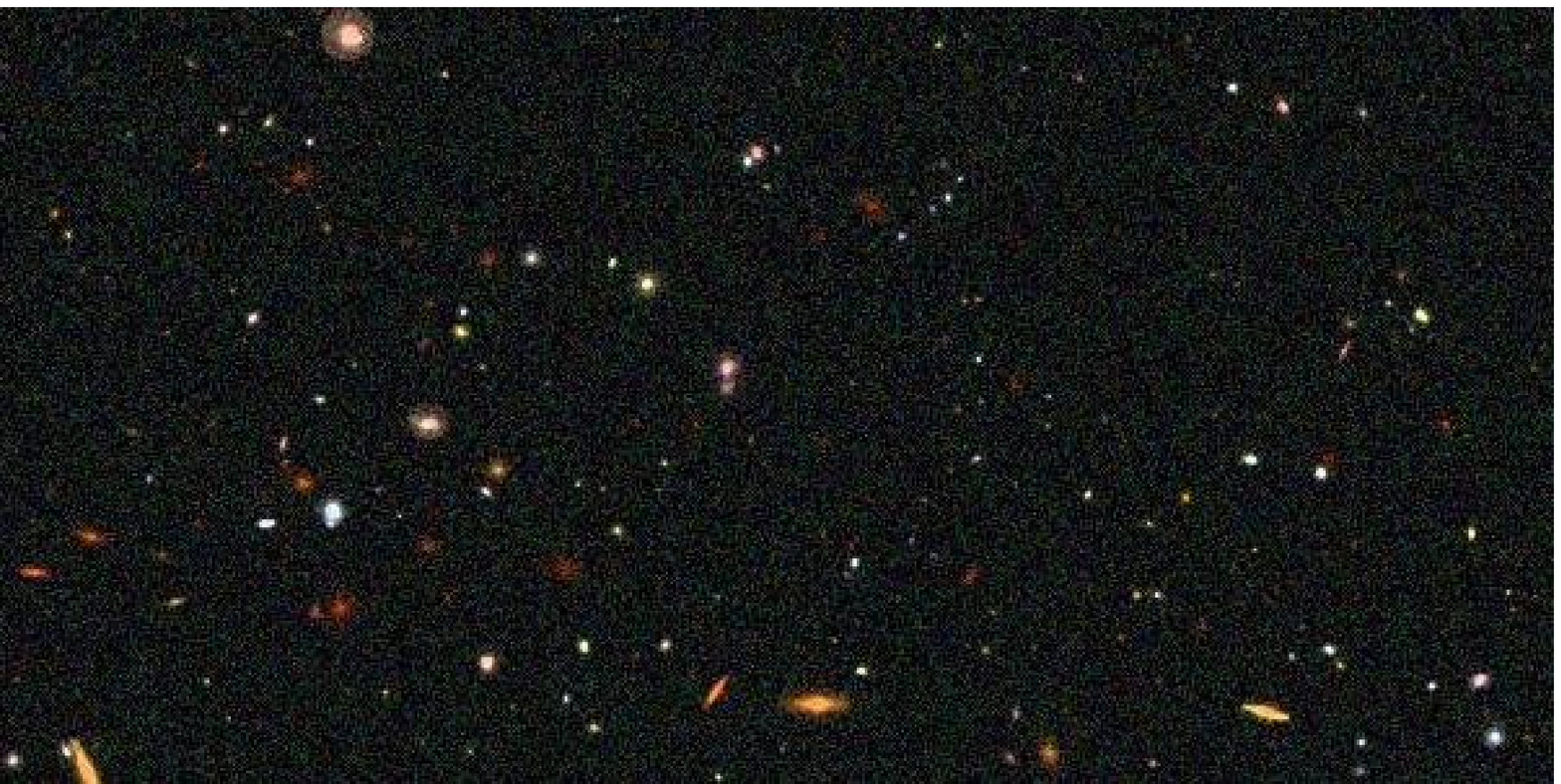}&
    \includegraphics[width=84mm]{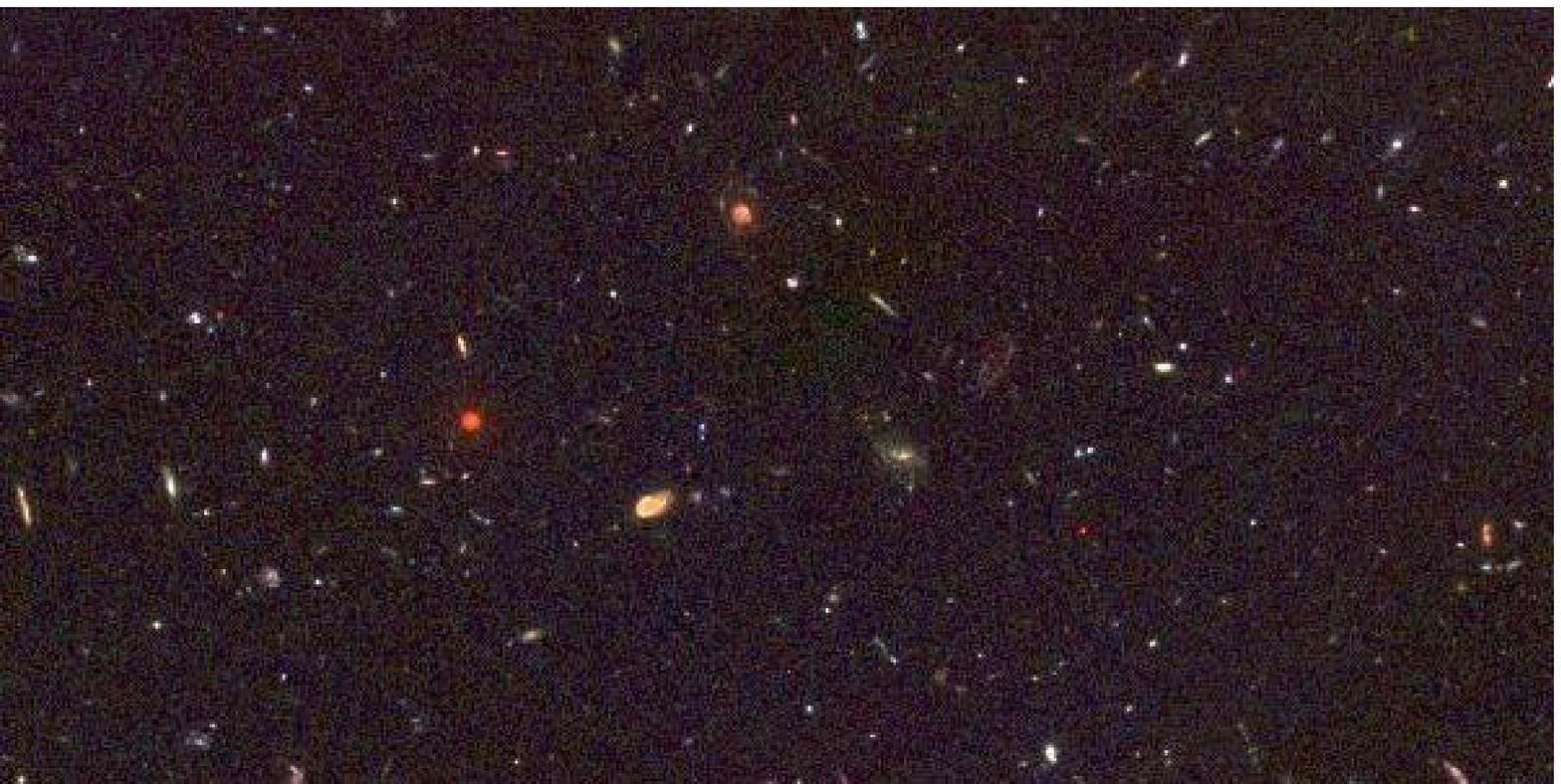}\\
Simulated GOODS Image & Real GOODS image  \\
Reference Model & \\[7mm]
    \includegraphics[width=84mm]{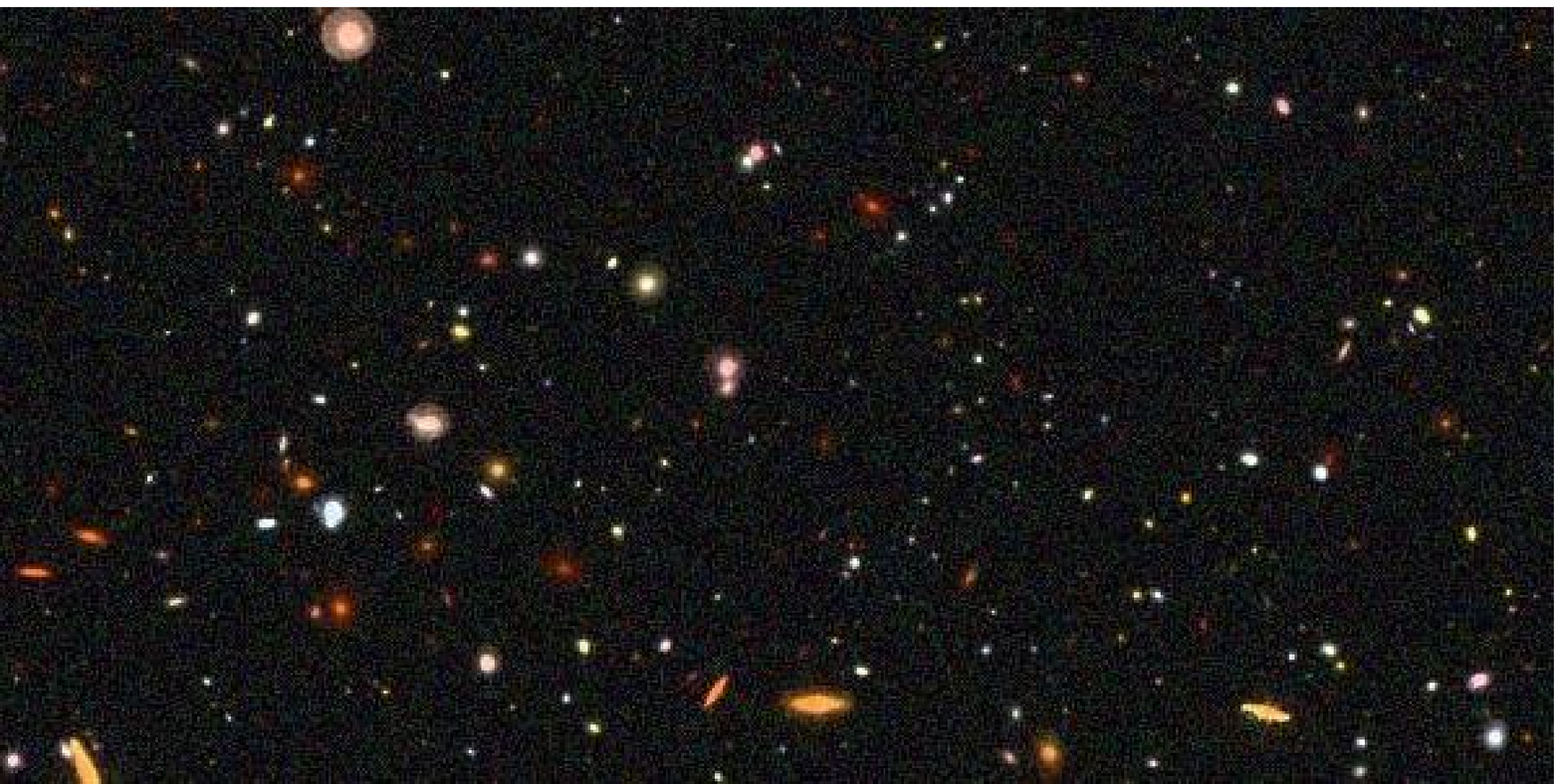}&
    \includegraphics[width=84mm]{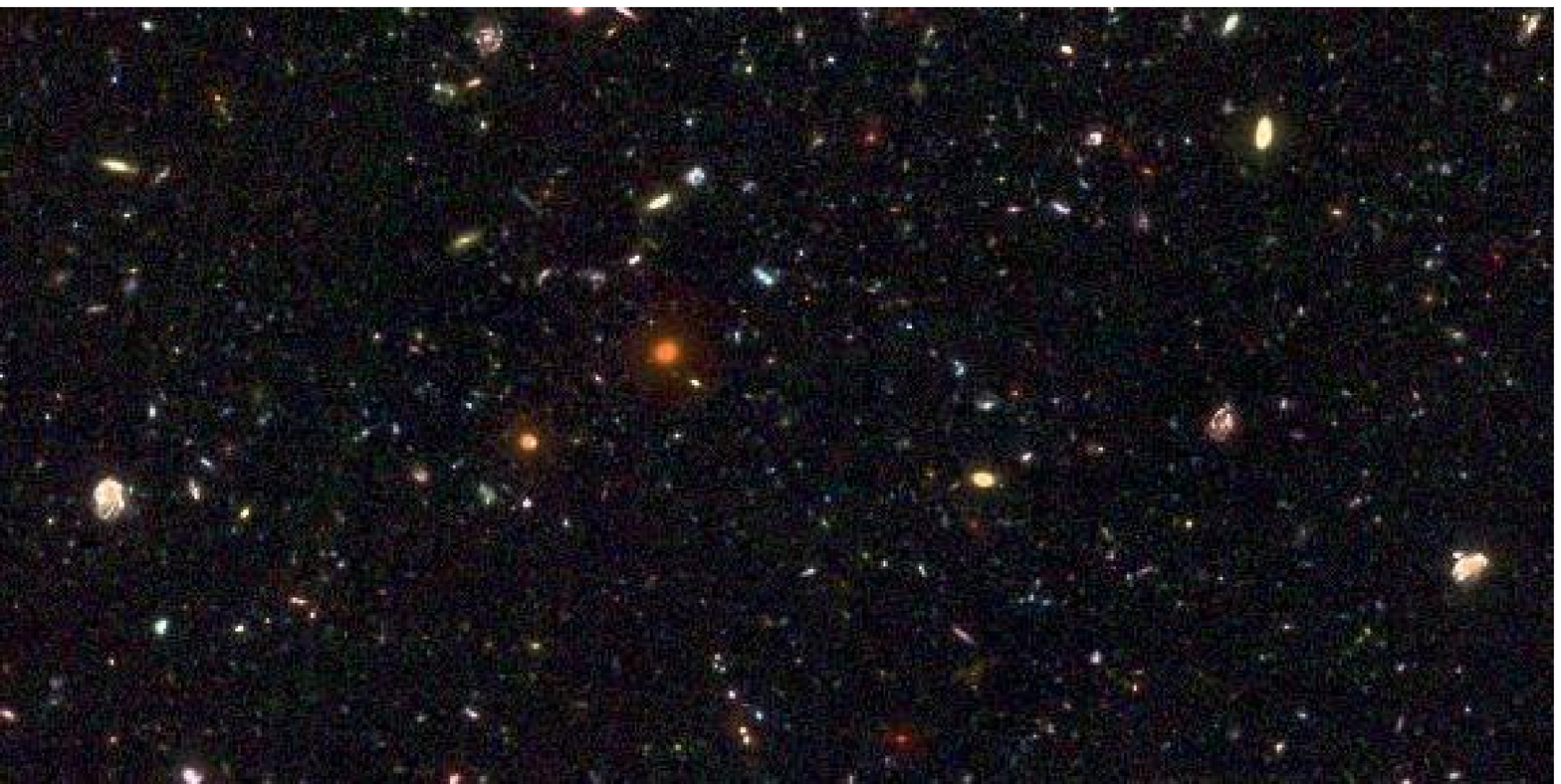}\\
Simulated HUDF Image & Real HUDF image  \\
Reference Model & \\[7mm]
  \end{tabular}
\caption{Simulated ACS/WFC F850LP+F606W+F435W images built from our
reference model (GOODS and HUDF depths), compared to the
equivalent real \textit{HST\/} images. The same exposure times and display
contrast are used for the comparison at each depth. The field of
view is $2400\times1200$ ACS pixels ($\sim1/6$ of the full ACS/WFC field of view).
}
\label{Fig:SMimages}
\end{figure*}

The perceptive reader may have noticed that even though semi-empirical modeling 
promises a perfect match of the predicted to observed universe, 
the global luminosity function predicted by our reference model does not agree perfectly with 
the observed luminosity function from SDSS at $z\sim 0$. 
The reason for this is that the $z\sim0$ SMHM relation from \cite{guo2010}
is not fully consistent with our method for converting 
stellar mass into light. They use the stellar mass function derived from 
\cite{li2009}, which is not directly observed but is inferred from the observed SDSS galaxy 
luminosity function at $z\sim 0$ and an assumed mass-to-light ratio. 
On the other hand, our mass-to-light ratio is computed
from the dark matter halo masses and the SMHM relation using
\cite{bruzual2003} spectral models.
We can in principle address this issue by
computing our own SMHM relation using an iterative process that compares our measured
luminosity functions to the observations.
This level of precision is not needed in our present exploratory study, but it will be
a useful longer-term goal for our approach.

\subsection{Simulated \textit{HST\/} Images and Derived Statistics}\label{Sec:SimImageAndStatistics}

In this section, we show simulated \textit{HST\/} images from our models and test the sensitivity
of statistics derived from the images to the model parameters.
We focus on the luminosity and size distributions, and we assess both biases in measured
parameters and source
detection incompleteness. For most tests we compare the perfectly known \textit{input} values of size and luminosity (as given by the models)
to the corresponding \textit{output} values measured by {\tt SExtractor} from the images.
Our data comprise simulated visible (ACS/WFC) and infrared
(WFC3/IR) images using filters and exposure times appropriate for the GOODS, CANDELS and HUDF surveys.
Image sizes and pixel scales are those of a single ACS/WFC exposure
(a $200\arcsec \times 200\arcsec$ field with 0.0495\arcsec\ pixels).  Note that this image
area is small compared with the areas surveyed by many \textit{HST\/} projects (e.g., GOODS and CANDELS),
which encompass many ACS/WFC fields; we considered it appropriate to begin with a more modest
sky area for this exploratory project. 

\subsubsection{Results from the Reference Model}\label{Sec:ResultsFromStandardModel}

Figure~\ref{Fig:SMimages} shows a comparison between reference-model simulated and real \textit{HST\/} ACS/WFC images.
At first glance, the spatial distribution of galaxies and associated
sizes seem to be very similar. As expected, the HUDF-depth images show that many
objects are hidden by noise in the GOODS-depth simulated images.

\begin{figure}
\begin{center}
\begin{tabular}{c}
   \includegraphics[width=70mm]{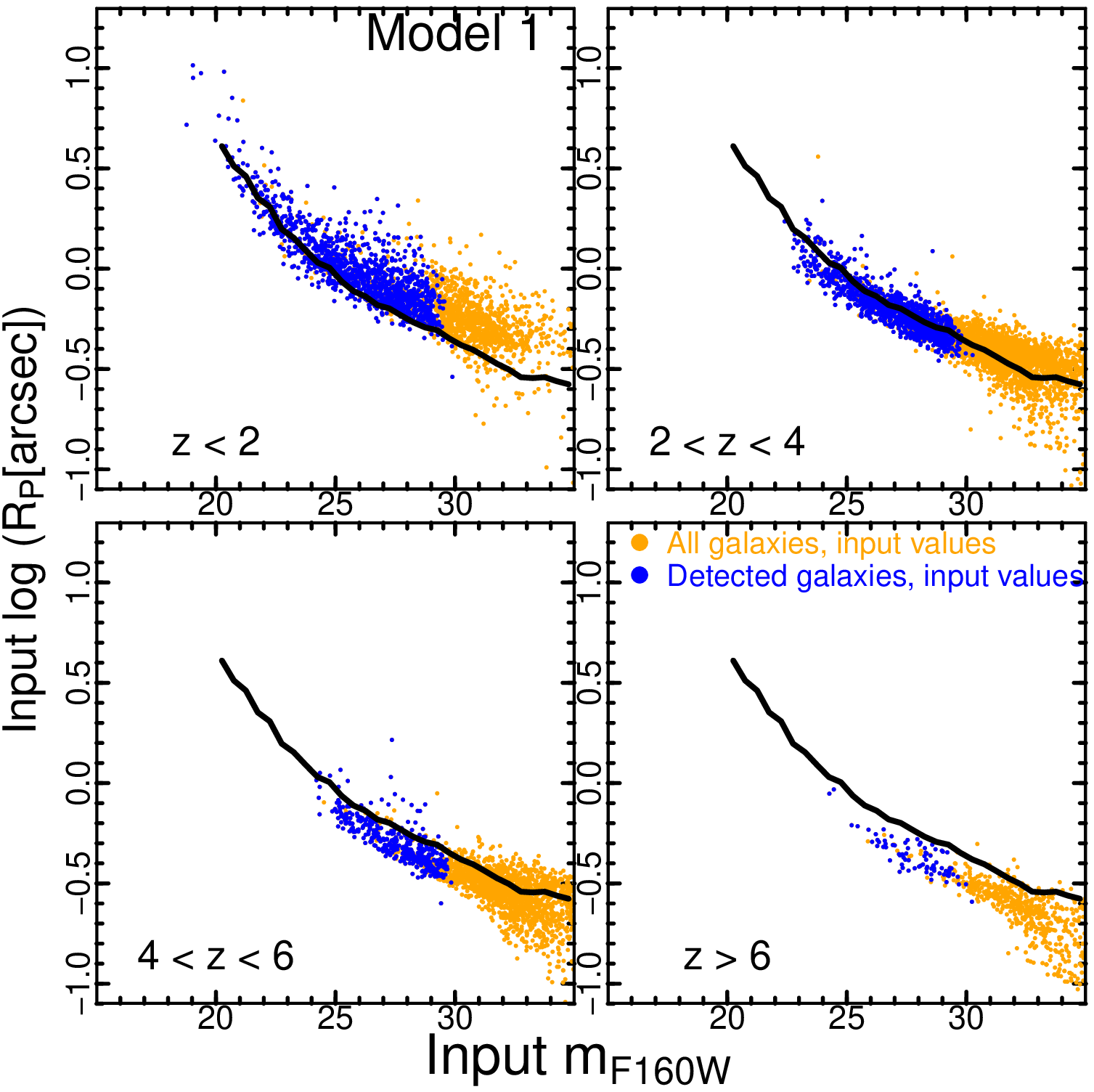}\\
   \includegraphics[width=70mm]{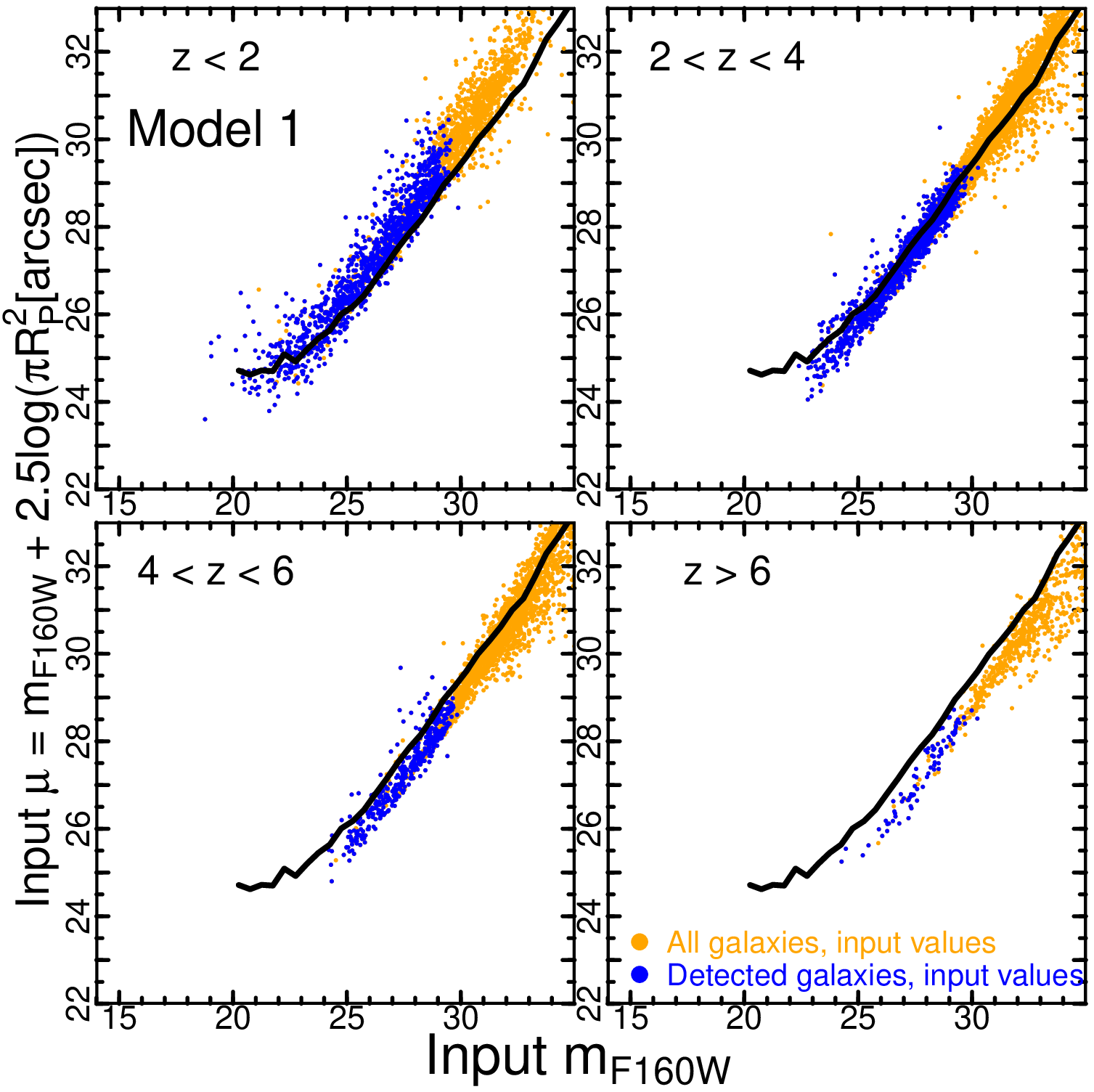}\\
\end{tabular}
\caption{
\textit{Top\/}: Log Petrosian radius versus apparent F160W magnitude for
the HUDF-depth simulated \textit{HST\/} image derived from our reference 
model (Model 1), separated in four redshift bins.
Blue points show galaxies detected by {\tt SExtractor};
orange points show undetected galaxies.
For all galaxies, the input (model) radii and magnitudes are shown
rather than values measured from the images.
The black line is the same in all four panels and shows the median
of the $\log R_{\rm P}$ versus magnitude distribution
integrated over all redshifts.
\textit{Bottom\/}: Apparent surface brightness $\mu$ versus magnitude
using the same color coding.}
\label{Fig:ApMagVsPetroRad}
\end{center}
\end{figure}

Figure~\ref{Fig:ApMagVsPetroRad} shows scatter plots of the \textit{input}
(true model) values of the apparent sizes and surface brightnesses
of both detected (using {\tt SExtractor}) and undetected
galaxies in the HUDF-depth image.
Most of the detected galaxies
can be selected via a cut at $m_{\rm F160W} \lesssim 29$,
which does not depend strongly on redshift.
In the lower panels of Figure~\ref{Fig:ApMagVsPetroRad}, which plot
surface brightness versus magnitude, it can be seen that there
is a small, redshift-dependent population of low surface brightness galaxies
that are brighter than the magnitude threshold but nonetheless are not detected.

\begin{figure}
\begin{center}
   \includegraphics[width=70mm]{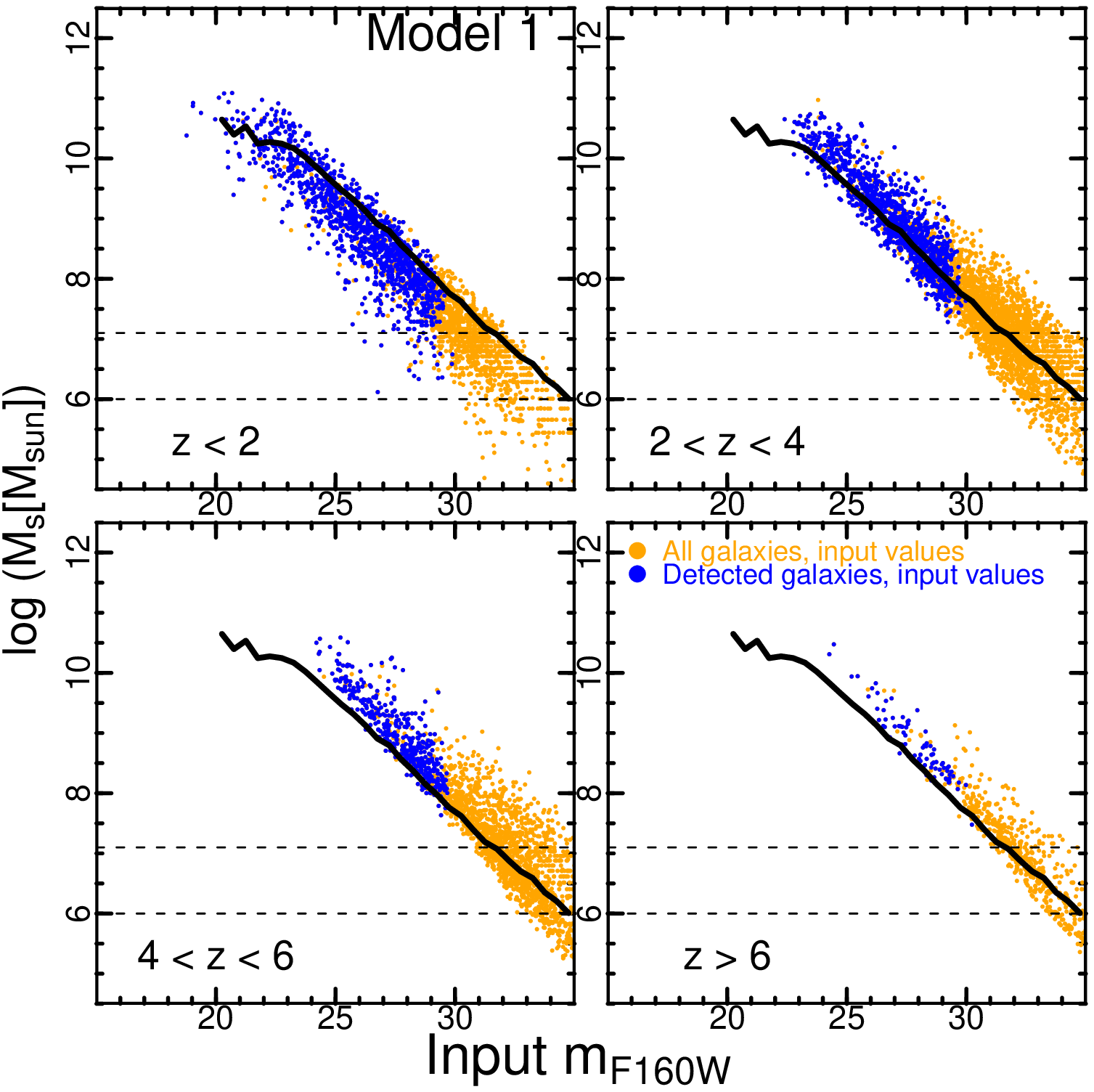}
\caption{
Stellar mass $\Ms$ versus apparent F160W magnitude for
the HUDF-depth simulated \textit{HST\/} image derived from our reference 
model (Model 1), separated in four redshift bins. Color coding is
the same as in Fig.~\ref{Fig:ApMagVsPetroRad}. The lower
dashed line shows the smallest stellar mass for SDSS galaxies
in the catalog of \cite{kauffmann2003} ($\Ms = 10^6\Msun$). The upper dashed line
shows the stellar mass threshold for halos resolved by the milli-Millennium
simulation ($\Ms=10^{7.1}\Msun$) using the
\cite{guo2010} SMHM relation.  Note that images
from our reference model observed at the HUDF depth are not
affected by either threshold, since
galaxies located below the thresholds are too faint to detect.} 
\label{Fig:ApMagVsStellMass} 
\end{center}
\end{figure}

\begin{figure}
\begin{center}
   \includegraphics[width=70mm]{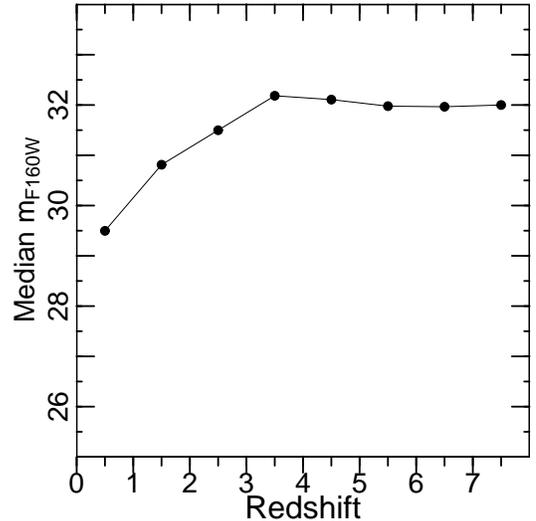} 
\caption{
Median apparent F160W magnitude as a function of redshift
using our reference model for
galaxies with stellar masses
near the milli-Millennium stellar mass limit of
$\Ms = 10^{7.1}\Msun$ (upper dashed line in
Fig.~\ref{Fig:ApMagVsStellMass}).
The curve flattens at $z>3$ where the youth and high star-formation
rates of galaxies compensate for cosmological dimming.
}
\label{Fig:MedianMagAt20Mass} 
\end{center}
\end{figure}

\begin{figure}[]
\begin{center}
\begin{tabular}{c}
    \includegraphics[width=65mm]{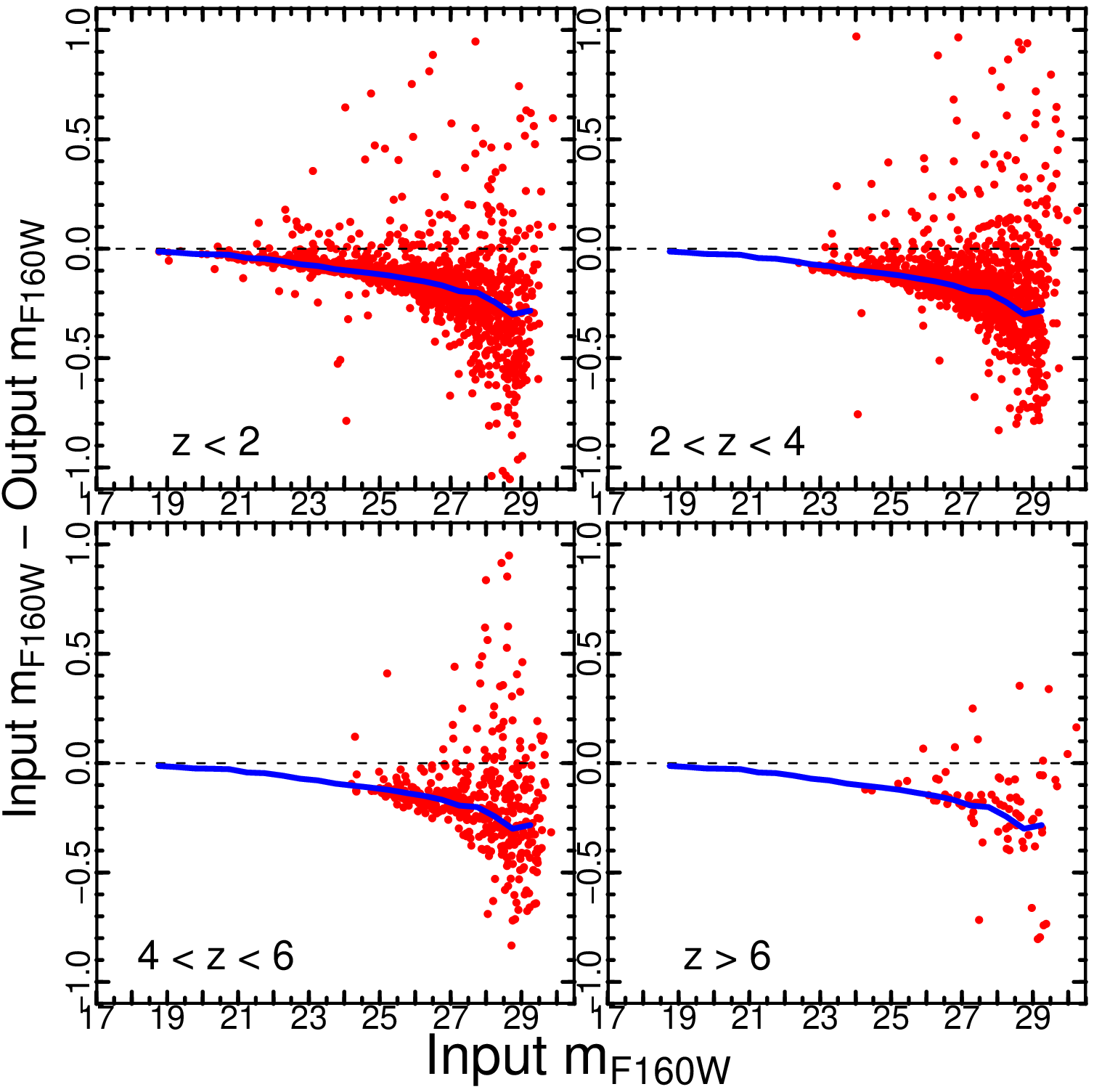}\\
    \includegraphics[width=65mm]{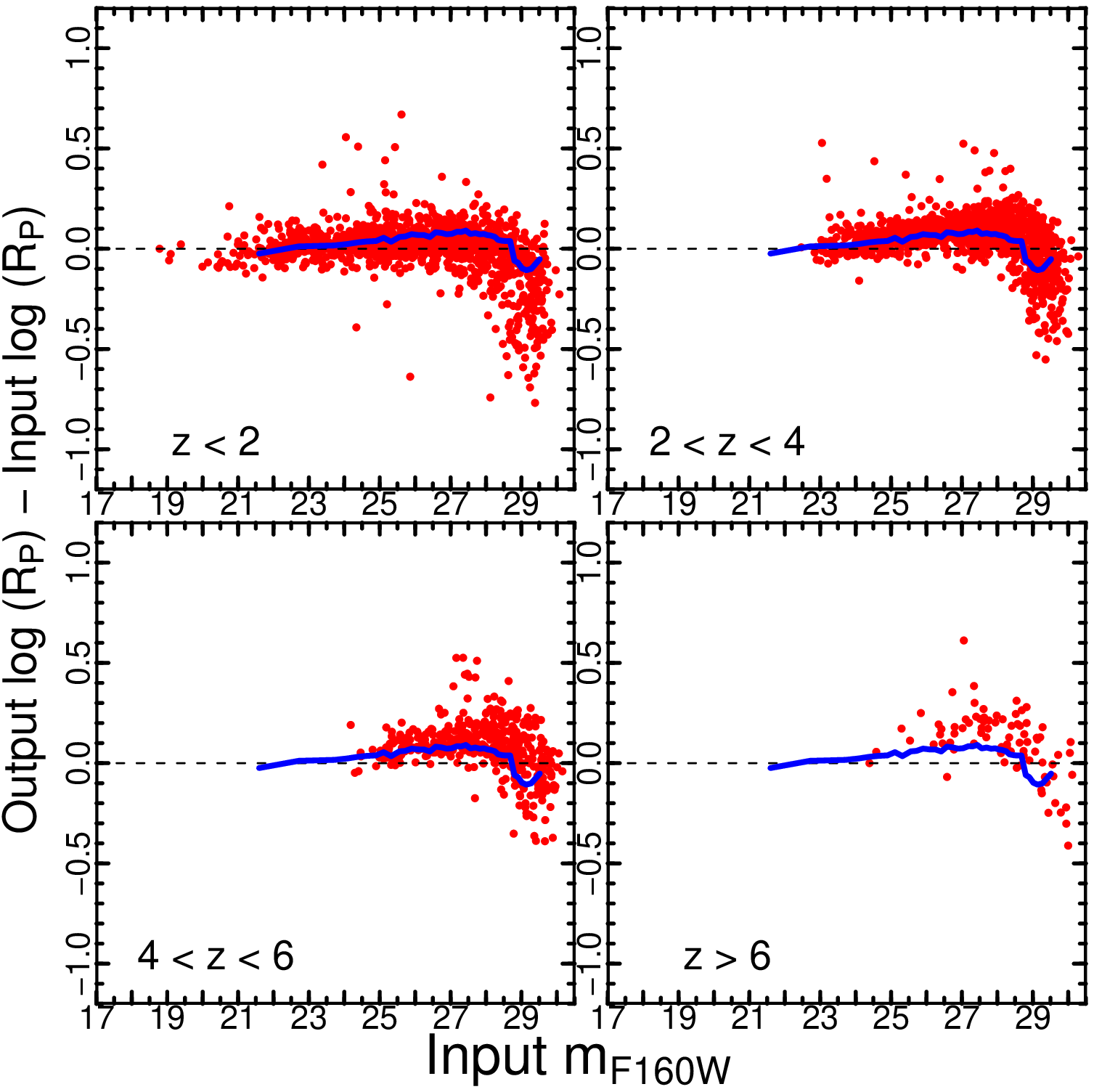}\\
\end{tabular}
\caption{
\textit{Top\/}: Magnitude biases: the difference between the true (input) $m_{\rm F160W}$
magnitude of galaxies and the {\tt SExtractor}-measured output
magnitude as a function of the input magnitude.
The results for four different redshift ranges in HUDF-depth simulations are shown.
Blue lines show the median magnitude differences over all redshifts
and are the same in each panel.  Points below the dashed line have measured magnitudes
that are fainter than the true magnitudes.  Fluxes are systematically
underestimated for fainter galaxies.
\textit{Bottom\/}: Size biases: same as \textit{top}, but showing the difference between the
true input $\log R_{\rm P}$ sizes and the {\tt SExtractor}-measured
sizes. There is a slight bias toward larger sizes for brighter galaxies, while fainter galaxies
have measured sizes that are substantially underestimated.  
Only the high-luminosity cores of faint galaxies are detected, while the remaining
extended emission is lost beneath the noise.}
\label{Fig:InputVsDeltaOutput} 
\end{center}
\end{figure}

\begin{figure}
\begin{center}
\begin{tabular}{cc}
   \includegraphics[width=40mm]{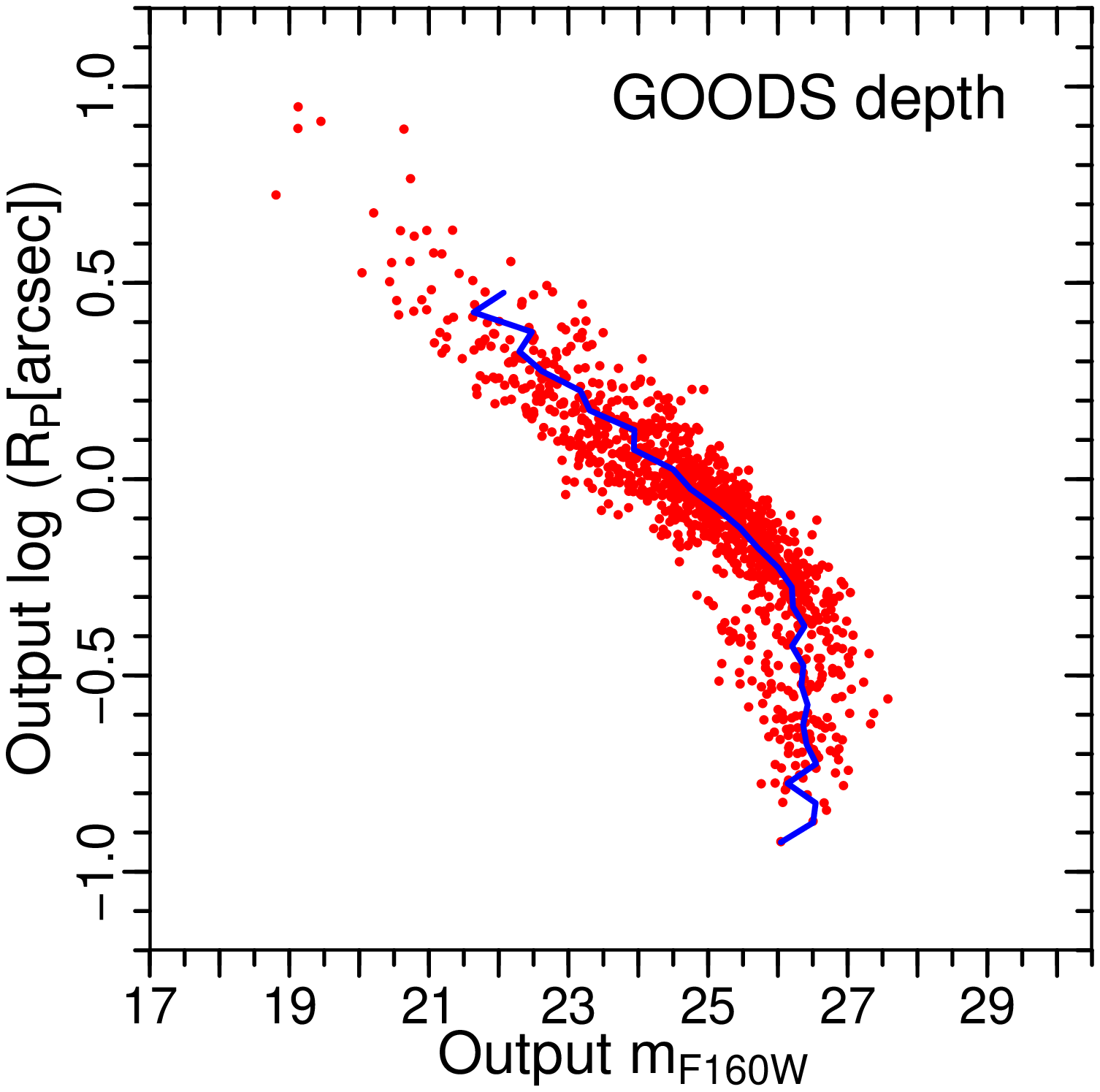}&
   \includegraphics[width=40mm]{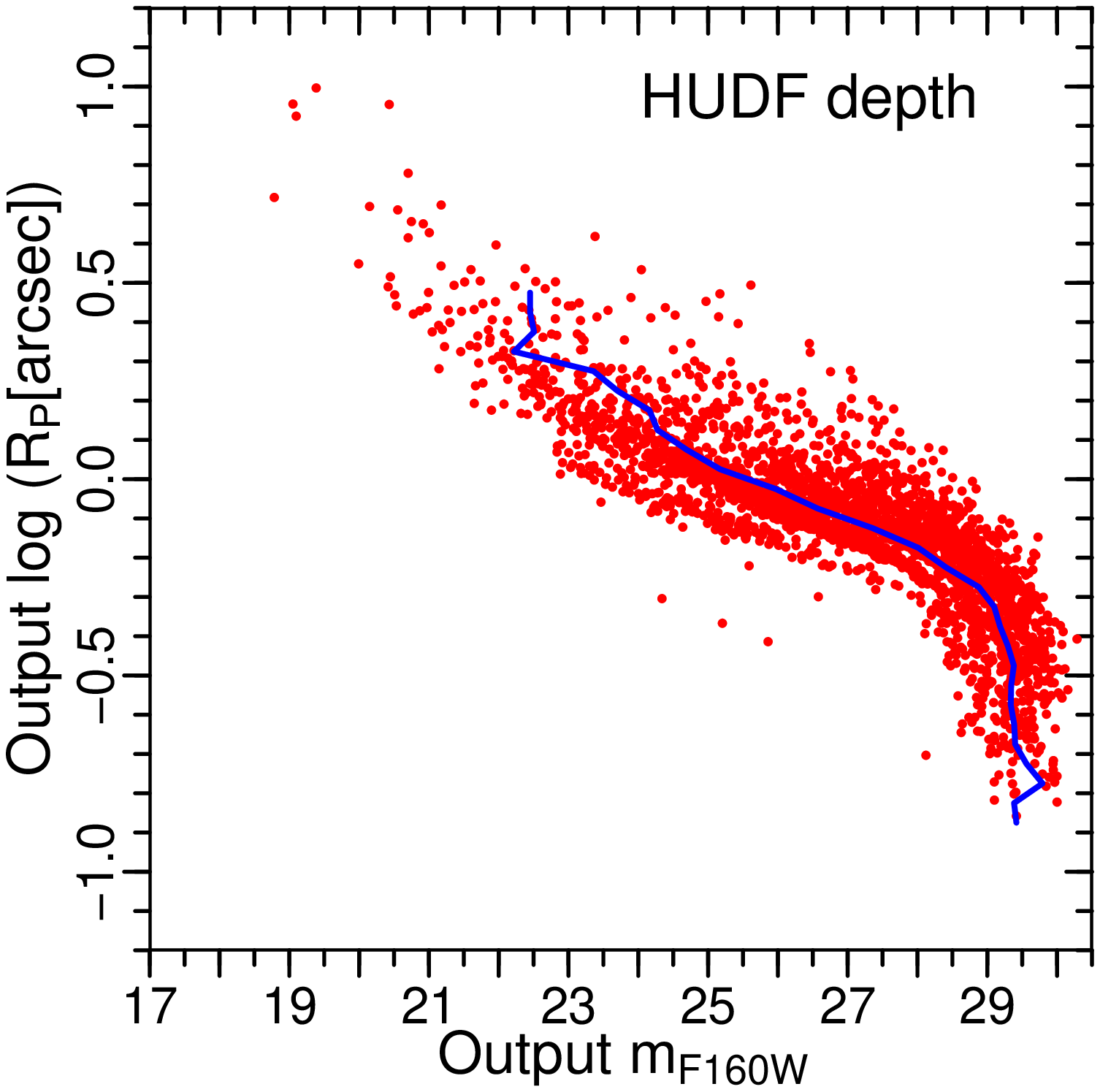}\\
\end{tabular}
\caption{
Measured (output) size $\log R_{\rm P}$ as a function of measured magnitude $m_{\rm F160W}$
for detected galaxies.
The left and right panels show results
from simulated images at GOODS and HUDF depth, respectively.
The blue lines show the median trend.
Note that the tendency to underestimate galaxy sizes
at the magnitude completeness limit is evident at both depths, which
demonstrates that it is not the product of any peculiarity in the reference 
model.
}
\label{Fig:OutputMagVsOutputRad} 
\end{center}
\end{figure}

\begin{figure}[h]
\begin{center}
\begin{tabular}{c}
    \includegraphics[width=79mm]{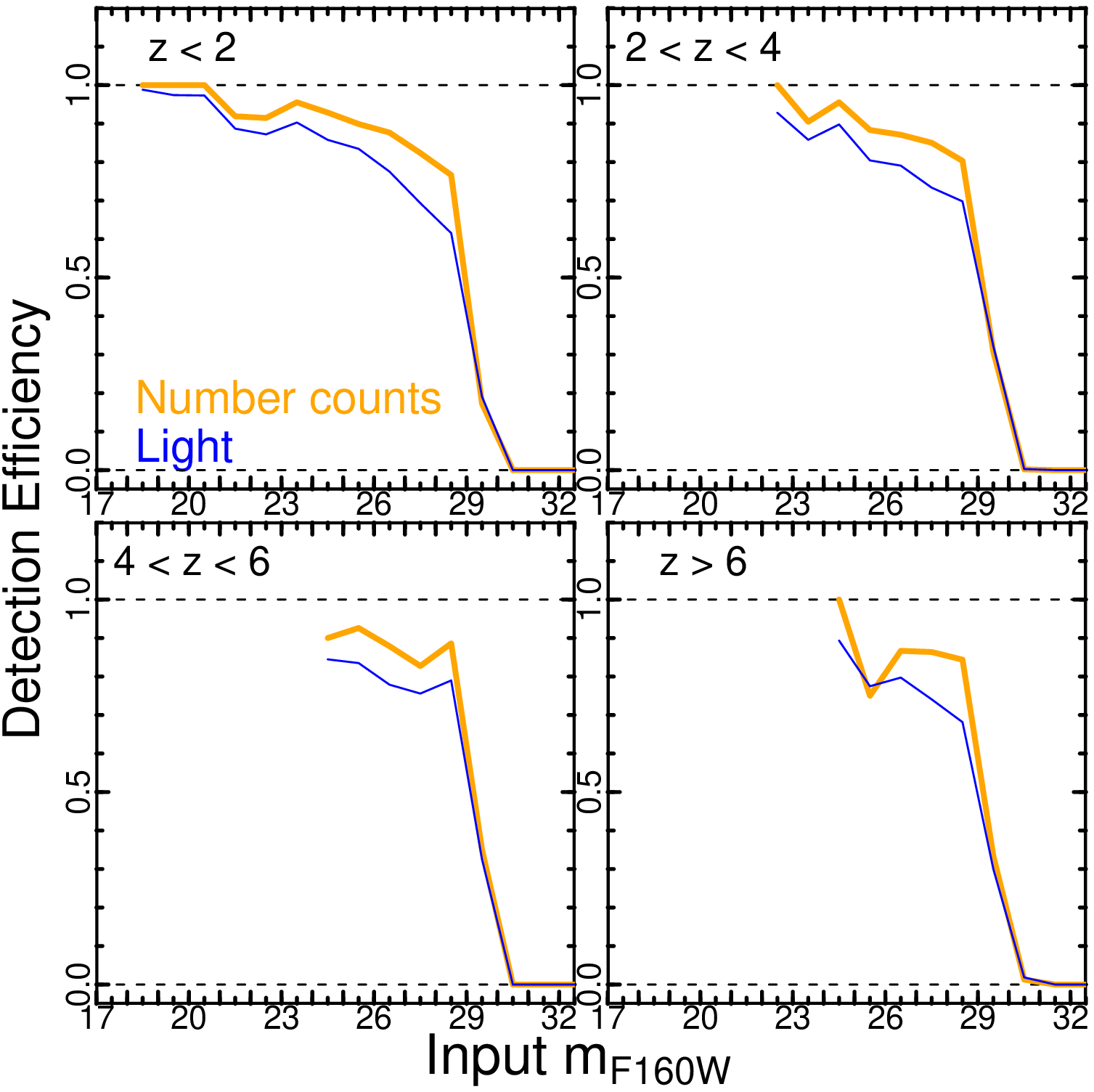}\\
    \includegraphics[width=79mm]{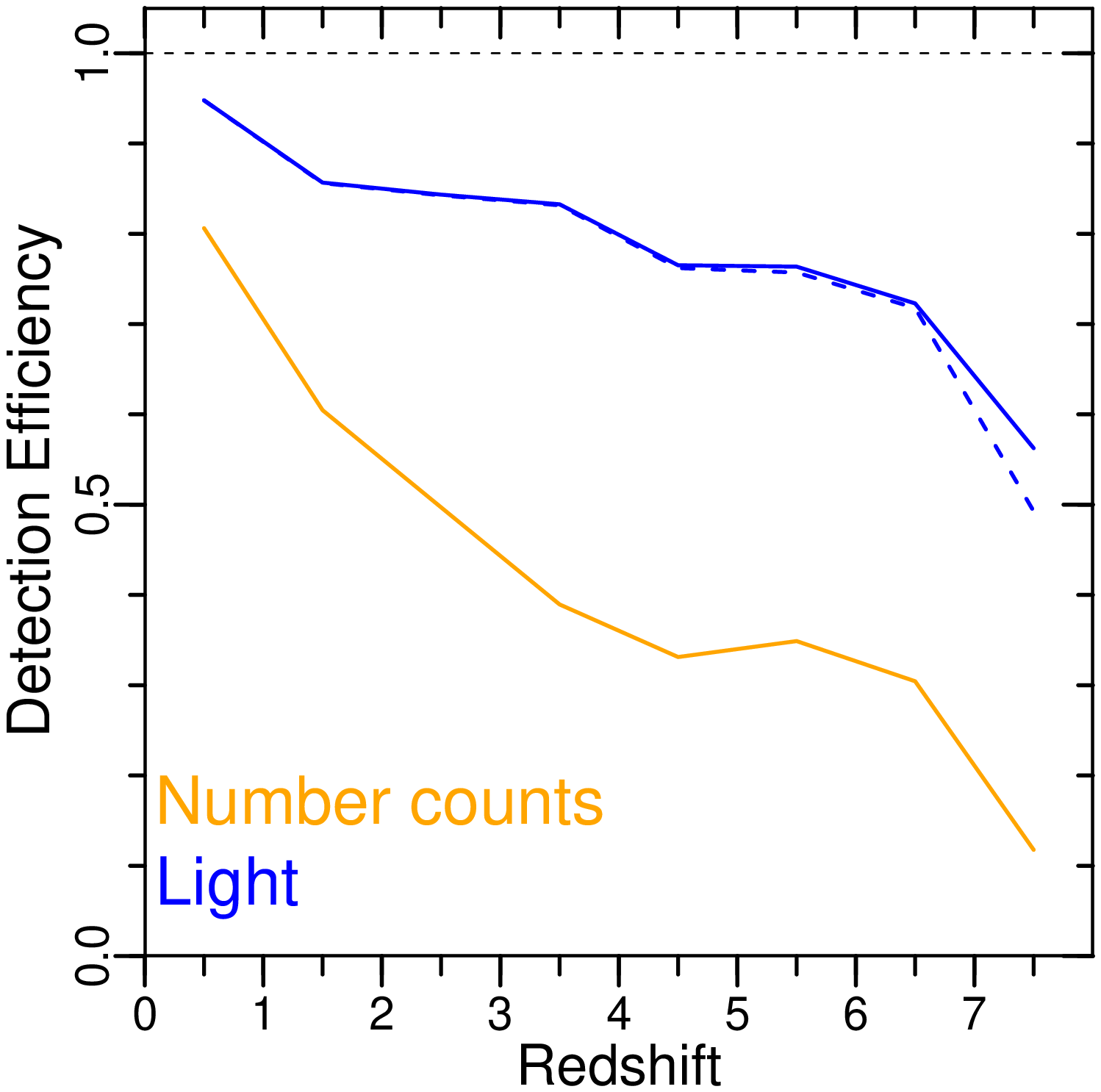}\\
\end{tabular}
\caption{
\textit{Top\/}: {\tt SExtractor} detection efficiency as a function
of input $m_{\rm F160W}$, measured on a simulated \textit{HST\/} image at HUDF
depth and built from our reference model (Model 1). The number
count detection efficiency is the ratio of the number of galaxies
detected by {\tt SExtractor} to that of all the galaxies placed in
the image. The light detection efficiency is the ratio of the output
fluxes (F160W band) of detected galaxies as measured by {\tt
SExtractor} to the corresponding model input fluxes of all galaxies
placed in the image. Note that light efficiency is smaller than
that for number counts because galaxies are not only missed but also
have their fluxes underestimated.
\textit{Bottom\/}: Efficiencies as a function of
redshift. The solid lines include only galaxies brighter than the
median $m_{\rm F160W}$ at the stellar mass incompleteness limit
(Fig.~\ref{Fig:MedianMagAt20Mass}). The dashed line uses a
power-law extrapolation beyond the mass limit to estimate the
light of faint galaxies (Fig.~\ref{Fig:LumFunSMredshiftBin}).
}
\label{Fig:EfficiencySM} 
\end{center}
\end{figure}

\begin{figure}
\begin{center}
   \includegraphics[width=85mm]{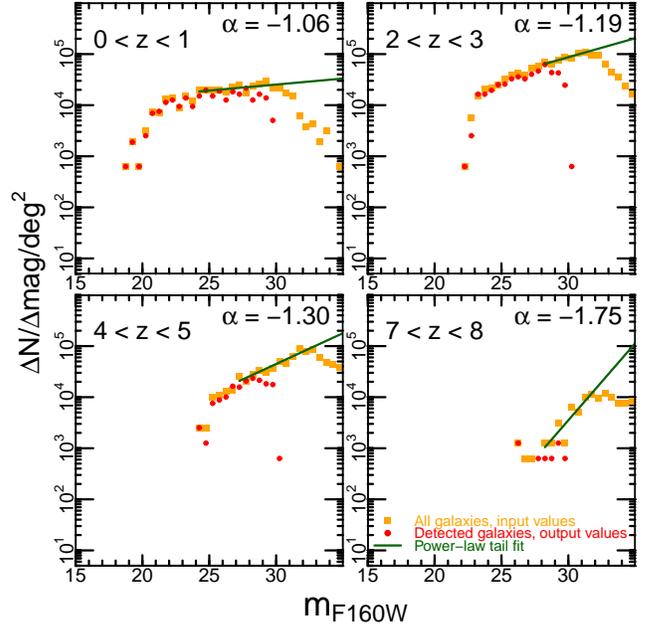}
\caption{
Apparent magnitude distribution of our reference 
model for HUDF depth 
separated in redshift bins. The green line shows a power-law low-luminosity
tail fitted with slope $\alpha$. The power-law slope at the faint
end tail becomes steeper with increasing redshift.
}
\label{Fig:LumFunSMredshiftBin} 
\end{center}
\end{figure}

\begin{figure}
\begin{center}
\begin{tabular}{cc}
    \includegraphics[width=40mm]{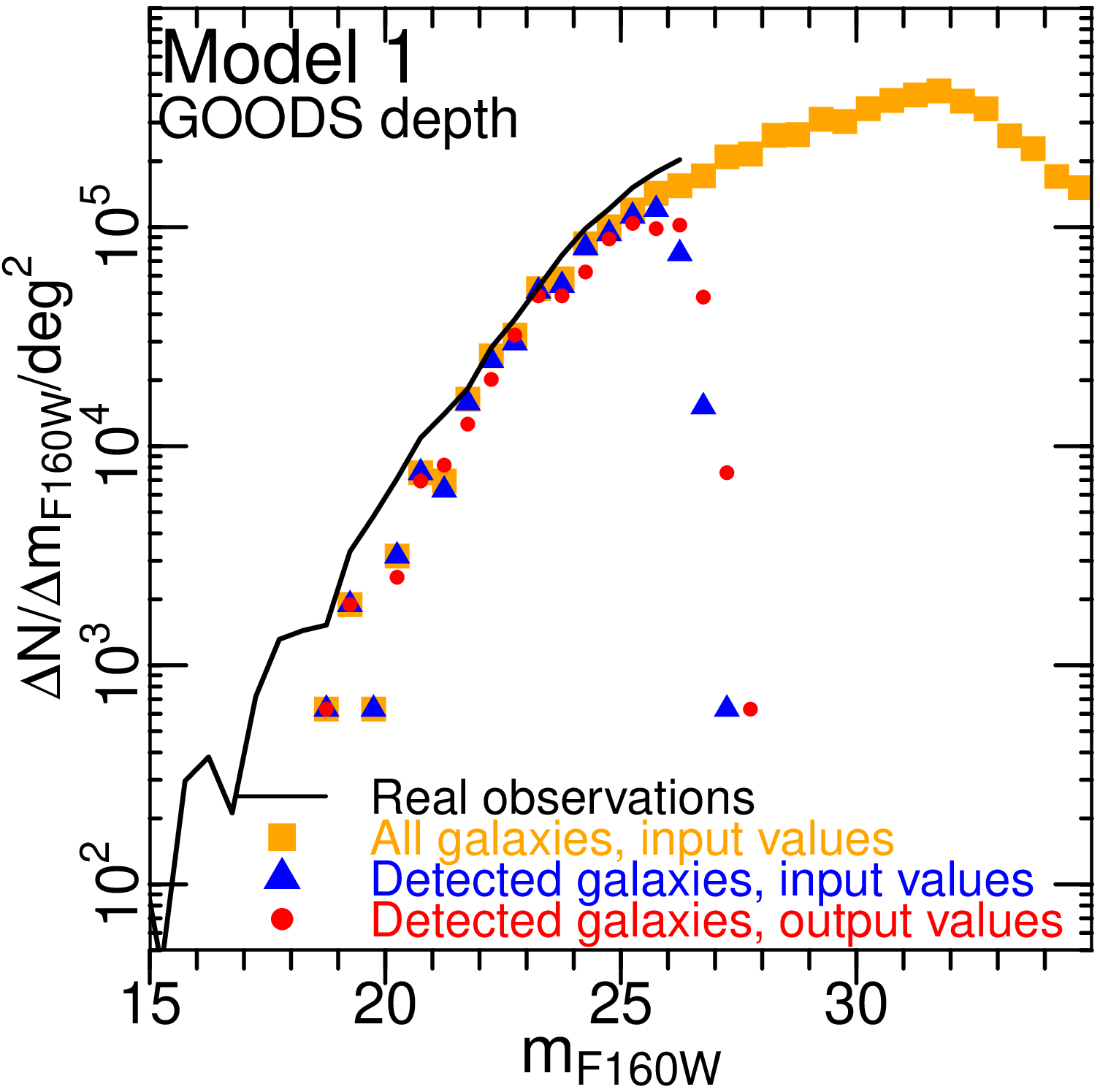}&
    \includegraphics[width=40mm]{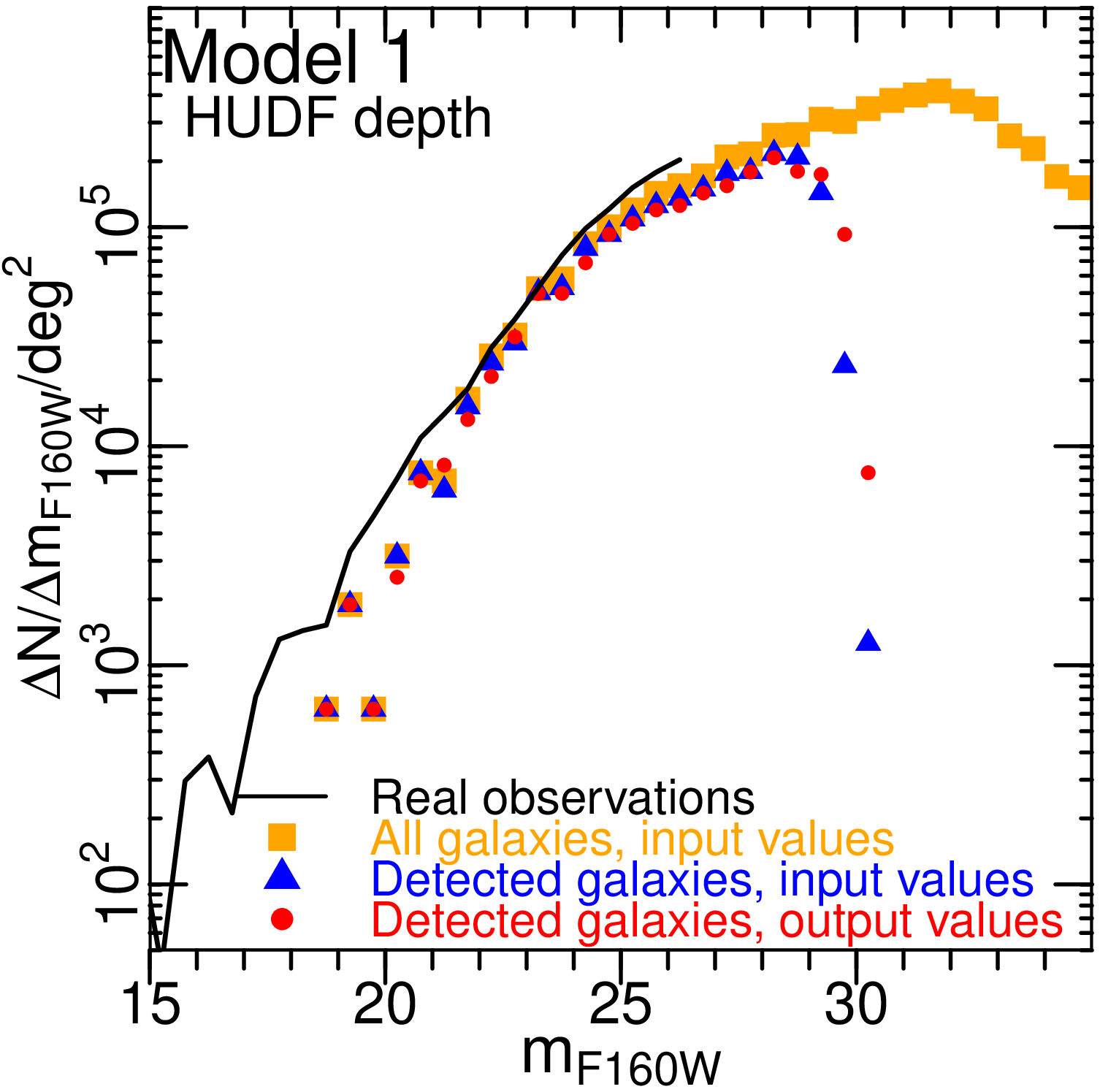}\\
[2mm]
    \includegraphics[width=40mm]{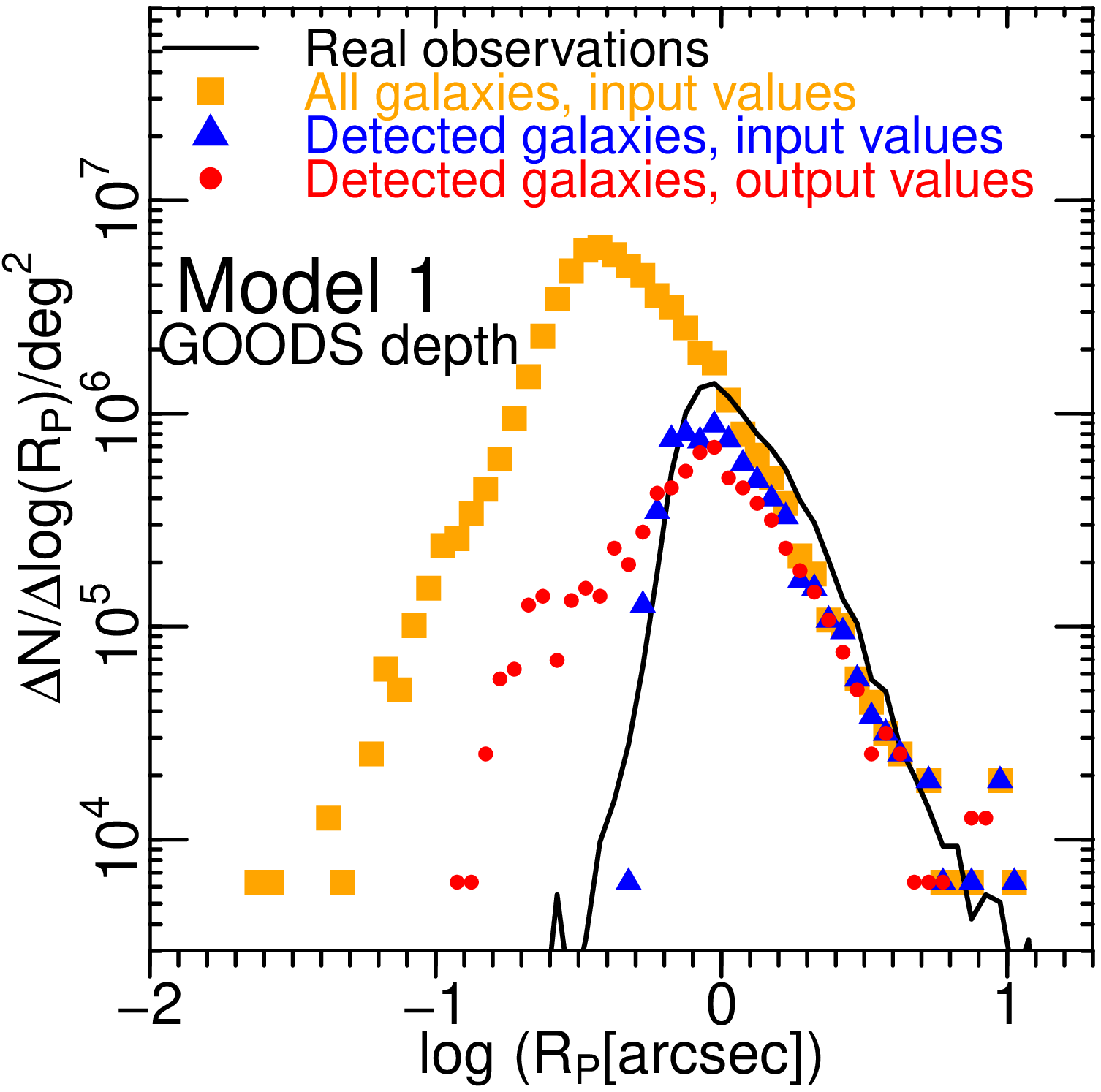}&
    \includegraphics[width=40mm]{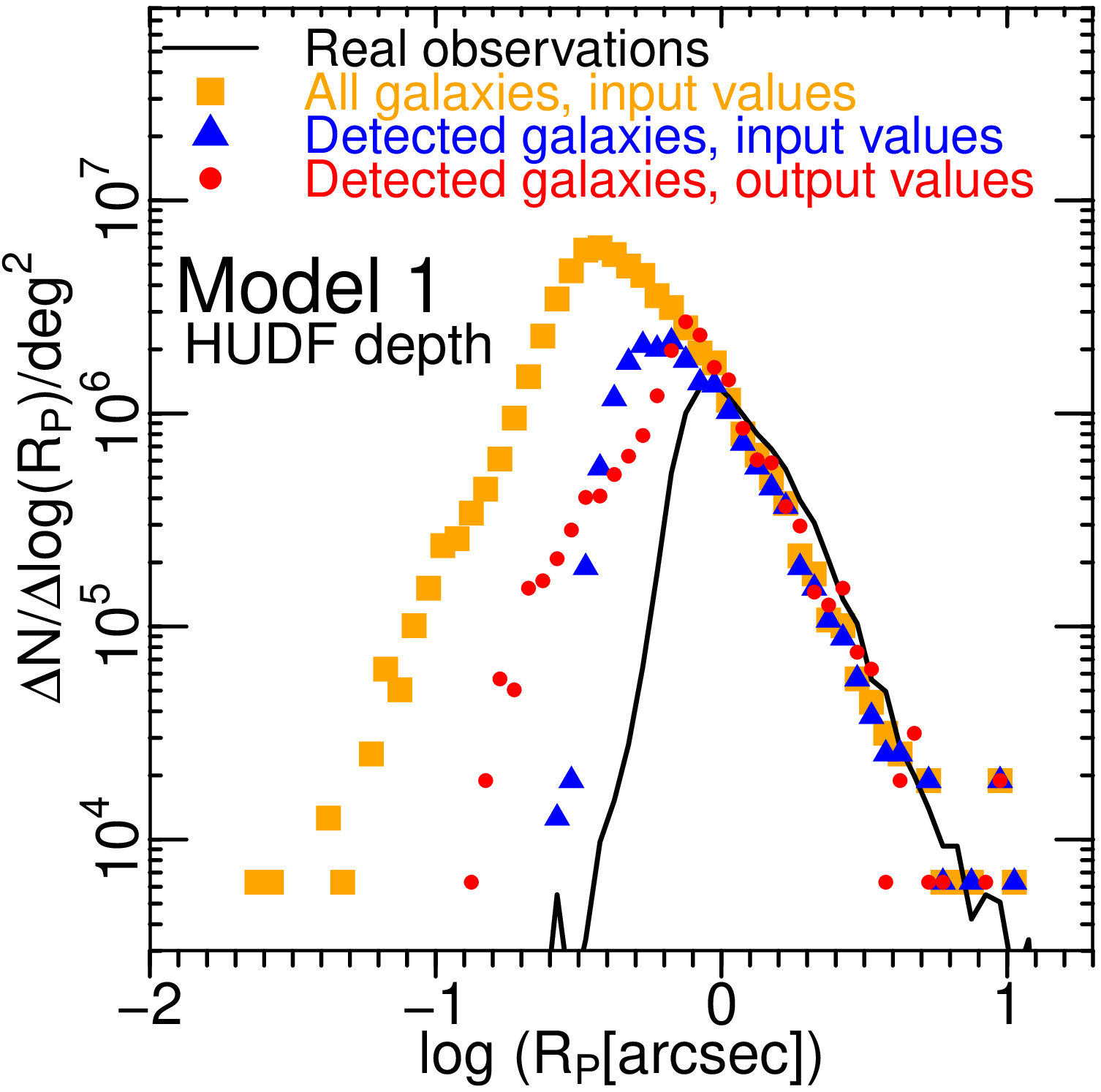}\\
[2mm]
\end{tabular}
\caption{
Galaxy number counts as a function of 
apparent F160W magnitude (top) and Petrosian radius
(bottom) from simulated \textit{HST\/} images at
both GOODS (left) and HUDF (right) depth using our reference 
model. The input values of all model galaxies placed in
the simulated image are shown as orange squares. The input values for galaxies
detected by {\tt SExtractor} are shown as blue triangles, while the output
values as measured by {\tt SExtractor} for those detected galaxies
are shown as red circles. {\tt SExtractor}-derived distributions are not
corrected for incompleteness. Black lines show the actual observed
distributions calculated from the CANDELS GOODS-S Multi-wavelength
Catalog \citep{guo2013b}. There is good agreement between the
simulated and observed distributions except for the smallest
size objects.
}
\label{Fig:LumAndRadFunSM} 
\end{center}
\end{figure}

Figure~\ref{Fig:ApMagVsStellMass} plots the stellar masses of
galaxies versus their input model apparent magnitudes. The
stellar mass decreases strongly with increasing redshift.
However, an important conclusion from this plot is that even the
least massive \textit{detected} galaxies are still generally above
the mass thresholds set by both the SDSS survey
\citep[$\Ms\gtrsim 10^{6}\Msun$,][]{kauffmann2003} and the milli-Millennium simulation
\citep[$\Ms\gtrsim 10^{7.1}\Msun$,][]{guo2010}.
Figure~\ref{Fig:MedianMagAt20Mass} shows 
the typical apparent magnitude values found at the mass completeness 
limit of the milli-Millennium simulation as a function of redshift.
Although the median magnitude initially dims as redshift increases,
the curve flattens out at $m_{\rm F160W} \sim 32$ for
$z>3$; although the galaxies at those redshifts have smaller masses,
they also have higher star formation rates and younger populations due to
the requirement that their stellar masses be assembled in a short time.
That approximately compensates for the cosmological dimming due to the increasing
luminosity distance.

\noindent\textit{Biases} ---
As we noted above, Figure~\ref{Fig:ApMagVsPetroRad} shows that a
relatively simple cut on the input (model) magnitude of galaxies
predicts with reasonable accuracy which galaxies will be detectable
in the images.  However, detection is not the whole story.  It is
also necessary to measure the galaxy properties, and those measurements
can be biased through complex effects related to the morphology and internal structure of
the galaxy.  Detection of an object does not ensure that its
magnitude, colors, or size can be measured correctly.  This is an
area where our forward-modeling approach provides more reliable results than previous approaches.

Figures~\ref{Fig:InputVsDeltaOutput} and~\ref{Fig:OutputMagVsOutputRad} show 
strong biases in the {\tt SExtractor}-derived
measurements of galaxy sizes and apparent magnitudes. From
galaxies detected in the HUDF-depth simulated image, we compare the 
differences between the true (\textit{input\/}) and {\tt SExtractor}-measured
(\textit{output\/}) values of $m_{\rm F160W}$ and $\log R_{\rm P}$. 
Our major conclusions are:

\begin{enumerate}

\item There is a significant magnitude-dependent luminosity bias, with
the measured $m_{\rm F160W}$ magnitudes on average
fainter than the true galaxy magnitudes
(Fig.~\ref{Fig:InputVsDeltaOutput}).
For fainter galaxies, only the compact, high surface brightness nuclei
are detectable above the image noise level, while the lower surface brightness
extended envelopes are hidden in the noise.
This bias is small for bright galaxies,
but reaches median values of $\sim0.4$ magnitude and extreme values of $\gtrsim1$~magnitude at 
the detection limit.
This effect does not depend strongly
on redshift.

\item There is also a strong bias in the measured sizes near the magnitude detection limit
(Fig.~\ref{Fig:InputVsDeltaOutput}).
The {\tt SExtractor}-measured sizes are smaller than the original input
sizes of model galaxies around $m_{\rm F160W}\simeq 29$. 
The explanation appears to be the same as for the luminosity bias: the detectable
part of a faint galaxy is significantly more compact than the true
Petrosian radius due to masking of the more extended component by noise.
The difference between \textit{output} and \textit{input} values of $\log R_{\rm P}$ has a median of
0.15~dex, with maximum values reaching even to 0.8~dex.
Note that there is also a Malmquist bias affecting these distributions: 
faint extended objects may be detectable only if noise fluctuations make their
surface brightness appear brighter than the detection limit.

\end{enumerate}

Figure~\ref{Fig:InputVsDeltaOutput} also reveals a
small bias toward larger measured output sizes for sources $\sim1$~mag brighter than the
detection limit. This bias
is largest in the $z>6$ redshift bin, where the difference
reaches around 0.2~dex, and might be related to intrinsic peculiarities in the 
way {\tt SExtractor} measures sizes.

These effects are certainly detection biases and not
a trend resulting from supposed evolutionary effects in the models.
Figure~\ref{Fig:OutputMagVsOutputRad} shows a scatter plot of the measured (output) sizes
versus measured magnitude for both GOODS and HUDF-depth
simulated images. The strong bias in the {\tt SExtractor}-measured
sizes at the magnitude detection limit are easily visible at both
depths, and the bias begins at a magnitude that is determined by the detection
limit rather than at any physically determined luminosity.
Note that the underlying images (before adding noise) are exactly the
same in these two cases: they are equivalent to observing the same sky
region twice with different exposure times.  This 
rules out any artifact introduced by the model. 
Note how the measured $R_{\rm P}$ extends down to values
close to the PSF size (FWHM=0.151\arcsec\ for the F160W band image)
at the magnitude detection limit.

We note that the true biases could be even larger than those measured
in our simulations.
Real galaxy light profiles have extended wings, whereas our galaxy image cutouts only 
include the light present within two Petrosian radii.  According to \cite{strauss2002}, 
82\% of the light from a de Vaucouleurs profile is contained within this radius, and
99\% of light from an exponential profile is included.
In addition to these moderate effects, there might be cases when the outer wings 
of the light profile are much more extended, with surface brightness falling close to $R^{-2}$.
Then the light in the extended halo would dominate the total luminosity, but such
extended emission would certainly not be detectable for faint galaxies, and the biases
would be increased.
The true radial dependence of the outer wings of galaxies is not currently well constrained, so the
amount of light missing from these wings remains an open issue.

\noindent\textit{Detection Efficiencies} ---
The detection efficiency is an important statistic, since it tells
us the amount of underlying information that we are losing when
observing with \textit{HST\/}. In this paper, we consider two different measures
of the efficiency: the number count detection efficiency is the
fraction of galaxies on the image that are detected in the simulated
data, and the light detection efficiency is the fraction of the
total galaxy flux in the observing band that is detected.  These
efficiencies are easily computed from the model images since we
know the properties of both the detected and undetected galaxies.
Both quantities can be computed as a function of other parameters
such as redshift or apparent magnitude.  The number count efficiency
is more commonly used but can be ill-defined when one considers the
possibility of numerous faint galaxies that are undetected but
contribute little stellar mass or light.  The light detection
efficiency is better behaved, since the integral of light from many
faint galaxies is typically a small correction to the light from
objects near the knee of the galaxy luminosity function.

The top
panel of Figure~\ref{Fig:EfficiencySM} shows the detection efficiencies
for number counts and for F160W-band light
at HUDF depth as a function of the model input magnitude of galaxies.
Both efficiencies drop sharply to zero at the magnitude
detection limit of $m_{\rm F160W}\simeq 29$.
Interestingly, the number count efficiency has the same shape in
all redshift bins, showing a slight decline toward fainter magnitudes
but remaining above 80\% before reaching the detection limit.
It reaches values closer to unity only at bright
magnitudes in the $z<2$ redshift bin.  The light detection
efficiency has a similar behavior but lower values:
it falls to $\sim60$\%--70\% just above the magnitude
completeness limit. The lower efficiency
for detecting F160W light is due to the magnitude bias discussed above:
not only are galaxies undercounted near the detection limit, but the
fluxes of detected galaxies near the detection limit are also significantly
underestimated (Fig.~\ref{Fig:InputVsDeltaOutput}). 

The bottom panel of Figure~\ref{Fig:EfficiencySM} shows the detection
efficiencies as a function of redshift.  These curves are a bit
more complicated to interpret because they are affected by the
finite mass resolution of the milli-Millennium simulation.  Very
low-mass galaxies ($<10^{7.1}\Msun$) cannot appear in the simulations,
so their contributions to the undetected counts and luminosity are
omitted.  That means that the computed efficiencies are only upper
limits to the real values.  This has little effect on the efficiency
as a function of magnitude since very few low-mass objects would
be brighter than 29th magnitude (Figs.~\ref{Fig:ApMagVsStellMass}
and~\ref{Fig:MedianMagAt20Mass}), but it has a significant effect
on the redshift-dependent efficiencies since faint galaxies can
appear at all redshifts.

Figure~\ref{Fig:EfficiencySM} shows two attempts
to correct the detection efficiencies for the
low-mass galaxies.  The solid curve simply omits all galaxies
fainter than the median magnitude of the lowest mass halos
in the milli-Millennium simulation (Fig.~\ref{Fig:MedianMagAt20Mass}); the efficiencies
plotted are then effectively the fraction of galaxies or light from galaxies brighter
than $\sim32$~mag, which is an upper limit to the true detection efficiency.
The dashed curve instead integrates the light in the extrapolated power-law tail of the faint galaxy
luminosity function (Fig.~\ref{Fig:LumFunSMredshiftBin}) and includes the light of the missing
faint galaxies as part of the total flux.
This produces a modest correction in the light detection efficiency.  Note that
a version with extrapolated number counts is not shown because the
number count correction factor is much larger (and far more uncertain).

The number count efficiency decreases 
rapidly from 80\% at low redshift to 20\% at $z\sim 7$. On the
other hand, the light detection efficiency drops more slowly, reaching
values of $\sim50$\%--70\% at $z\sim 7$. The light detection efficiency is greater
than the number count efficiency because every redshift bin includes
numerous faint galaxies, and the detected ones (at fixed
redshift) are generally the brightest and carry most of the light
content of the image. 

The correction from including the extrapolated light from faint low-mass
galaxies increases at higher redshifts.
This is can be seen directly in the steepening of the slope of the fitted power-law tail at increasing redshifts
which changes from $\alpha\simeq -1$ at $z<1$ to $\alpha=-1.75$ at $z=7-8$ (Fig.~\ref{Fig:LumFunSMredshiftBin}).
Interestingly, the steep slopes for young galaxies at high redshifts are
also found in the present-day universe. \cite{Taghizadeh-Popp2012} isolated similar populations
of small, blue galaxies with rapid star formation at $z=0$ and also
found faint-end luminosity
slopes close to $\alpha \sim-1.6$.

\begin{figure*}[htp]
  \centering
  \begin{tabular}{cccc}
    \includegraphics[width=42mm]{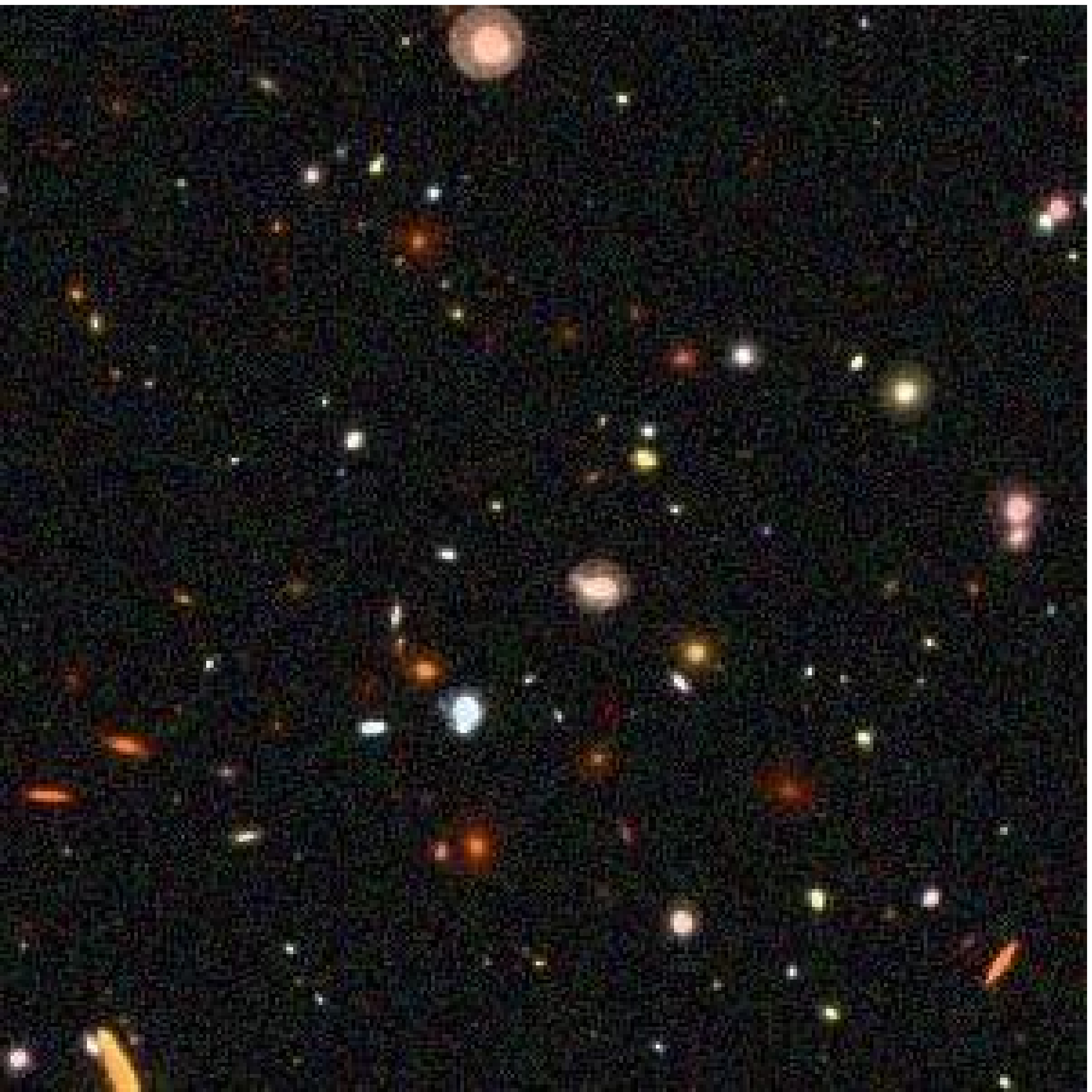}&
    \includegraphics[width=42mm]{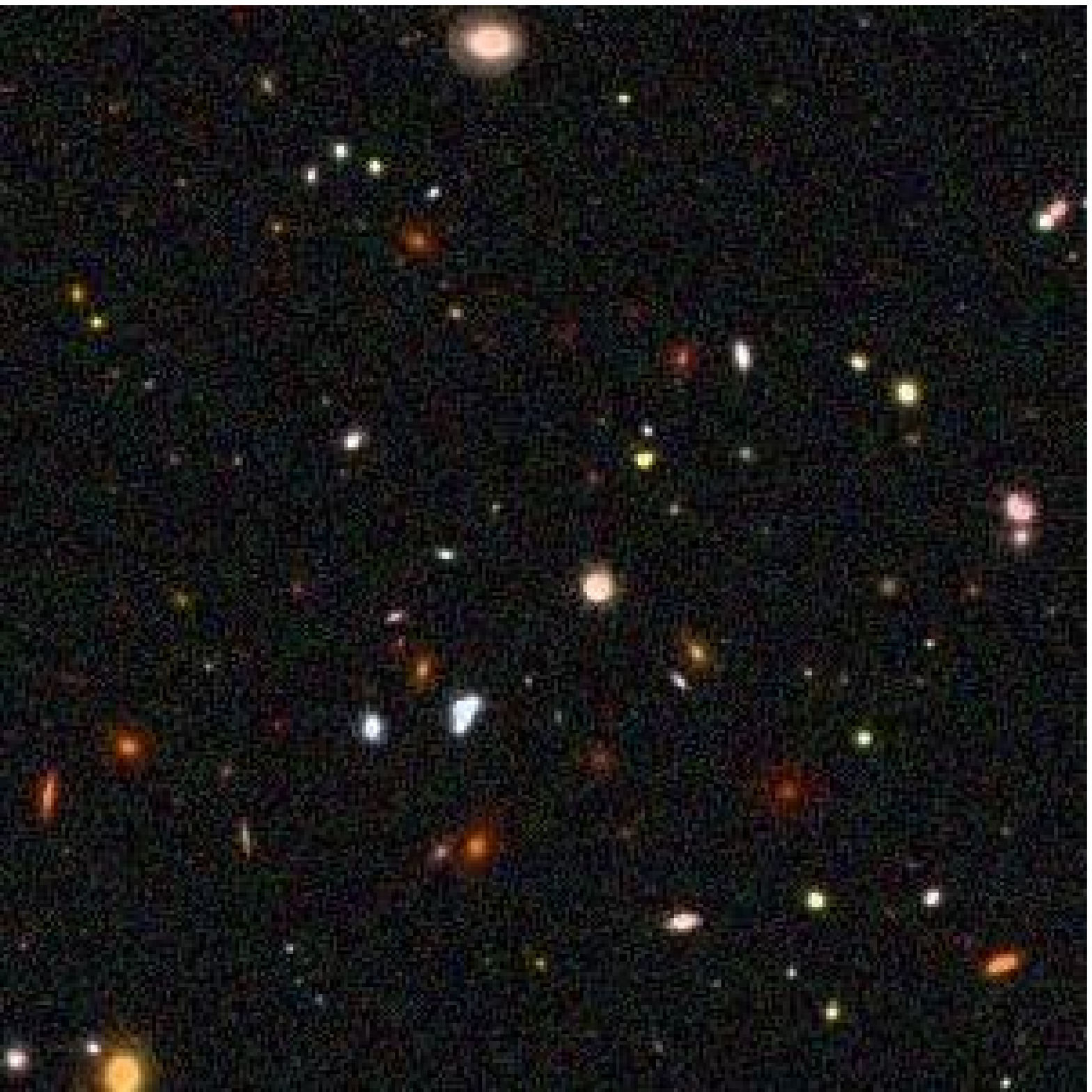}&
    \includegraphics[width=42mm]{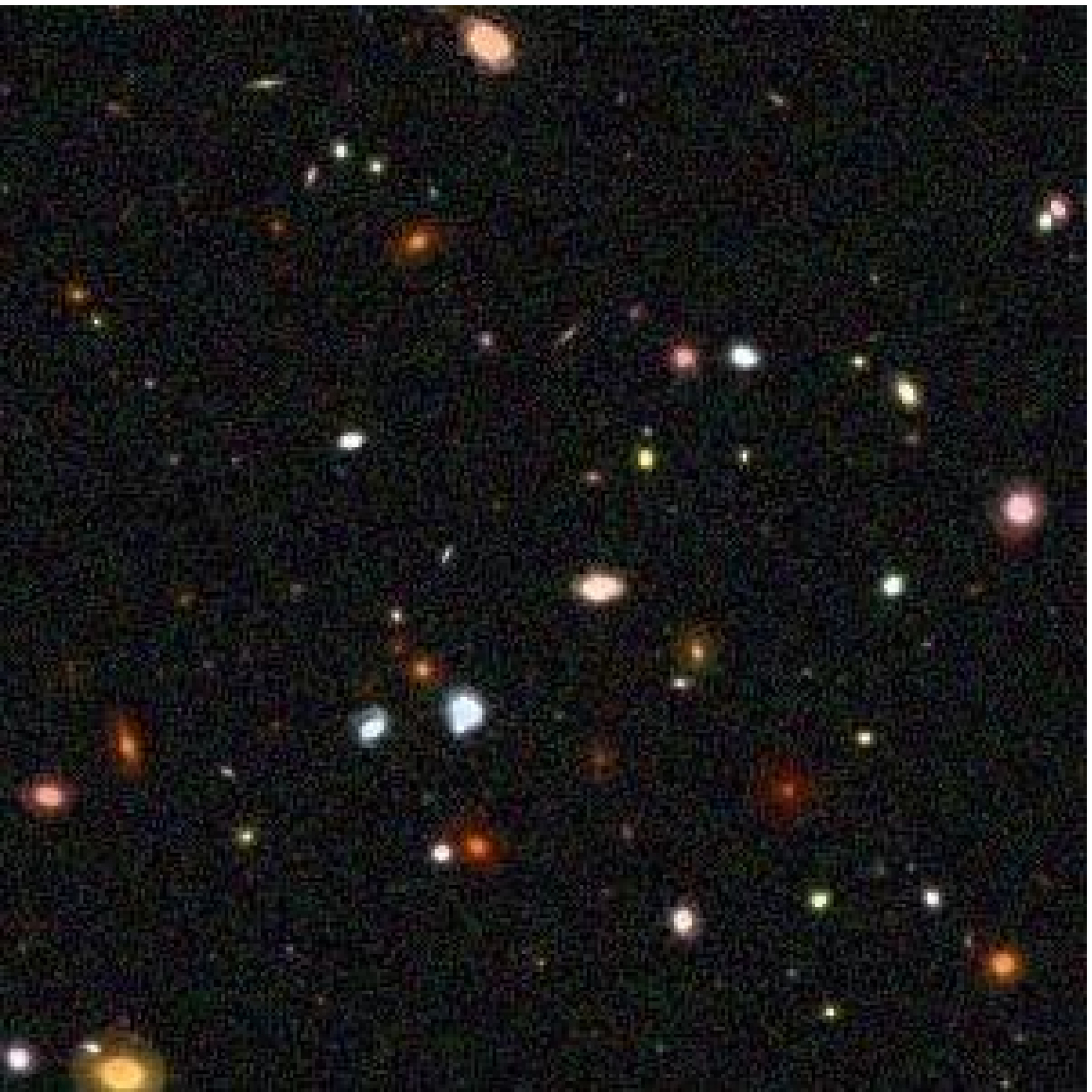}&
    \includegraphics[width=42mm]{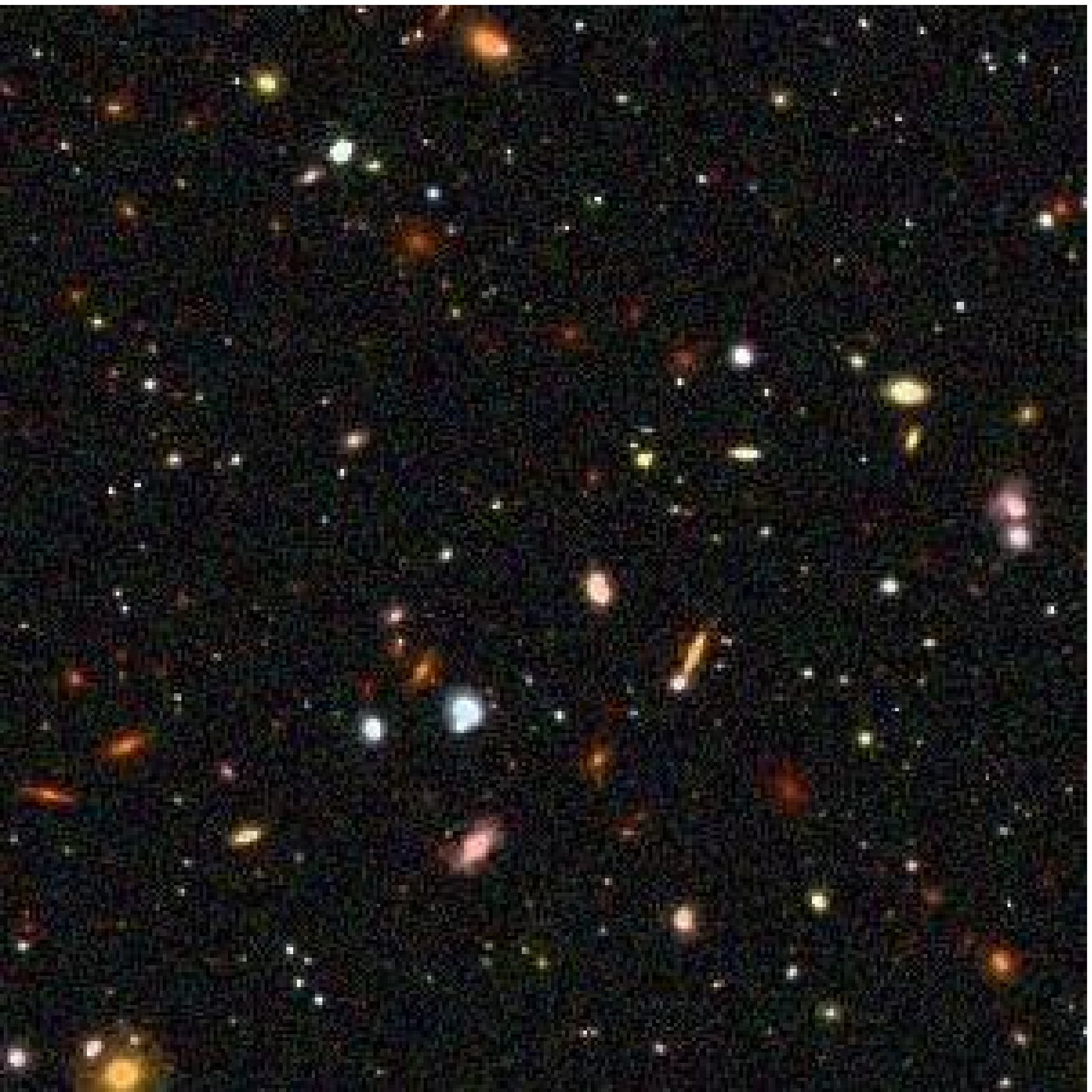}\\
Model 1 & Model 2 & Model 3 & Model 4  \\[4mm]
    \includegraphics[width=42mm]{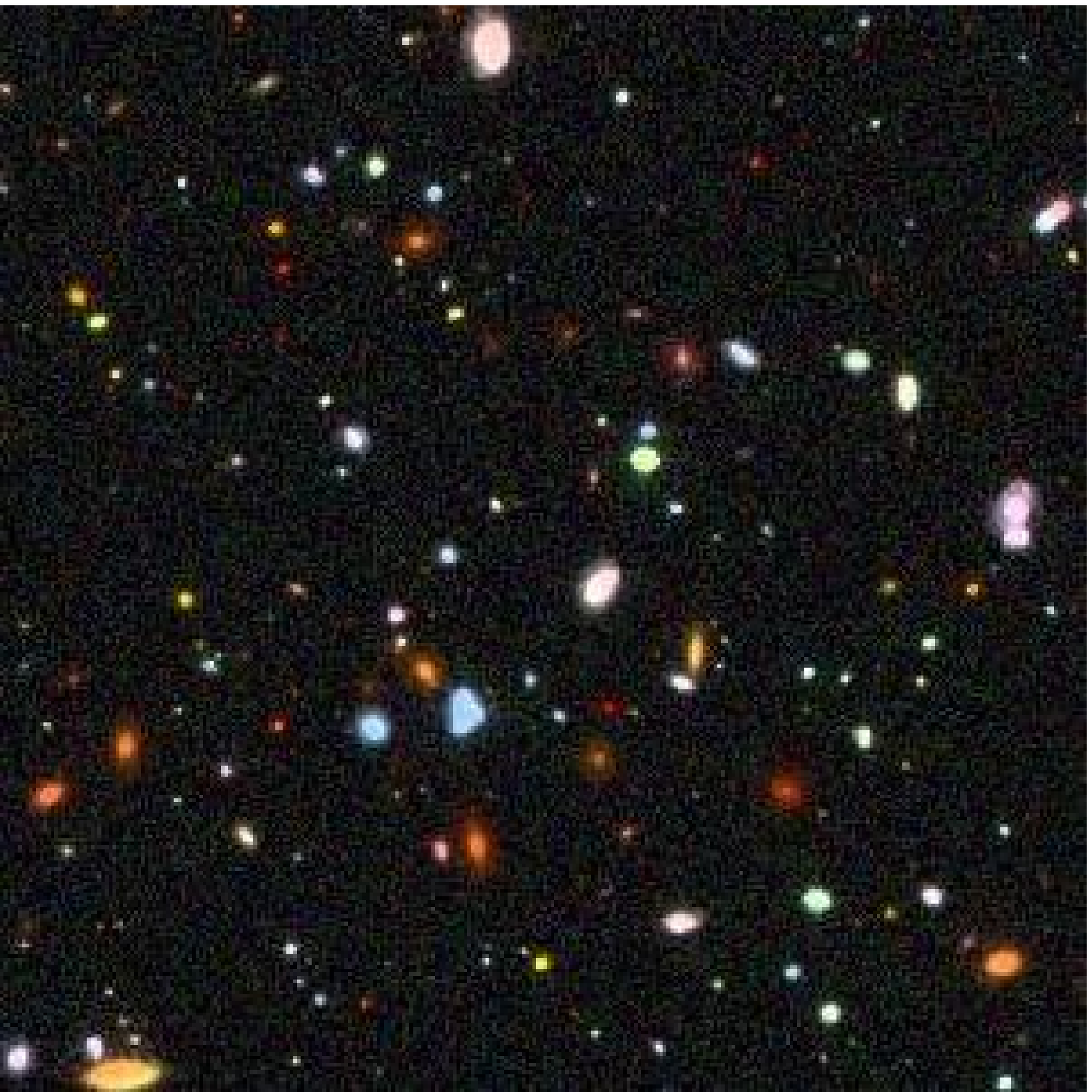}&
    \includegraphics[width=42mm]{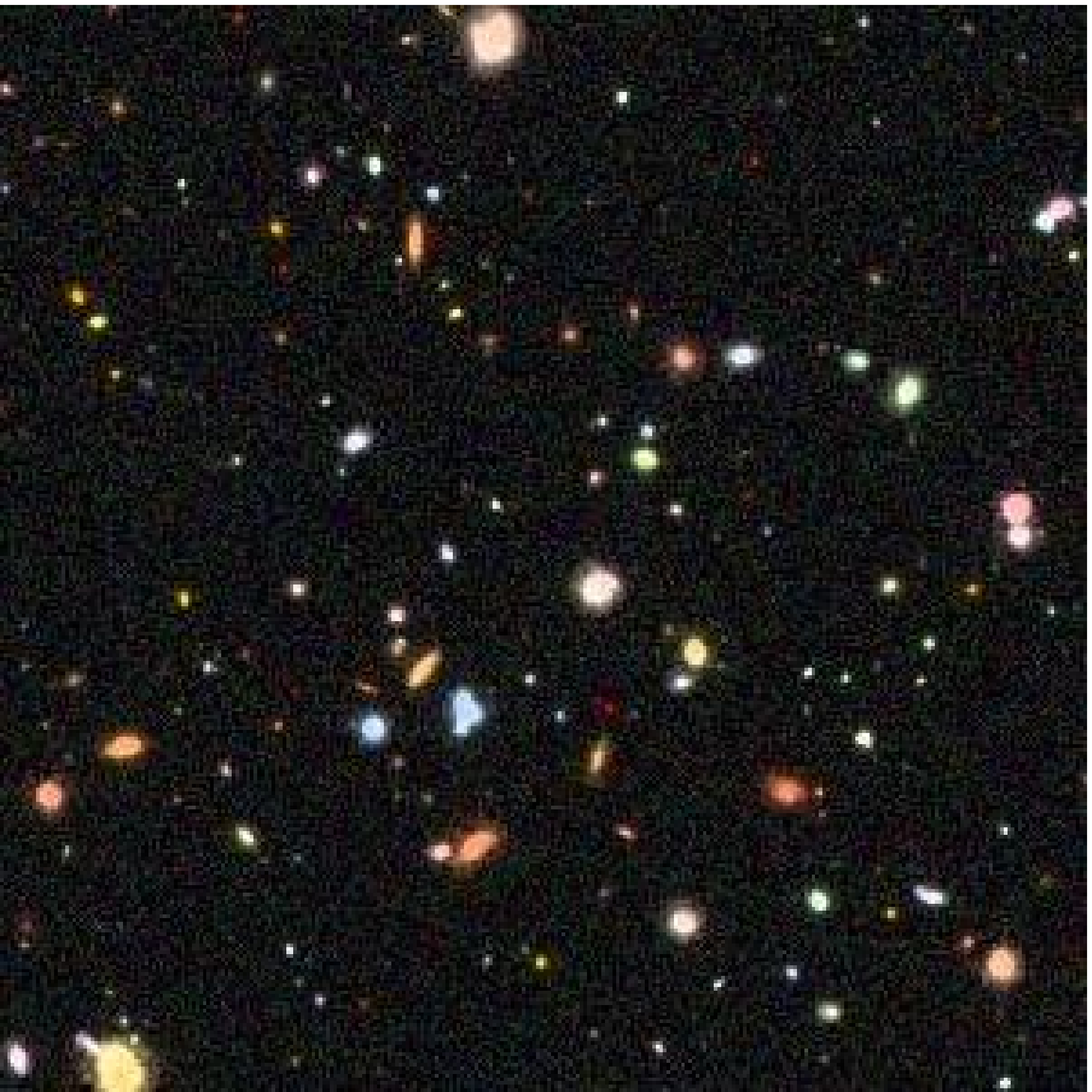}&
    \includegraphics[width=42mm]{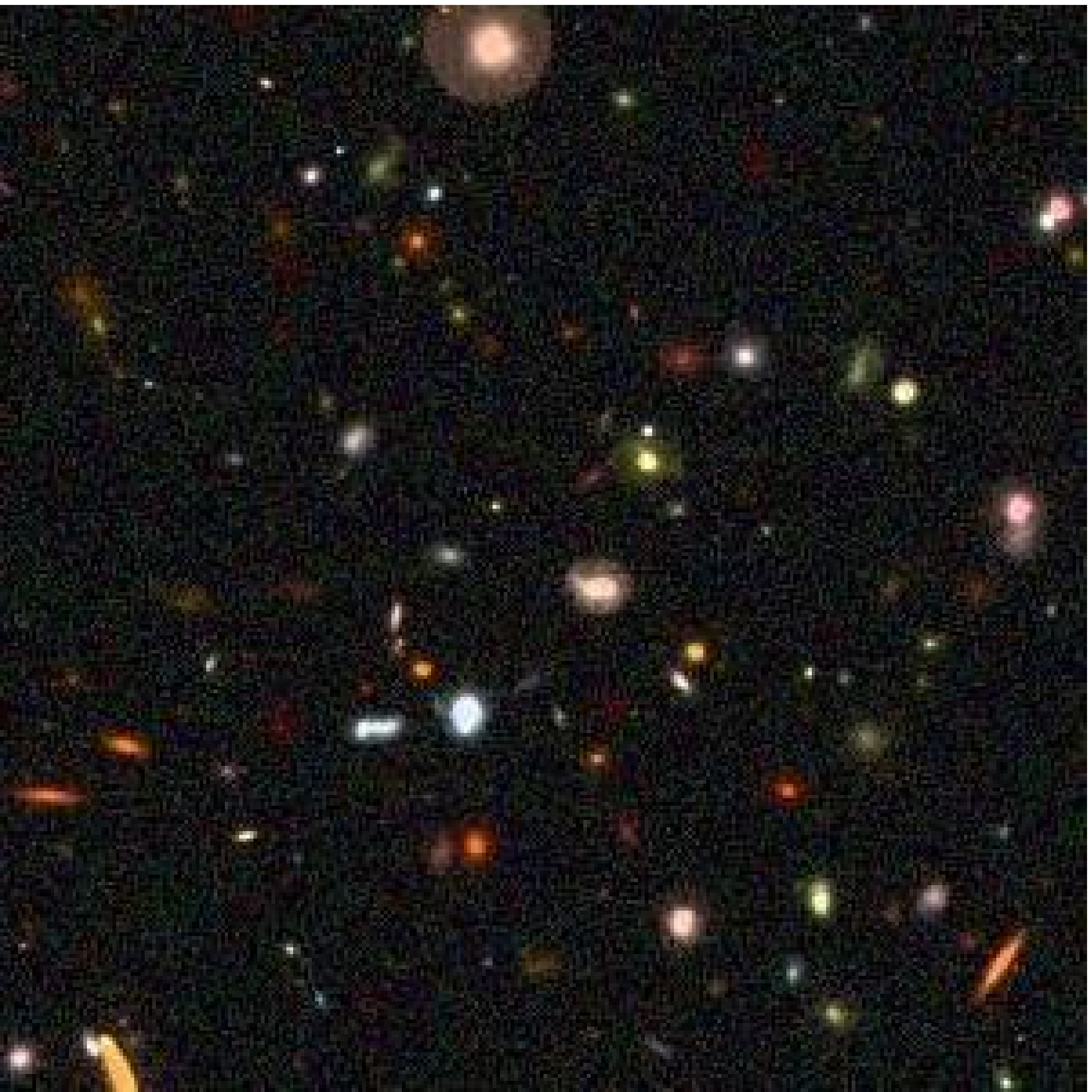}&
    \includegraphics[width=42mm]{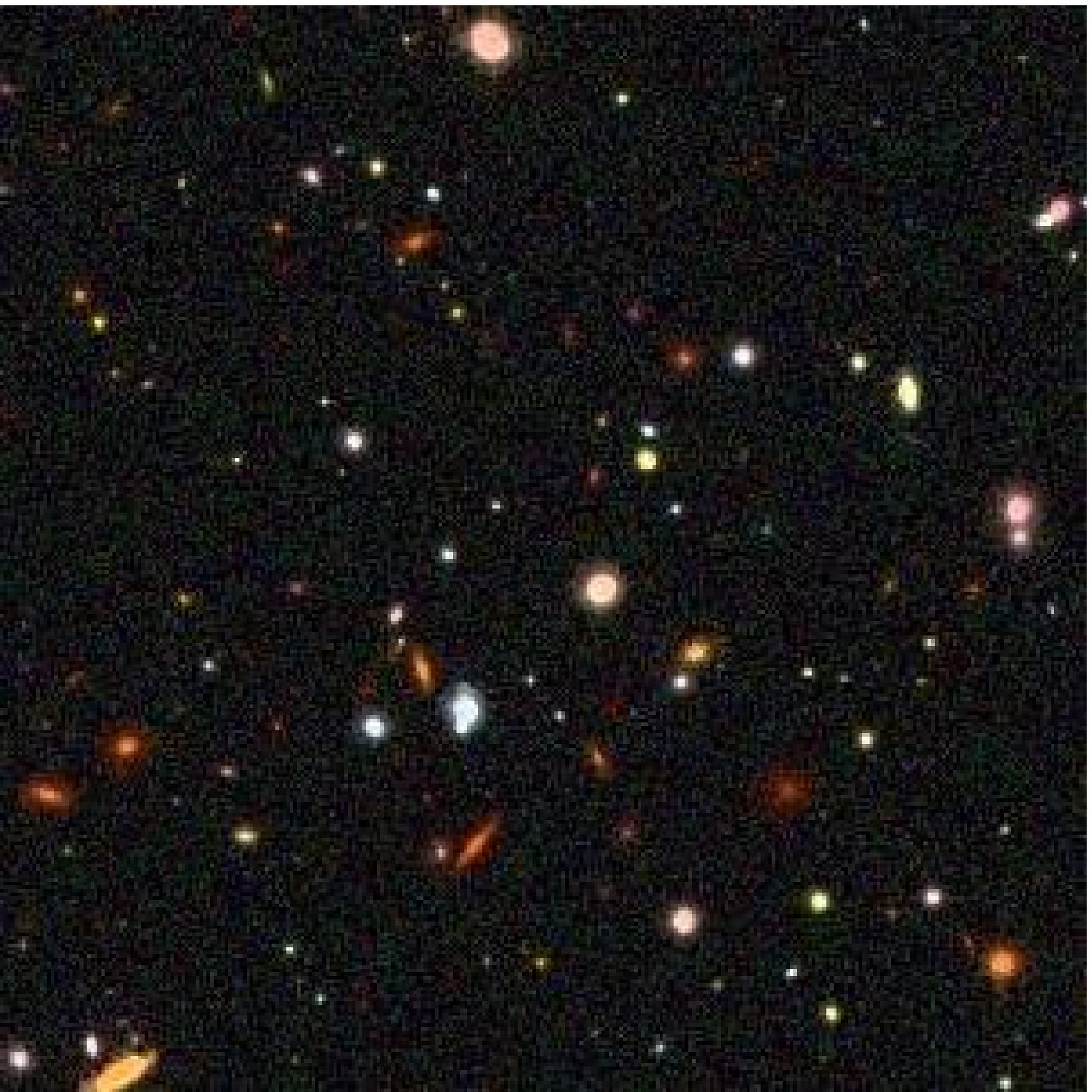}\\
Model 5 & Model 6 & Model 7 & Model 8 \\[4mm]
	\end{tabular}
\caption{Simulated ACS/WFC F850LP+F606W+F435W images (HUDF depth) derived from for all models explored in this paper. The same exposure times and display contrast are used for the comparison at each depth. The field of view is $1200\times1200$ ACS pixels ($\sim 1/12$ of the full ACS/WFC field of view). Changes in the distribution of luminosity and sizes are apparent for different models.}
\label{Fig:ImagesAllModels} 
\end{figure*}

\begin{figure}[htp]
  \centering
  \begin{tabular}{c}

	\includegraphics[width=70mm]{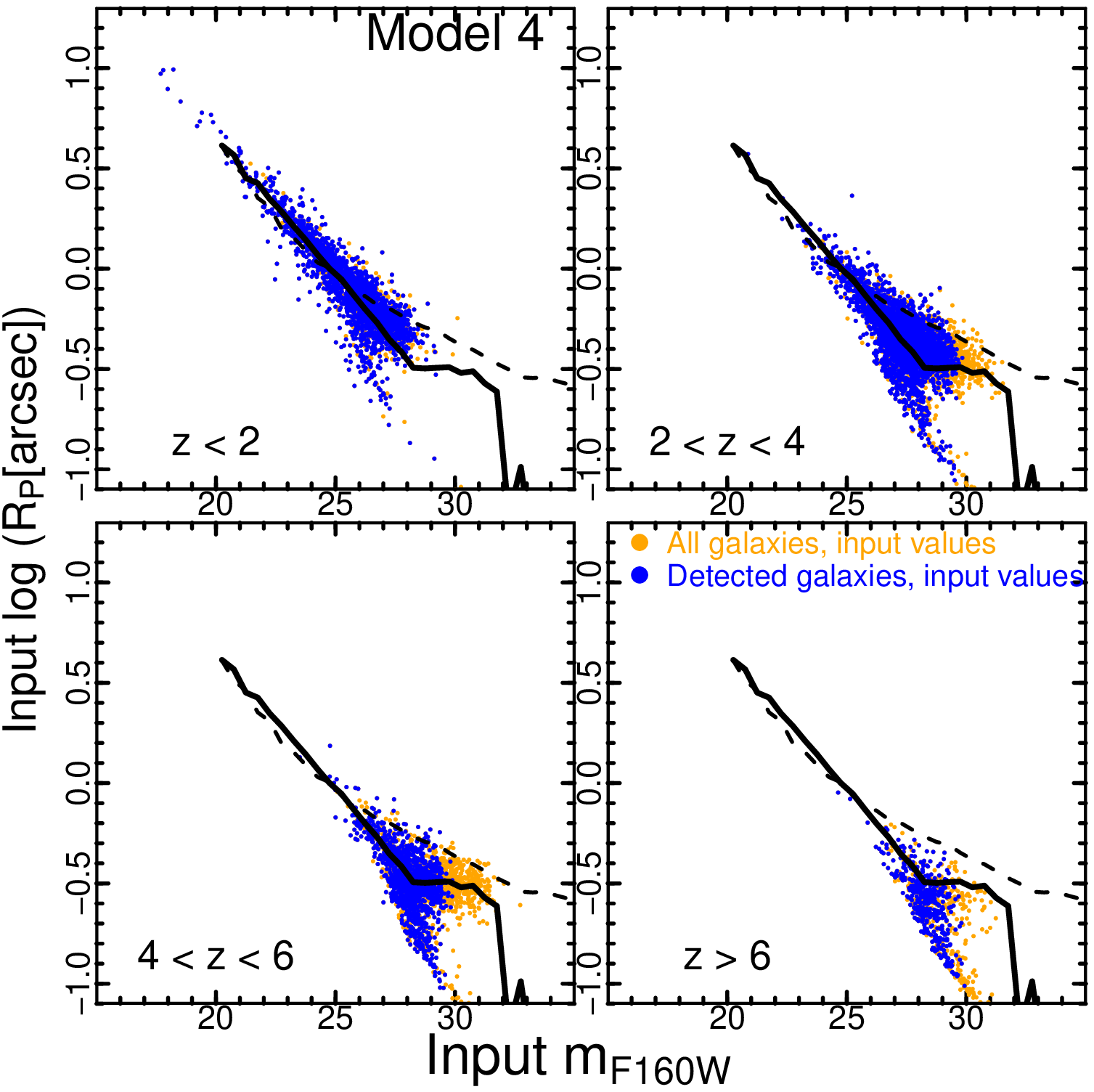}\\
	\includegraphics[width=70mm]{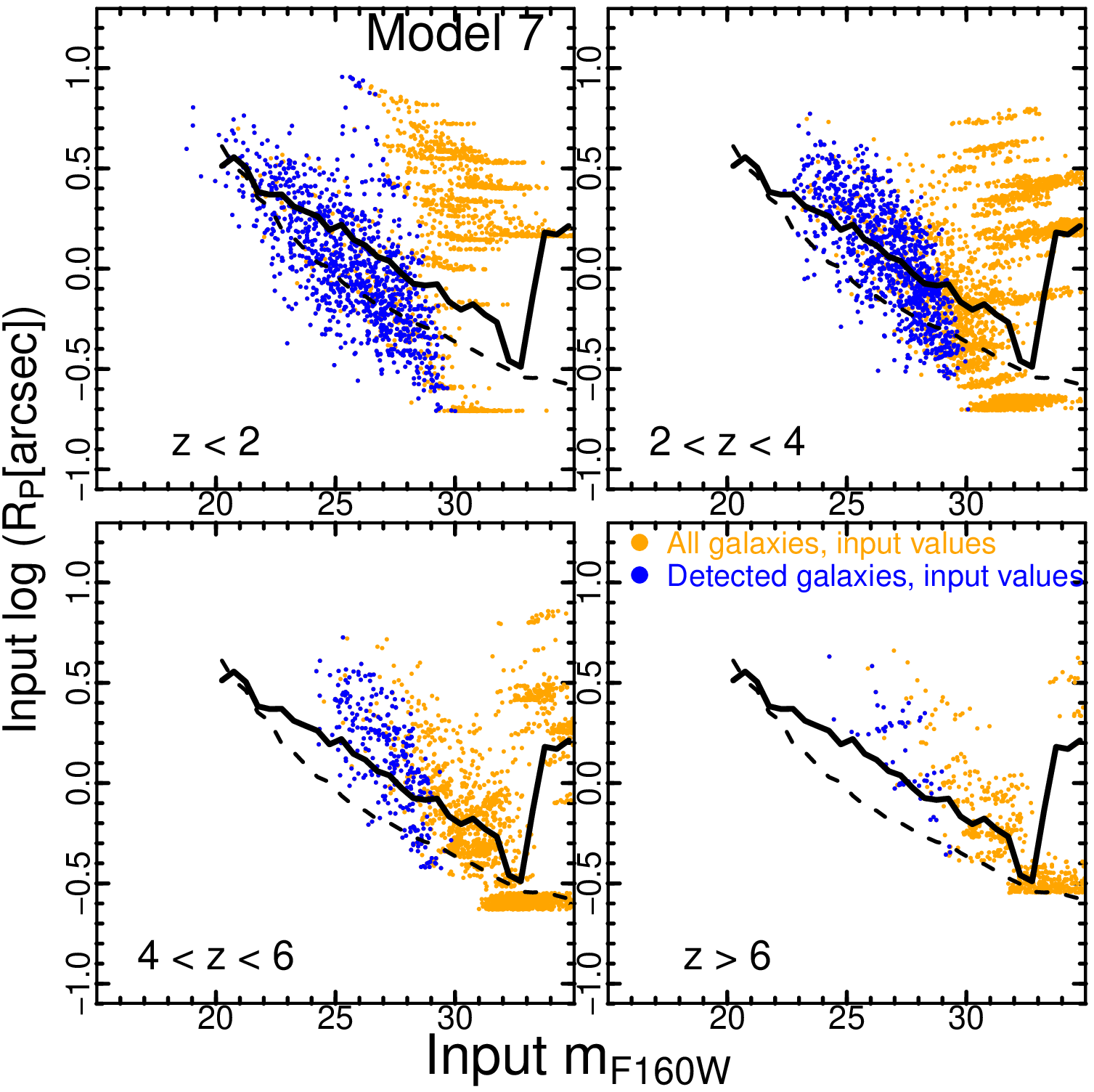}\\[2mm]

\end{tabular}
\caption{
Petrosian radius versus apparent F160W magnitude for the
HUDF-depth simulated \textit{HST\/} image derived from Models 4 (top) and 7
(bottom), separated in four redshift bins.
Colors are the same as
Fig.~\ref{Fig:ApMagVsPetroRad}. The solid line shows the median
values for these models, while the dashed line
shows the median for the reference model
for comparison. Small galaxies are given a higher
luminosity in Model 4; the large galaxy sizes in Model
7 lead to many galaxies being undetected due to low
surface brightnesses.
}
\label{Fig:ApMagVsPetroRadAllModels}
\end{figure}

\begin{figure}[htp]
  \centering
	\includegraphics[width=70mm]{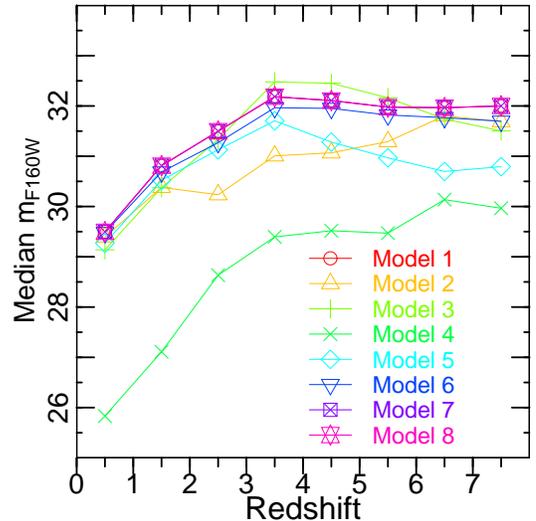}
\caption{
Median apparent F160W magnitude of galaxies having stellar masses
near the simulation mass incompleteness limit of
$\log \Ms/ \Msun =7.1$, shown for galaxies in different redshift bins.
Data is shown for all the models explored in this paper.
}
\label{Fig:MedianMagAt20MassAllModels} 
\end{figure}

\begin{figure*}[t]
\begin{center}
\begin{tabular}{cc}
    \includegraphics[width=80mm]{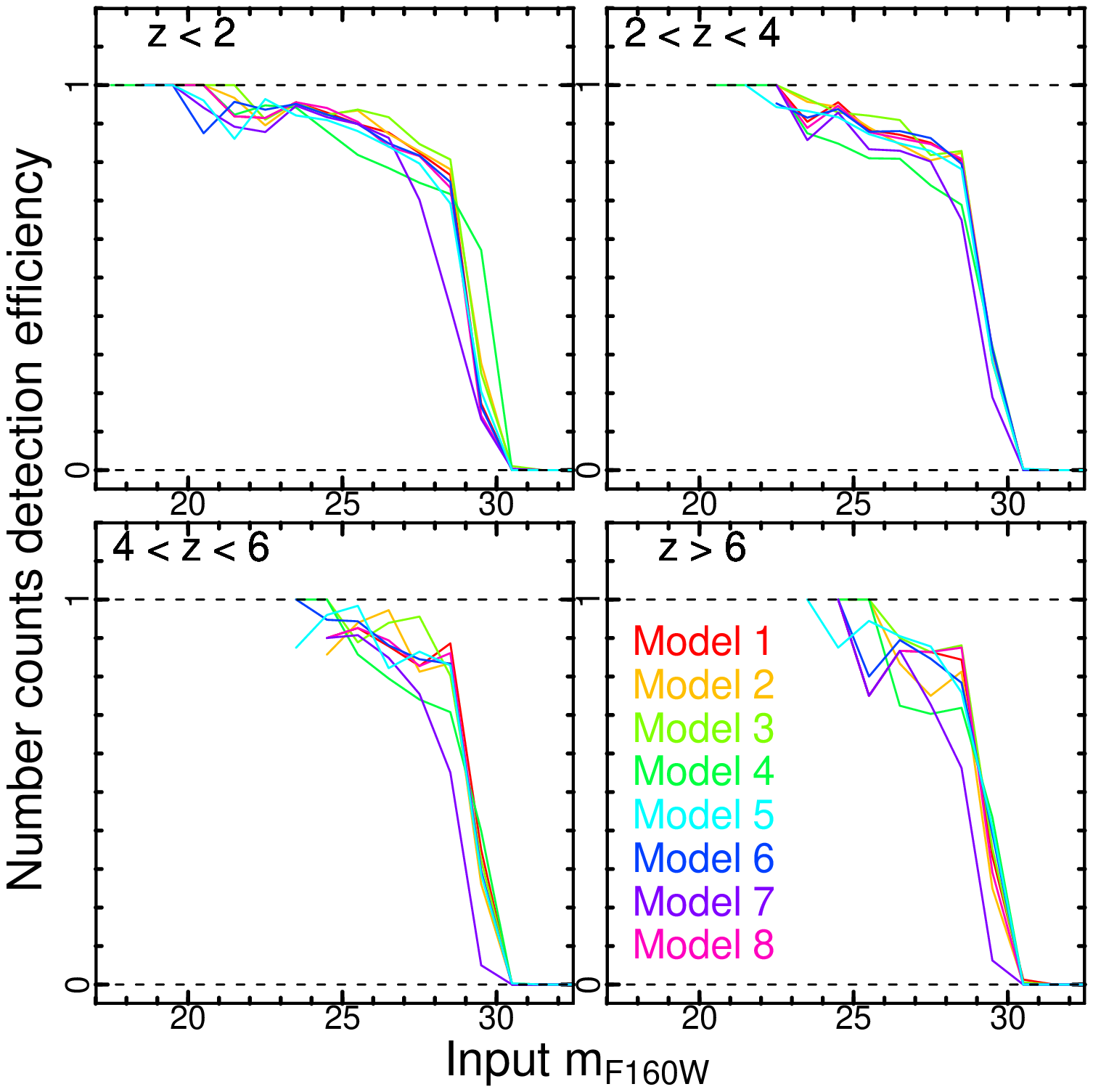}&
    \includegraphics[width=80mm]{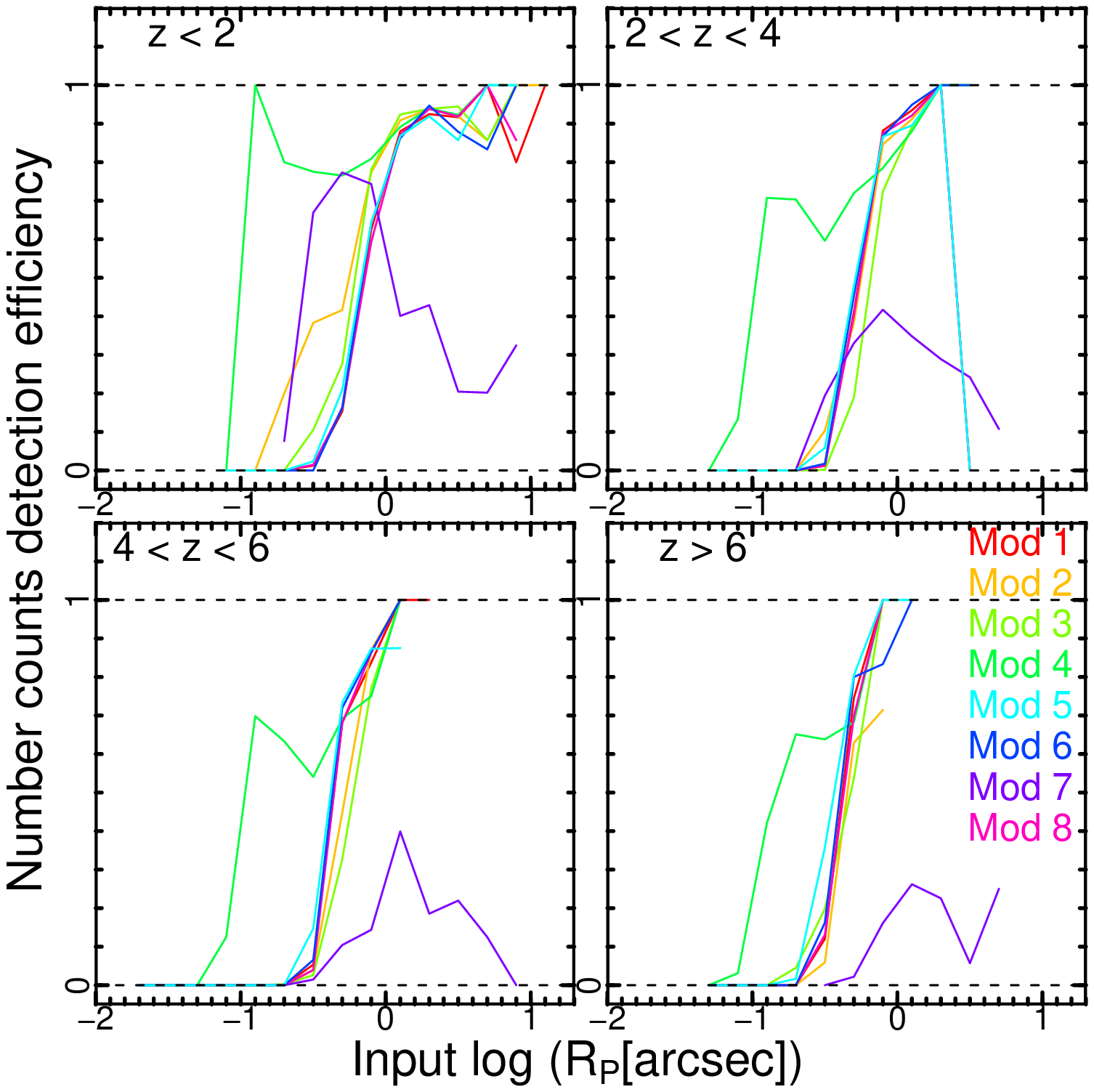}\\[4mm]
    \includegraphics[width=80mm]{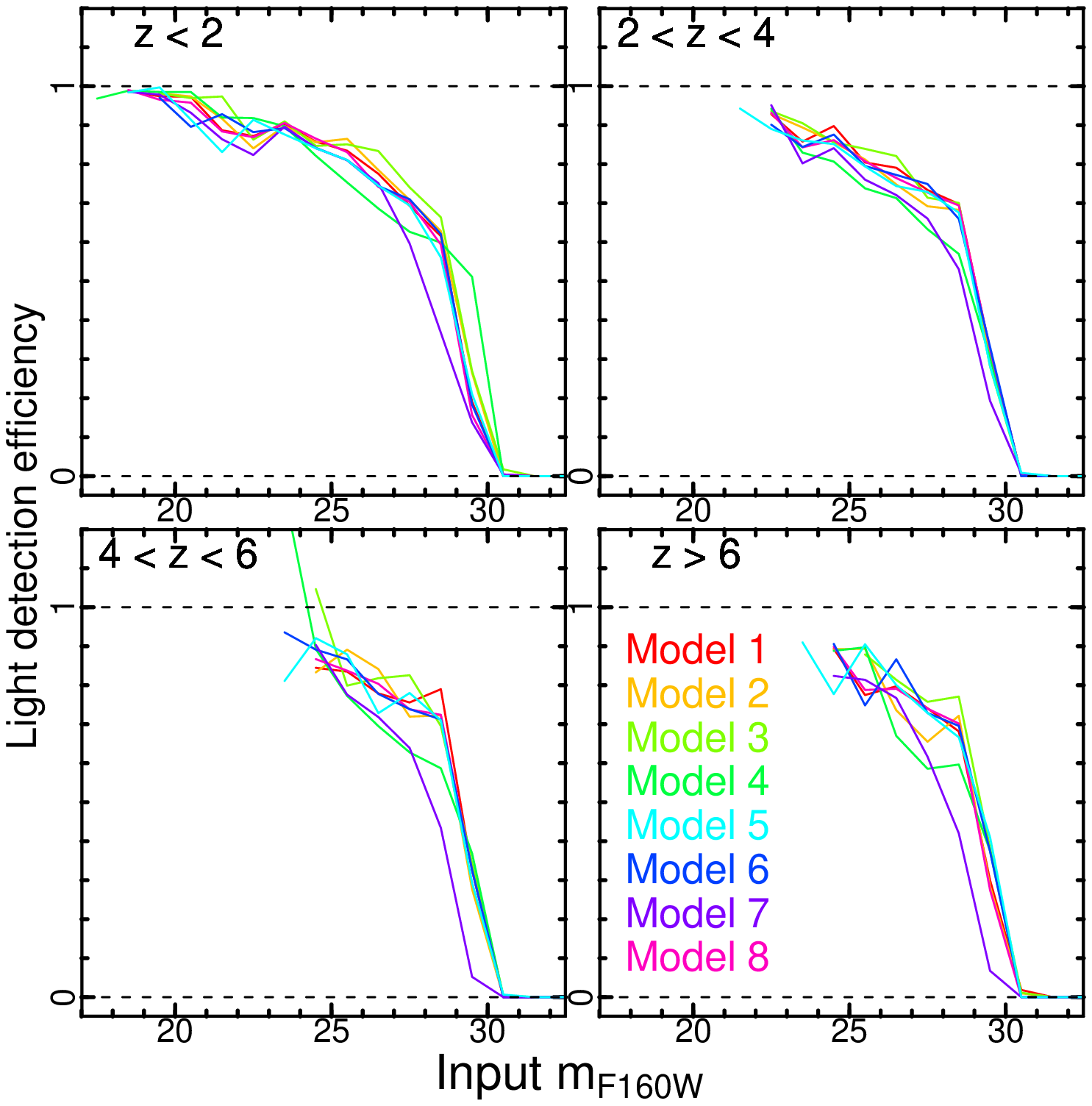}& 
    \includegraphics[width=80mm]{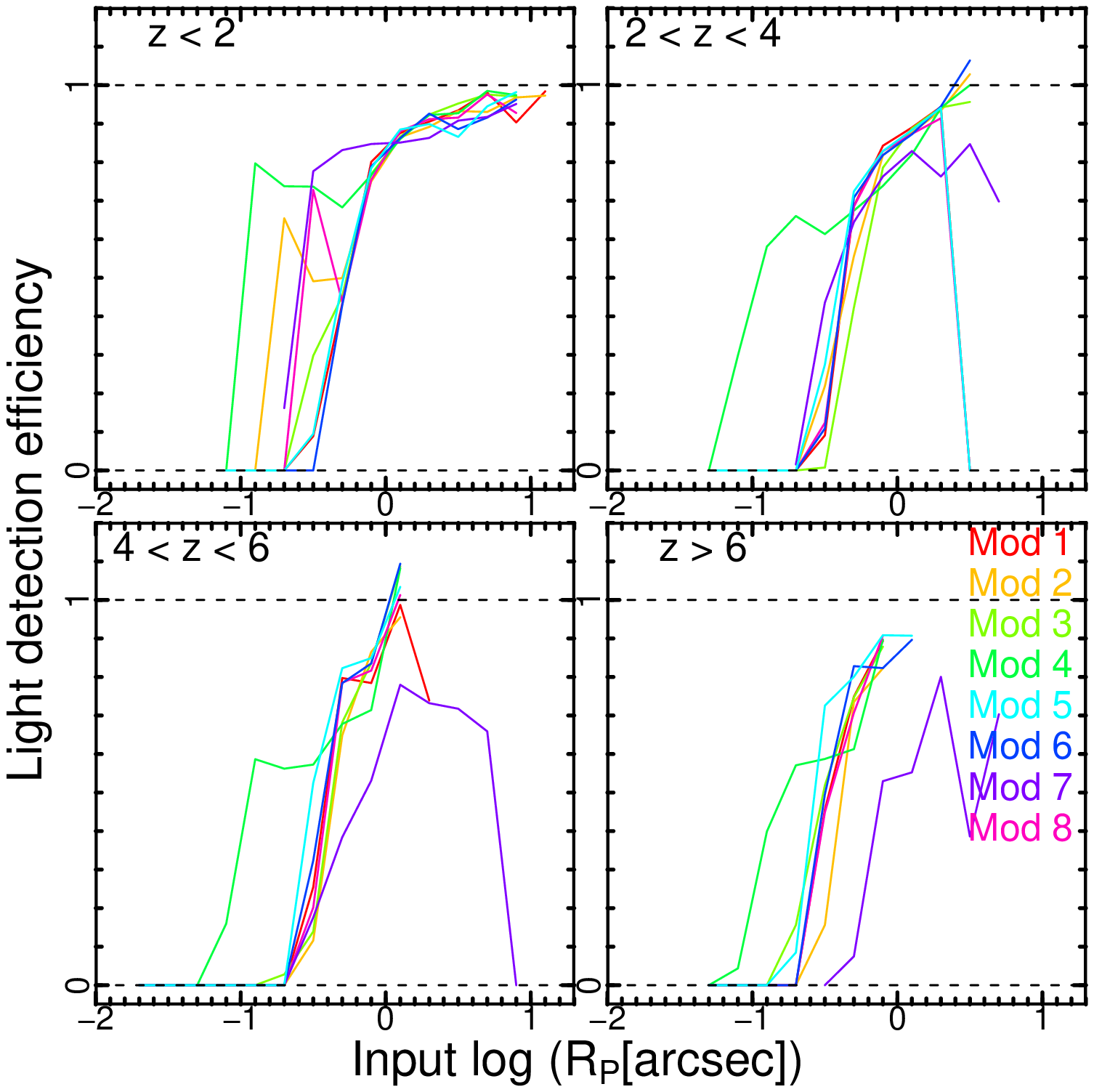}\\[4mm]
\end{tabular} 
\end{center}
\caption{
{\tt SExtractor} number count (top row) and light (bottom row)
detection efficiencies as a function of the input
$m_{\rm F160W}$ magnitudes (left) and Petrosian radii (right) of detected
galaxies in the HUDF-depth simulated \textit{HST\/} image, shown for each of
the models explored in this paper.  Note that the efficiency 
is lower for small galaxies because smaller galaxies are also
fainter.
}
\label{Fig:EfficiencyAllModels}
\end{figure*}

\begin{figure*}[t]
\begin{center}
\begin{tabular}{cc}
    \includegraphics[width=70mm]{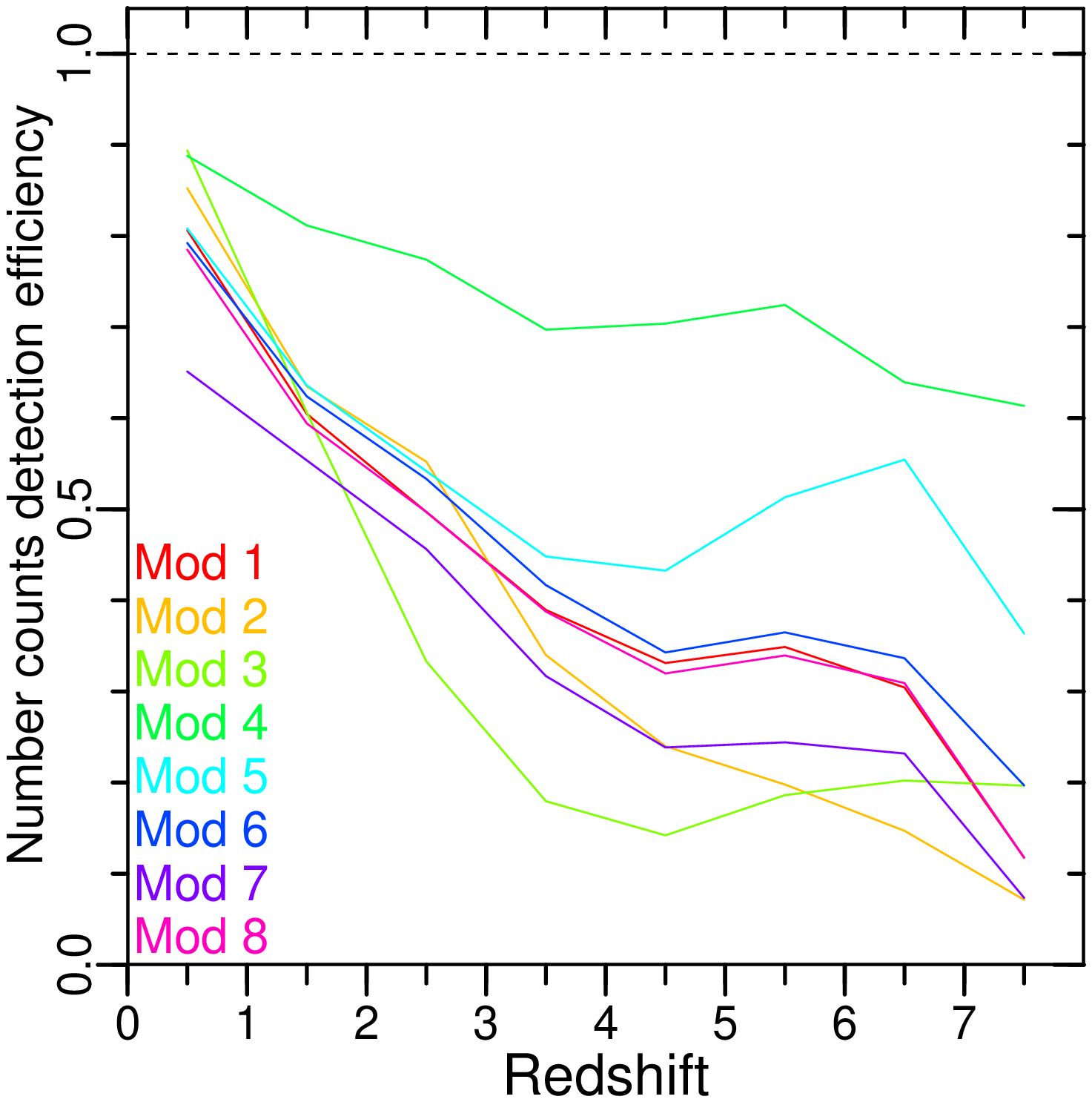}&
    \includegraphics[width=70mm]{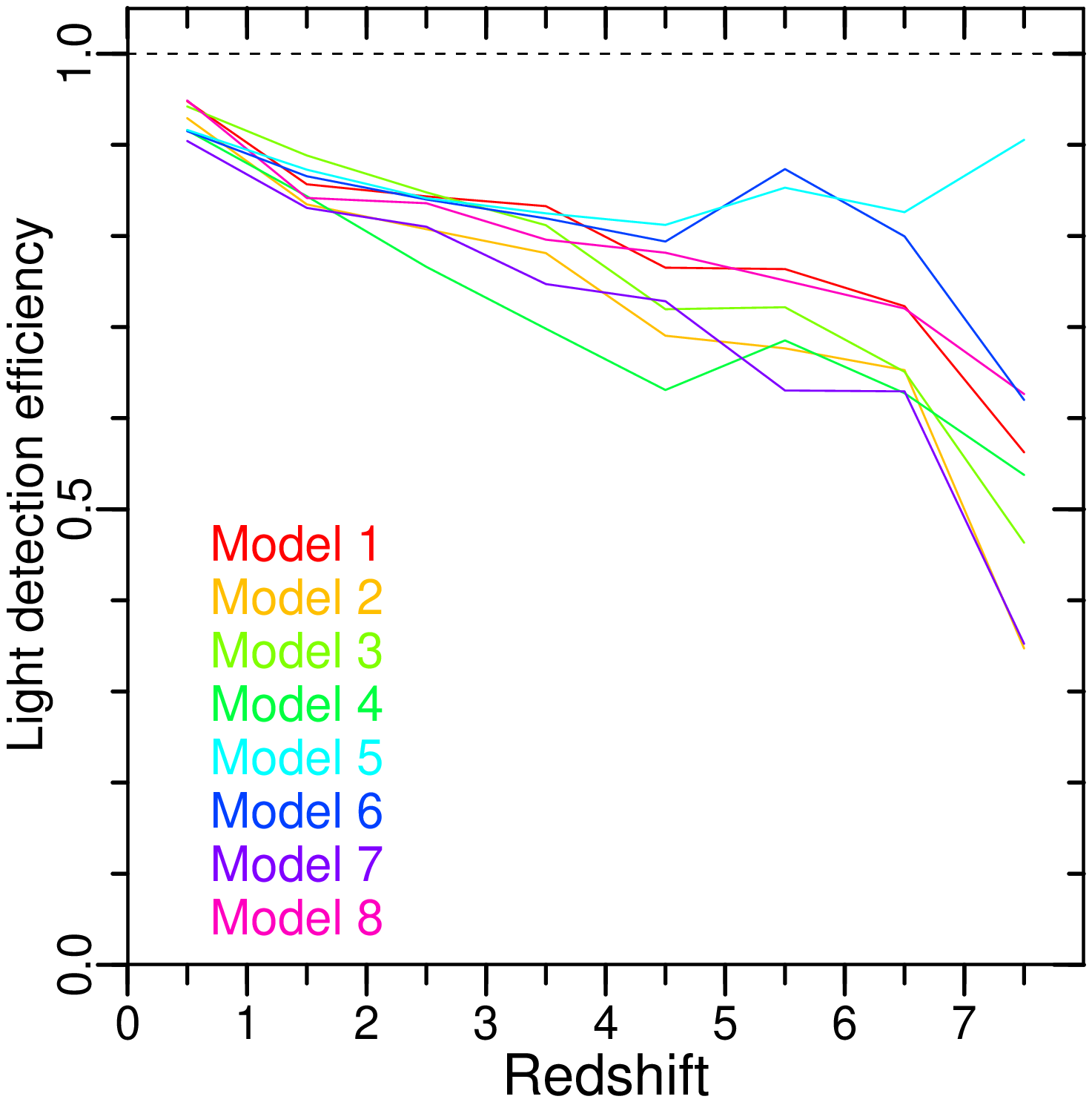}\\
\end{tabular}
\end{center}
\caption{
Detection efficiencies as in Fig.~\ref{Fig:EfficiencyAllModels}, but
plotted as a function of redshift. We include only galaxies brighter
than the median $m_{\rm F160W}$ at the stellar mass
incompleteness limit, as shown in
Fig.~\ref{Fig:MedianMagAt20MassAllModels}.
}
\label{Fig:EfficiencyRedshiftAllModels}
\end{figure*}

\begin{figure*}[htp]
  \centering
  \begin{tabular}{cccc}
    \includegraphics[width=42mm]{ApparentLumFun_udf_wfc3ir_f160w_0016_1.0_0.3_0.02_m1_c1_SizeEvol_Match1.eps}&
    \includegraphics[width=42mm]{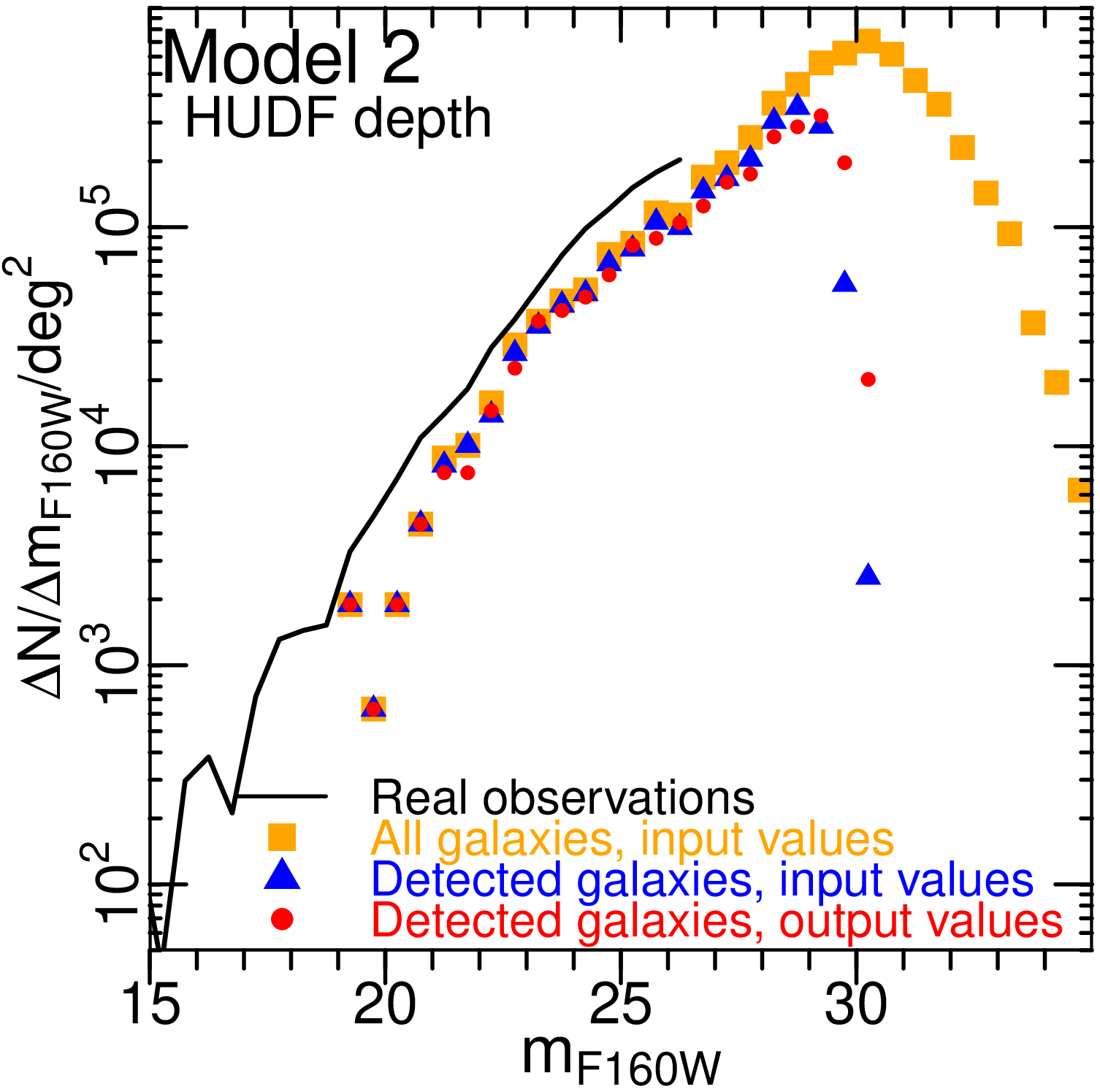}&
    \includegraphics[width=42mm]{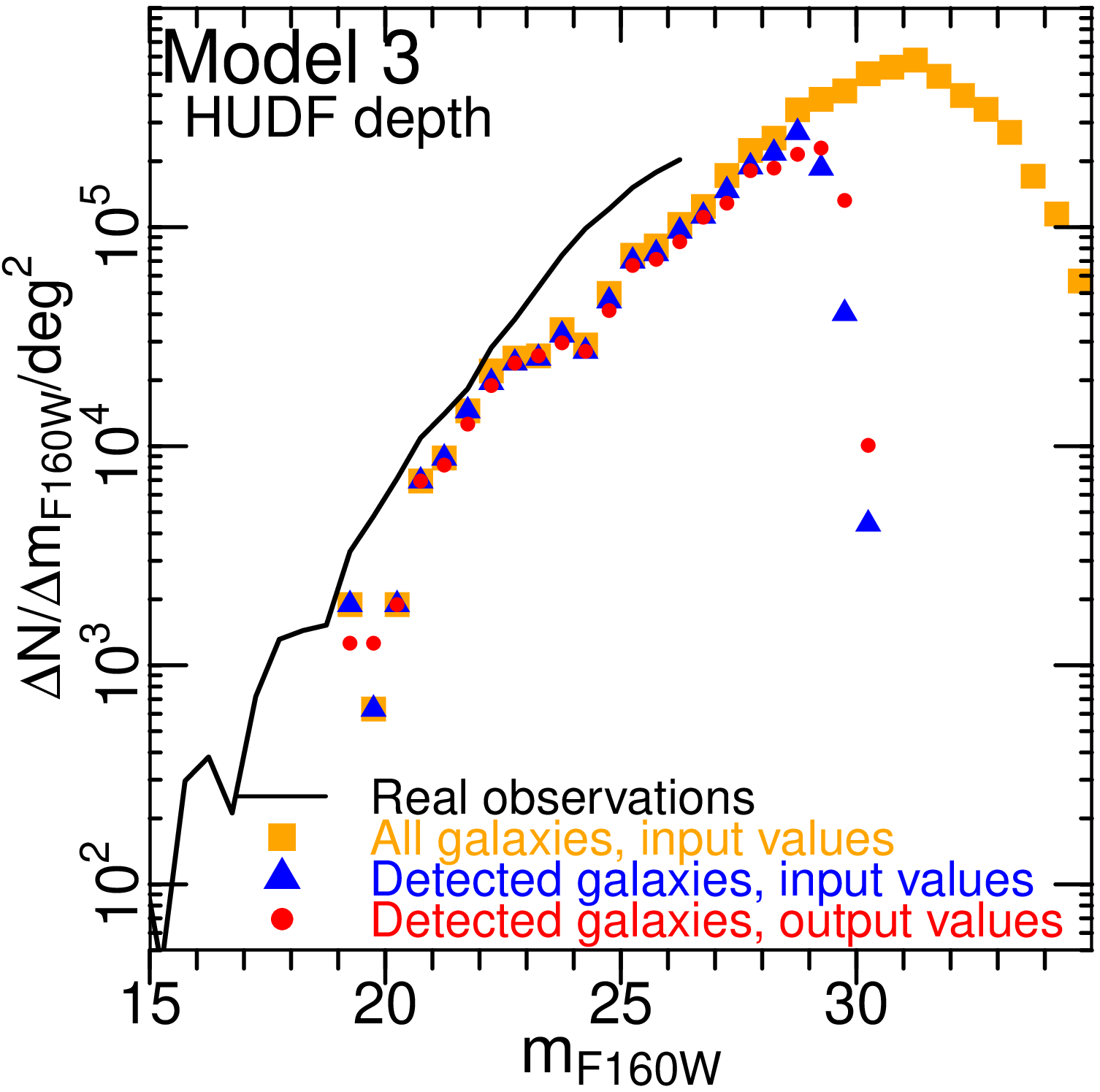}&
    \includegraphics[width=42mm]{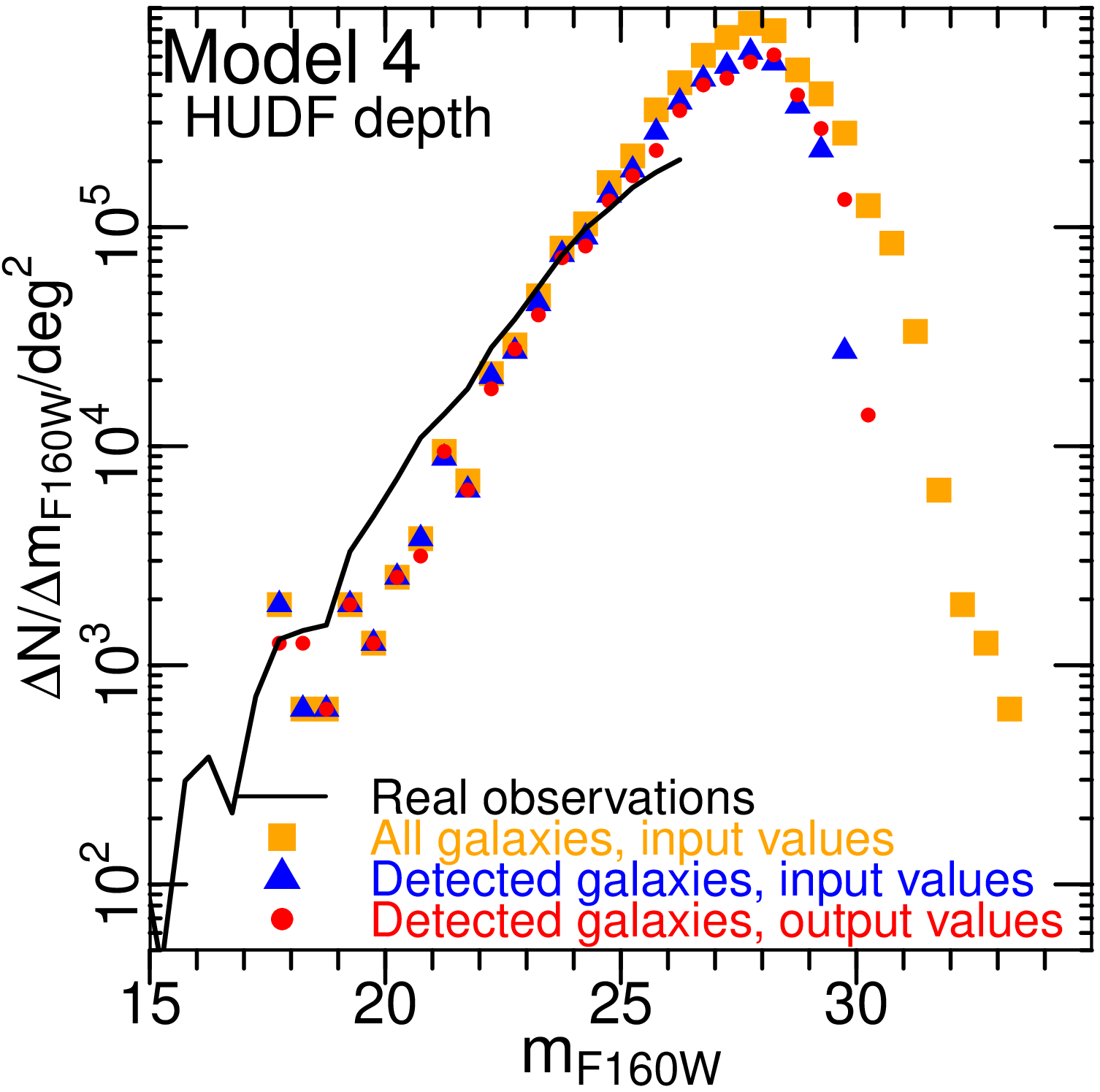}\\[2mm]

    \includegraphics[width=42mm]{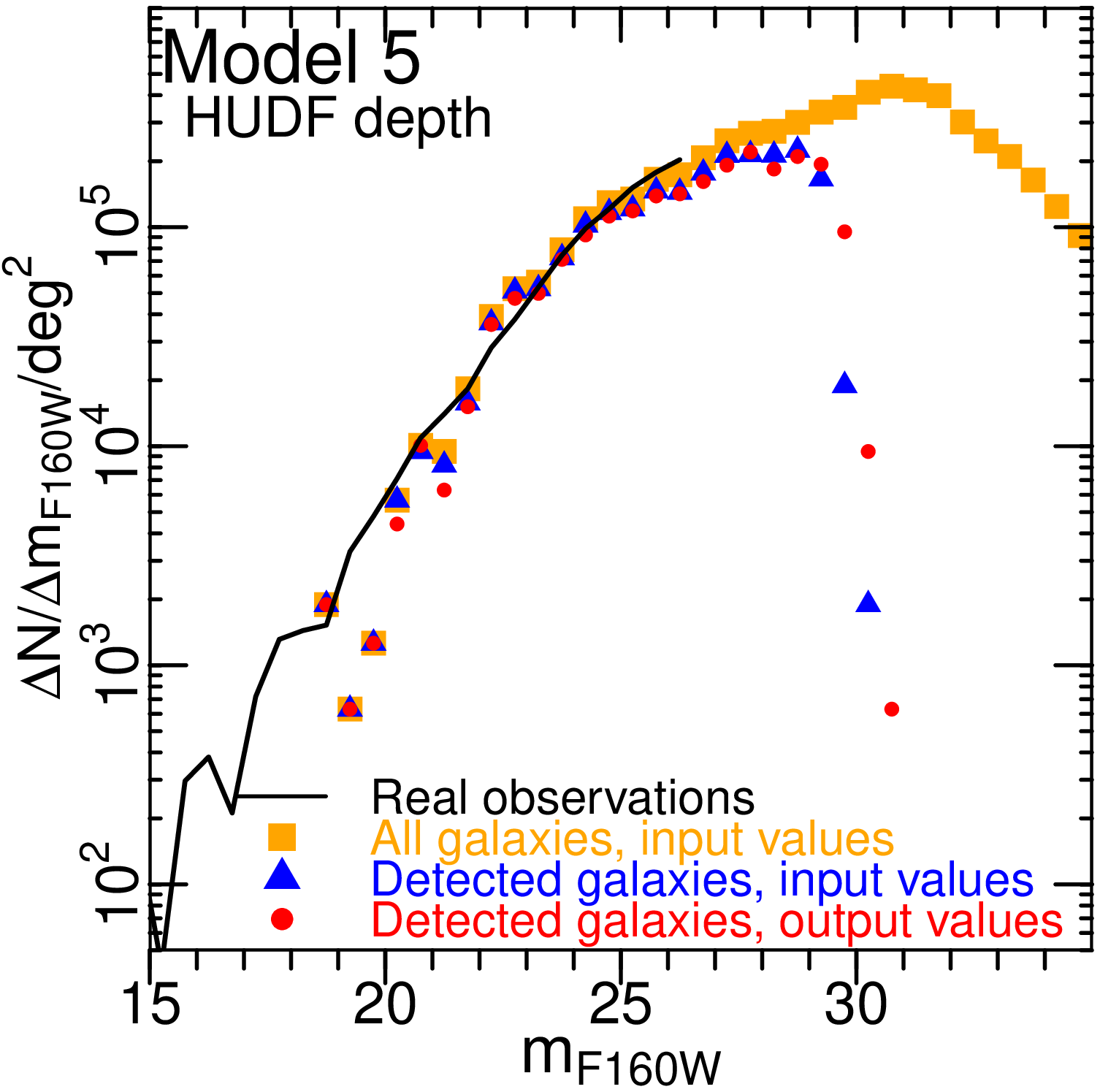}&
    \includegraphics[width=42mm]{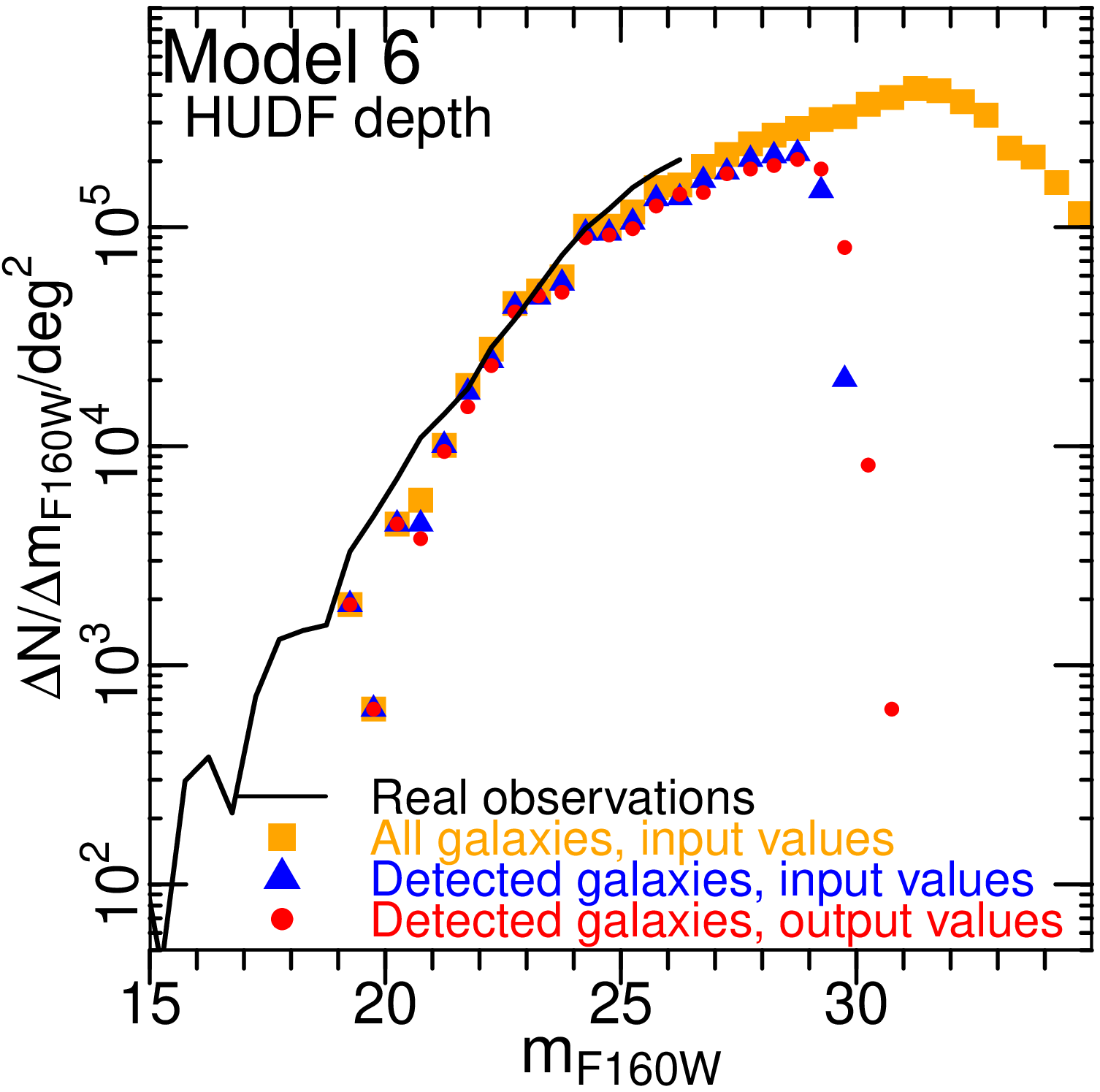}&
    \includegraphics[width=42mm]{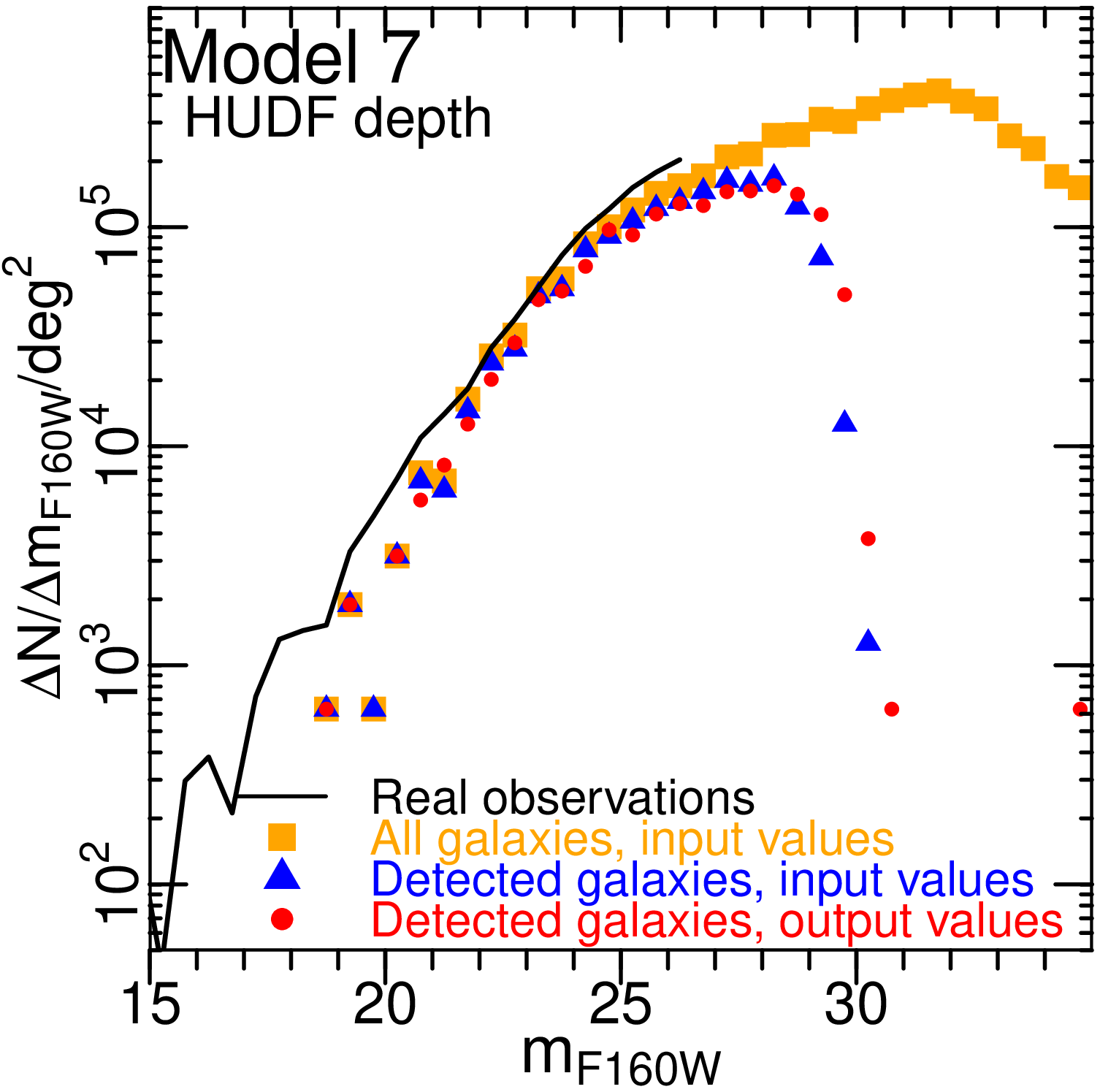}&
    \includegraphics[width=42mm]{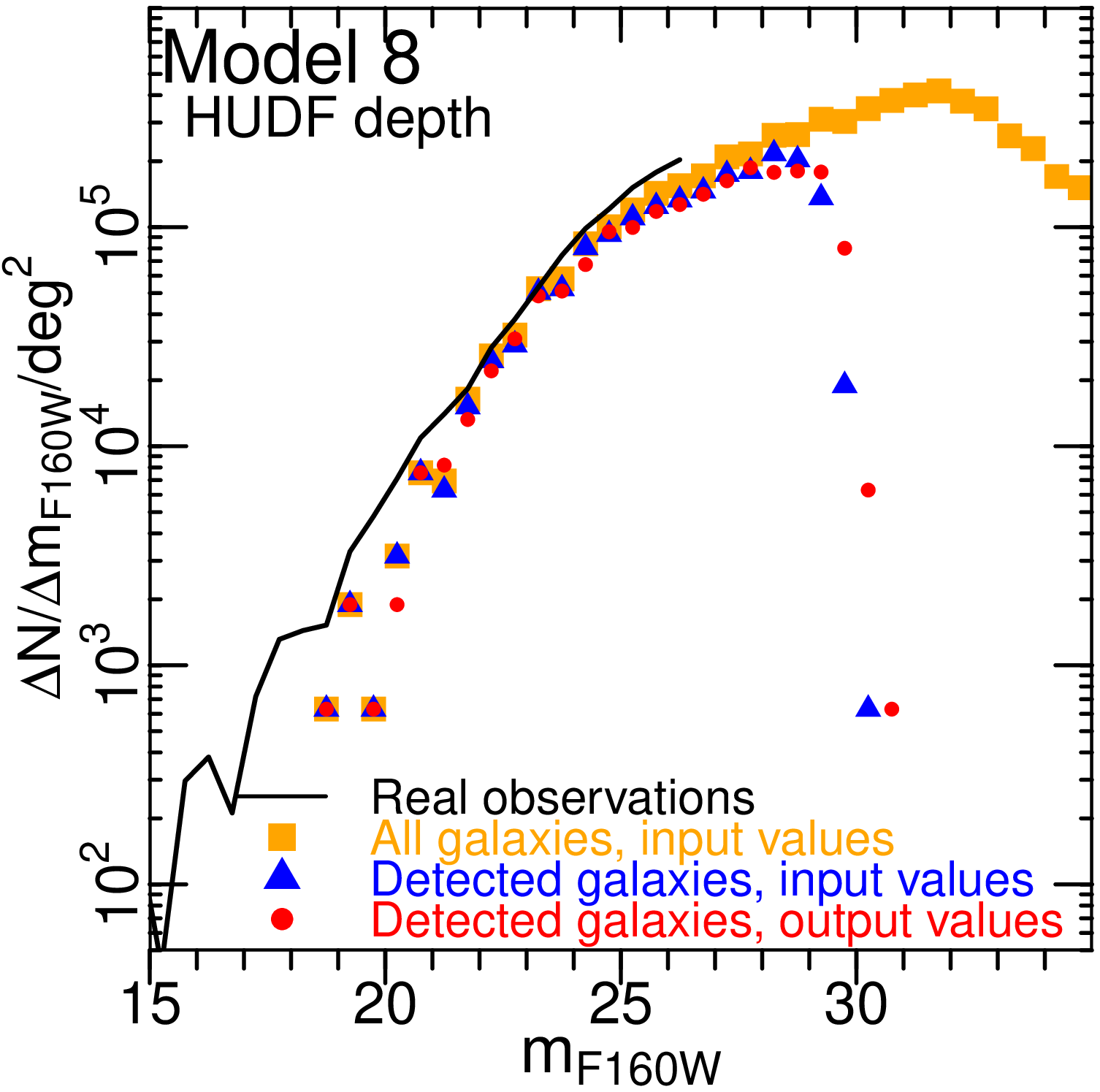}\\[2mm]
\end{tabular}
\caption{
Apparent F160W magnitude distribution from simulated \textit{HST\/} images at
HUDF depth for all models tested in this paper. 
Symbols are the same as in Fig.~\ref{Fig:LumAndRadFunSM}.
The biggest differences in the shapes of the
simulated distributions come from the use of different stellar mass-halo
mass relations (Models 1--4). Removing dust or metal content (Models 5 and 6) mostly shifts the
distribution toward brighter magnitudes.
}
\label{Fig:LumFunAllModels}
\end{figure*}

\begin{figure*}[htp]
  \centering
  \begin{tabular}{cccc}
    \includegraphics[width=42mm]{RadFun_udf_wfc3ir_f160w_0016_1.0_0.3_0.02_m1_c1_SizeEvol_Match1.eps}&
    \includegraphics[width=42mm]{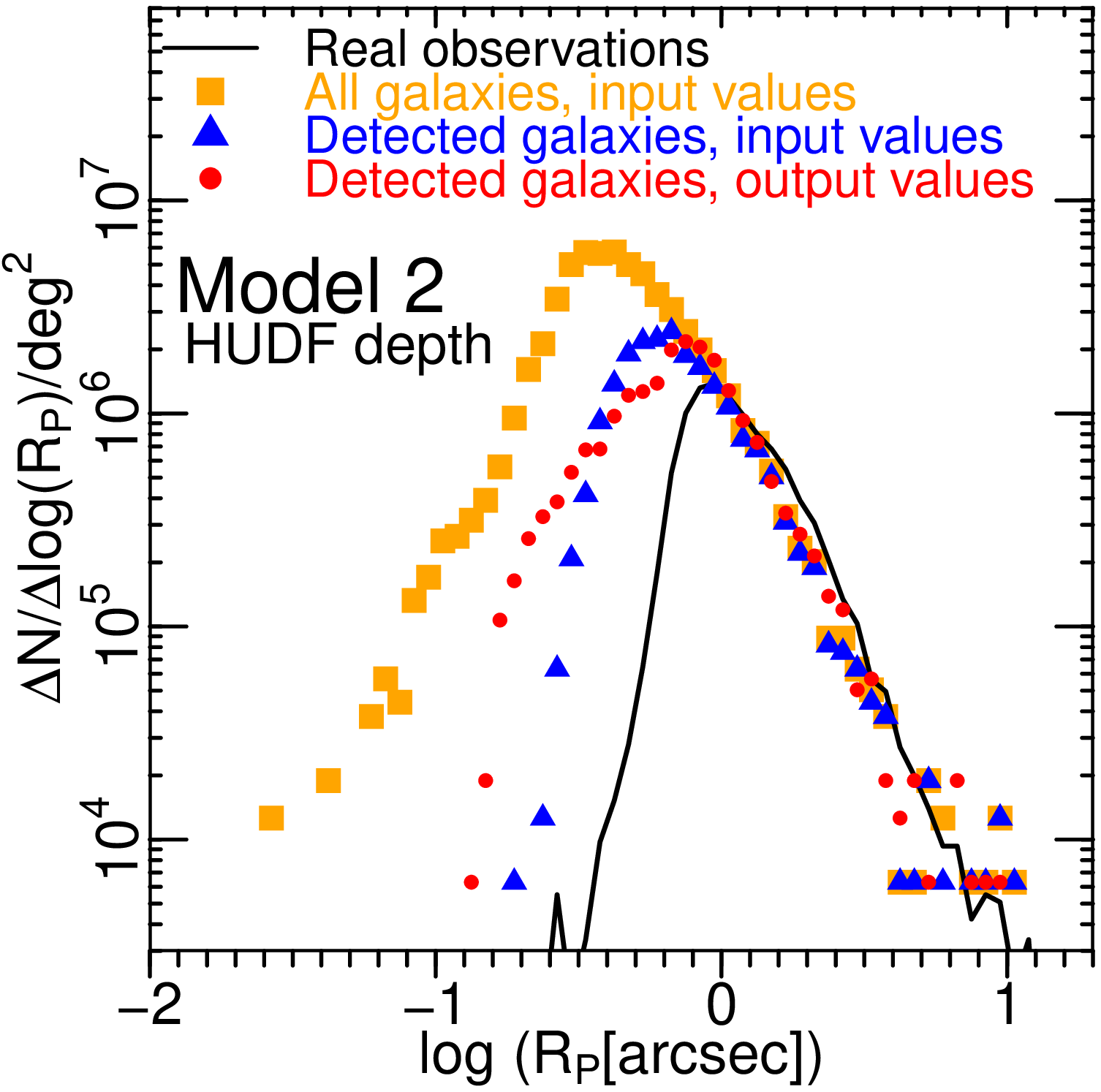}&
    \includegraphics[width=42mm]{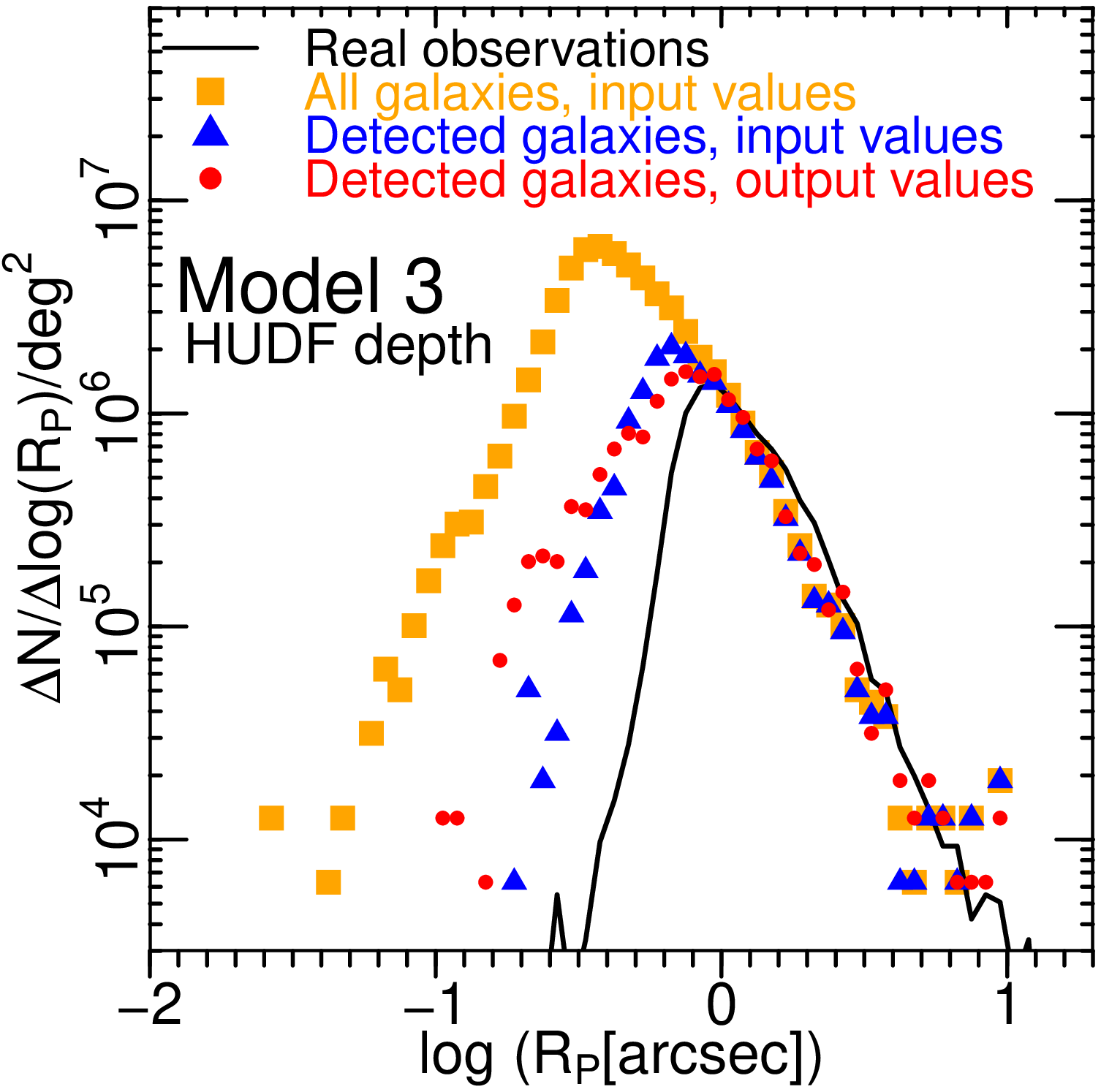}&
    \includegraphics[width=42mm]{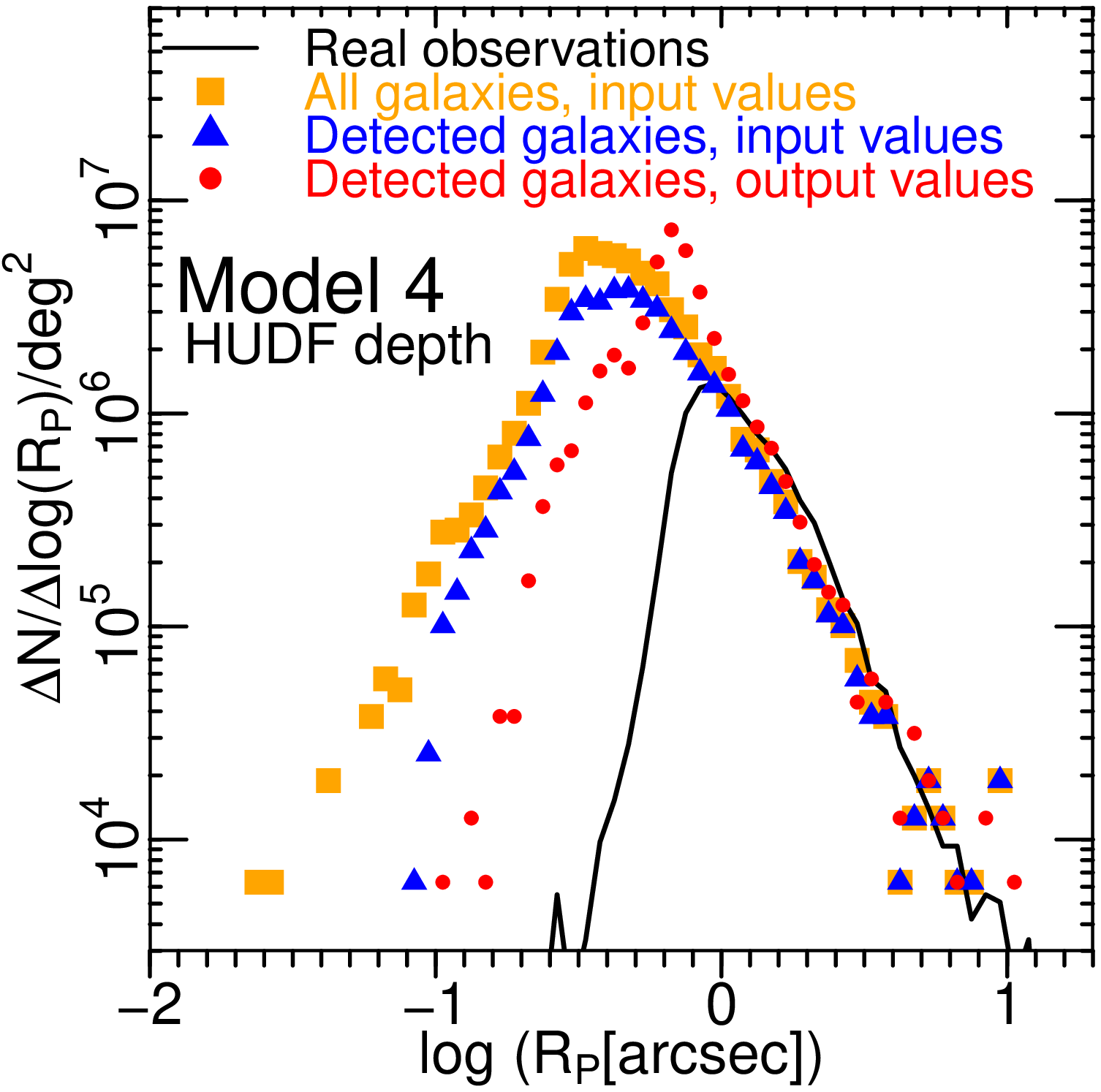}\\[1mm]

    \includegraphics[width=42mm]{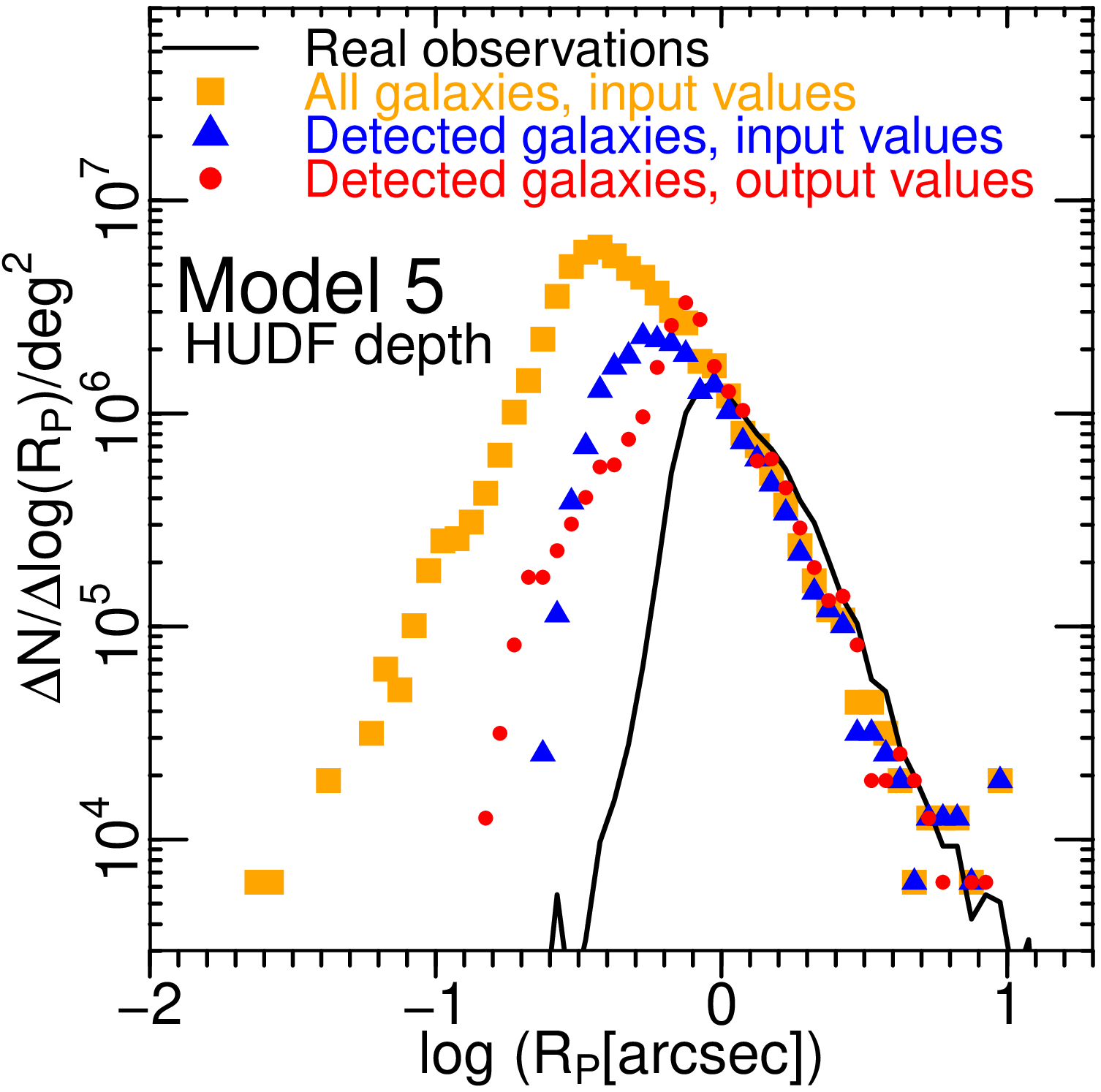}&
    \includegraphics[width=42mm]{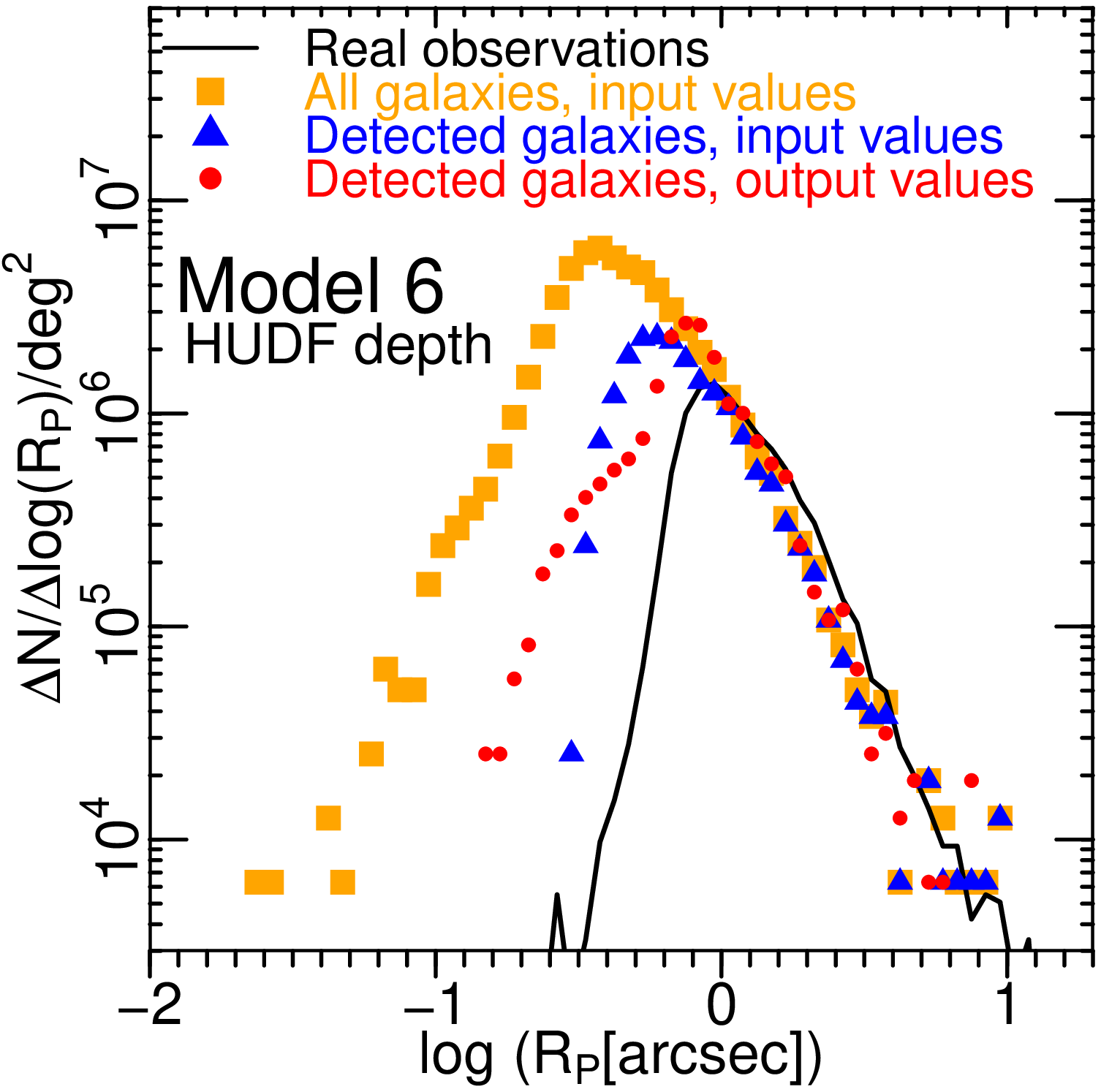}&
    \includegraphics[width=42mm]{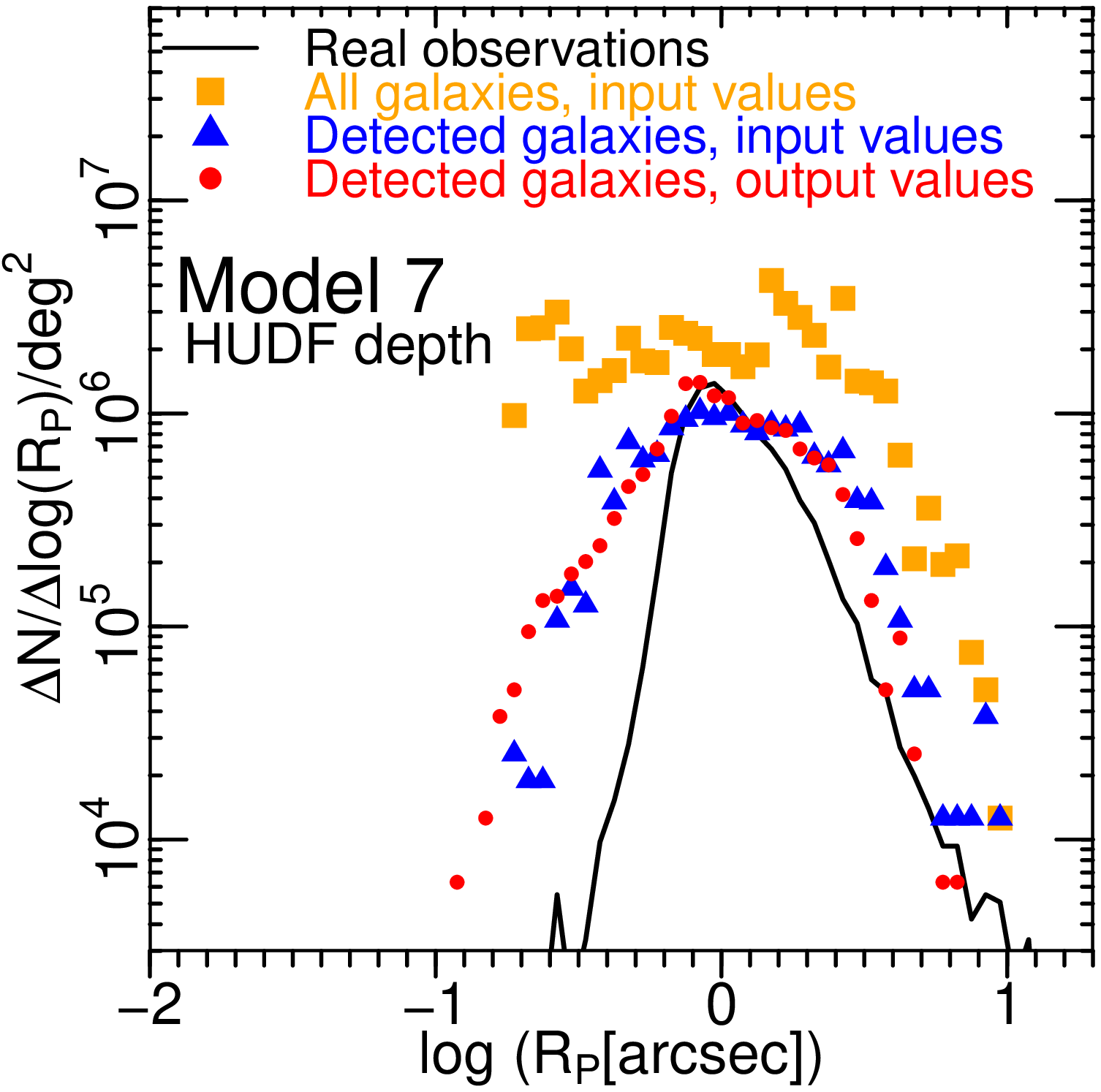}&
    \includegraphics[width=42mm]{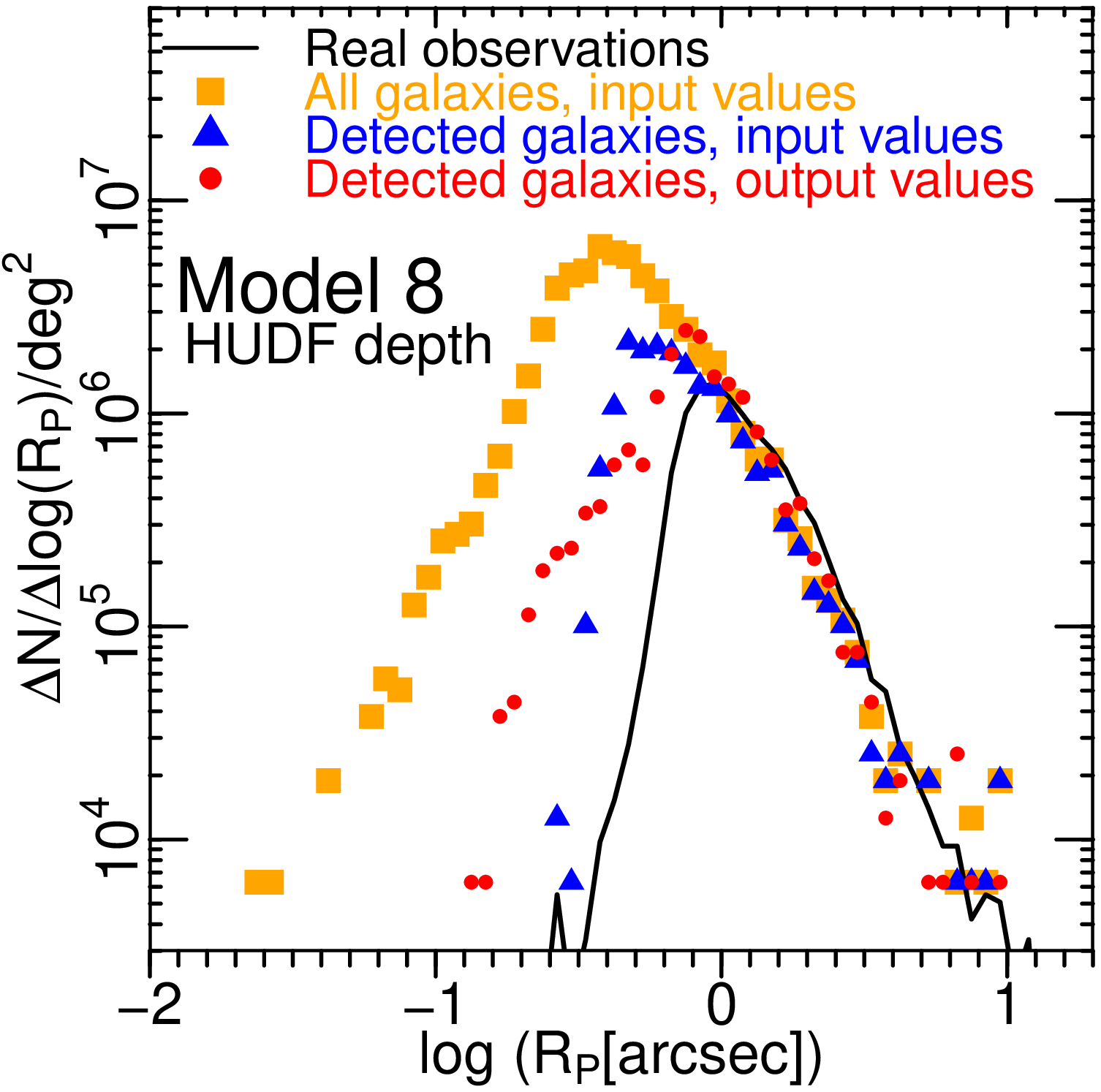}\\[1mm]
\end{tabular}
\caption{
Size functions from simulated \textit{HST\/} images at HUDF depth for
all models explored in this paper. Symbols are the
same as those in Fig.~\ref{Fig:LumFunAllModels}. Reasonably good agreement
for larger objects
is found
with the real observations (black line) except 
for Model 7, which has larger galaxy
sizes that do not scale with the halo size evolution.
}
\label{Fig:RadFunAllModels}
\end{figure*}

\noindent\textit{Luminosity and size distributions} ---
The luminosity and size distributions for all galaxies 
detected on simulated images from the reference model
are shown in Figure~\ref{Fig:LumAndRadFunSM}, together with
the respective distribution of model input values and {\tt
SExtractor}-measured output values. When compared to the 
observed luminosities and sizes derived from the
CANDELS GOODS-S Multi-wavelength Catalog \citep{guo2013b}, the {\tt SExtractor} output
values from the reference model (red) agree surprisingly
well. For the luminosity distribution, a good fit is found
at intermediate magnitudes, but the model predicts slightly
too few galaxies
at the bright end (possibly affected by sample variance
due to the small simulated image area). The drop in the input
luminosity distribution at $m_{\rm F160W}\sim 32$ is an artifact of
the finite mass resolution in the dark matter
simulation. Since the detection limit of the HUDF-depth image ($m_{\rm F160W}\sim 30$)
is 2 mag brighter, we are safe from this artificial incompleteness.
The
{\tt SExtractor} size distribution also agrees with the observations,
although the model seems to be slightly shifted toward smaller
sizes. The bias between the input and output distributions
of detected galaxies is also visible in this figure.
The HUDF-depth image shows that the peak value of the output $R_{\rm
P}$ distribution is shifted toward bigger radii by $\sim 0.1$~dex
with respect to the input distribution. Since this bias is not as
strong in the GOODS-depth image, we conclude that the amplitude of the
bias depends on the properties of galaxies near the detection limit, which
differs in the GOODS and HUDF-depth images.
Another interesting
feature is that the output size distribution departs from the input
size distribution at small values of $R_{\rm P}$ (down to the PSF
size), where measured sizes for small galaxies are much less than their
true sizes.
This effect was also discussed above (Fig.~\ref{Fig:OutputMagVsOutputRad}).

\subsubsection{Results from Other Models}\label{Sec:ResultsFromAllModels}

The statistics derived from simulated images built with other models
are qualitatively similar to those of the reference model and follow similar trends. However, 
the simulated universe is sensitive to changes in the model parameters,
and we can easily distinguish variations among the derived statistics
for different models.

Figure~\ref{Fig:ImagesAllModels} shows the simulated images from all models. 
At first glance it is easy to recognize
differences between the reference model and the others. For
example, some images show a deficit of small galaxies (e.g., Models 2
and 3) or an excess of them (e.g., Model 4). Models 5 and 6
with no dust and low metallicity look brighter, and Model 7
with no size scaling has much larger galaxies.

Figure~\ref{Fig:ApMagVsPetroRadAllModels} shows the size versus apparent
magnitude for two models that stand out from the others.
Model 4, with the linear SMHM relation, predicts higher stellar
masses in small-mass halos than the reference Model 1,
especially in the high-redshift universe where halos are just starting
to assemble. The net effect is to make compact galaxies brighter.
Compared with
the reference model in Figure~\ref{Fig:ApMagVsPetroRad}, a
much higher fraction of the galaxies are detected by {\tt SExtractor},
especially in the $z<2$ redshift range, since most of the galaxies on
the image are brighter than the magnitude detection limit.
Model 7, with a non-evolving size-mass relation, has a very different
distribution of apparent sizes compared with that of the reference
Model 1 and the observed distribution.  This is a powerful demonstration
that galaxies were smaller in the past, even at fixed mass.

The median apparent magnitudes at the mass completeness limit 
have different redshift dependences among the
models. In Figure~\ref{Fig:MedianMagAt20MassAllModels},
most of the models follow the same behavior as Model
1, with the exception of Models 4 and 5. Model 4 has median
magnitudes at the mass completeness limit 1.5--3 mag brighter
than those of Model 1, since the linear SMHM relation assigns more
stellar mass content per unit halo dark matter mass. Model 5,
with no dust, brightens
at $4<z<7$ in comparison with Model 1, indicating that dust typically
dims galaxies about by $\sim1$~mag in the redshifted filter bandpass. All the models, however, 
flatten at high redshift as discussed for the reference model.

Figure~\ref{Fig:EfficiencyAllModels} shows detection efficiencies for all models
as a function of both magnitude and size. 
The number count efficiency as a function of magnitude varies for all models
similarly to that of the reference model,
declining slowly to 80\% just above the magnitude detection
limit and then dropping dramatically, irrespective of redshift.
The light detection efficiency as a function of magnitude also shows a close
similarity among the models, dropping to 60\%--70\% just above the magnitude
detection limit.

Regarding the number count detection efficiency as a function of size, 
most models have high efficiency (80\%--90\%) for larger objects with a sharp drop 
at input radii of $\sim 1\arcsec$. The drop is due to smaller objects also having
smaller fluxes so that they are closer to the detection limit.
There are, however, two models that stand apart.
The linear SMHM relation in Model 4 makes compact, low-mass galaxies that are
more massive and brighter
than those in Model 1, and this greatly enhances the detectability of smaller objects.
As a consequence, the
detection efficiency stays above 50\% at input sizes down
even to 0.1\arcsec\ for $z>4$. At the other extreme, Model 7 (with the non-evolving size-mass
distribution) has larger sizes and much lower detection efficiency at high redshift
compared to the other models.  The larger size distribution is seen because
galaxy sizes are not scaled in proportional to the halo sizes.
Consequently high redshift galaxies are large fuzzy objects that are strongly
suppressed in the images by the $(1+z)^4$ cosmological surface brightness dimming,
and very few of them (10\%--20\%) are detected.  The smaller sizes of galaxies at high redshift
in most of the models are essential to making them detectable.
The light detection efficiency as a function of size shows similar effects,
although the differences for Models 4 and 7 are less dramatic.

The detection efficiencies as a function of redshift in
Figure~\ref{Fig:EfficiencyRedshiftAllModels} are only upper limits to
the real values, since we are missing halos smaller than the mass completeness limit in the dark matter simulation. 
We calculate these efficiencies using only galaxies that are above the median magnitude
set by the simulation mass resolution.  The distributions for different models
are qualitatively similar although there are some
significant variations.
The efficiency amplitudes depend strongly on the model
and on how compact and luminous the galaxies are, especially in the case
of the number count efficiency.  For example, Model 4, with its more
compact galaxies, shows by far the highest number count detection efficiencies,
$>60$\% at all redshifts.
The light detection efficiency is higher than the number count
efficiency for all models.  It is noteworthy that Model 5, with no dust, shows the
highest light detection efficiencies at $z>4$.  These experiments show that the
efficiencies depend on details of the models and that understanding them is
key to inferring the physics of galaxy formation from observations.

Figure~\ref{Fig:LumFunAllModels} shows luminosity functions, which are
especially suitable for exploring the effects of our different models
on the simulated images. Changing the SMHM relation
modifies the stellar mass distribution across redshift, which can
be seen directly as a change in the apparent luminosity functions.
According to
Figure~\ref{Fig:HaloMassVsStellarMassAndMassFunAtZ0},
the linear SMHM relation (Model 4) predicts
a much higher $\Ms$ at a given $\Mhalo$ than the Guo
et al.\ relation, especially for halos with $\Mhalo\lesssim 10^{11}
\Msun$. That
shifts the peak of the luminosity function toward brighter
magnitudes compared with the reference model. Since the
galaxies are brighter, more of them are
detected by {\tt SExtractor} or seen by eye (e.g.,
Figs.~\ref{Fig:ImagesAllModels} or~\ref{Fig:ApMagVsPetroRadAllModels}).
With the Behroozi et al.\ SMHM relation, $\Ms$ is higher than with the
Guo et al.\ SMHM relation at small values of $\Mhalo$, but lower
at high $\Mhalo$ (for any redshift). The net effect is a shift
of the peak to brighter magnitudes, but at the expense of fewer
galaxies just brighter than the peak. At $z=0$, the
Behroozi et al.\ SMHM relation is similar to that of Guo et al., so the luminosity
function of Model 2 is closer in shape to that of Model 1. Galaxies
in Model 5 (with no dust) are, as expected, more luminous than in
the reference model, in some cases $\sim 0.7$ magnitudes brighter in the
range $22\lesssim m_{\rm F160w} \lesssim 25$. Model 6 (with low
metallicity) behaves similarly to Model 5, but the magnitude shift is not as large. 
These 2 models also give good fits to the observations. The last 2 models
share the same input $z=0$ luminosity function as the reference  
model, since only the size of galaxies are modified.
Therefore, Model 8, which uses $R_{50}$ in the matching between model
and SDSS galaxies, has a luminosity function at $z=0$ similar to that of Model 1.
However, Model 7 is slightly different at the
onset of incompleteness, since here galaxy sizes are not scaled and
are therefore larger than in the reference model. The galaxy sizes
affect the surface brightnesses and so are
influential in determining the detectability of faint galaxies.

The size distributions in Figure~\ref{Fig:RadFunAllModels} mostly
behave in the same way as in the reference  
model, except for Models 4 and 7. For Model 4 (linear
SMHM relation), the peak of the output size distribution shows a
larger bias toward bigger sizes (by $\sim 0.2$~dex) compared
to the input distribution. Note also that
the output size distribution for smaller detected galaxies always falls below that of the
input distribution. The input sizes of the smallest detected galaxies extend well below the FWHM of the PSF
(0.151\arcsec), but those bright enough to be detected have {\tt SExtractor} 
sizes comparable to the FWHM.  (We have not modified the {\tt SExtractor} sizes to
remove the effects of the PSF.)
The size distribution for Model 7 is, not surprisingly,
very different from the other distributions,
since its galaxies lie on a non-evolving size-mass relation rather
than having their sizes rescaled by the evolving sizes of their
halos.
That results in many large, faint galaxies at high
redshift that are difficult to detect.

\begin{figure*}
  \centering
\includegraphics[width=0.4\linewidth]{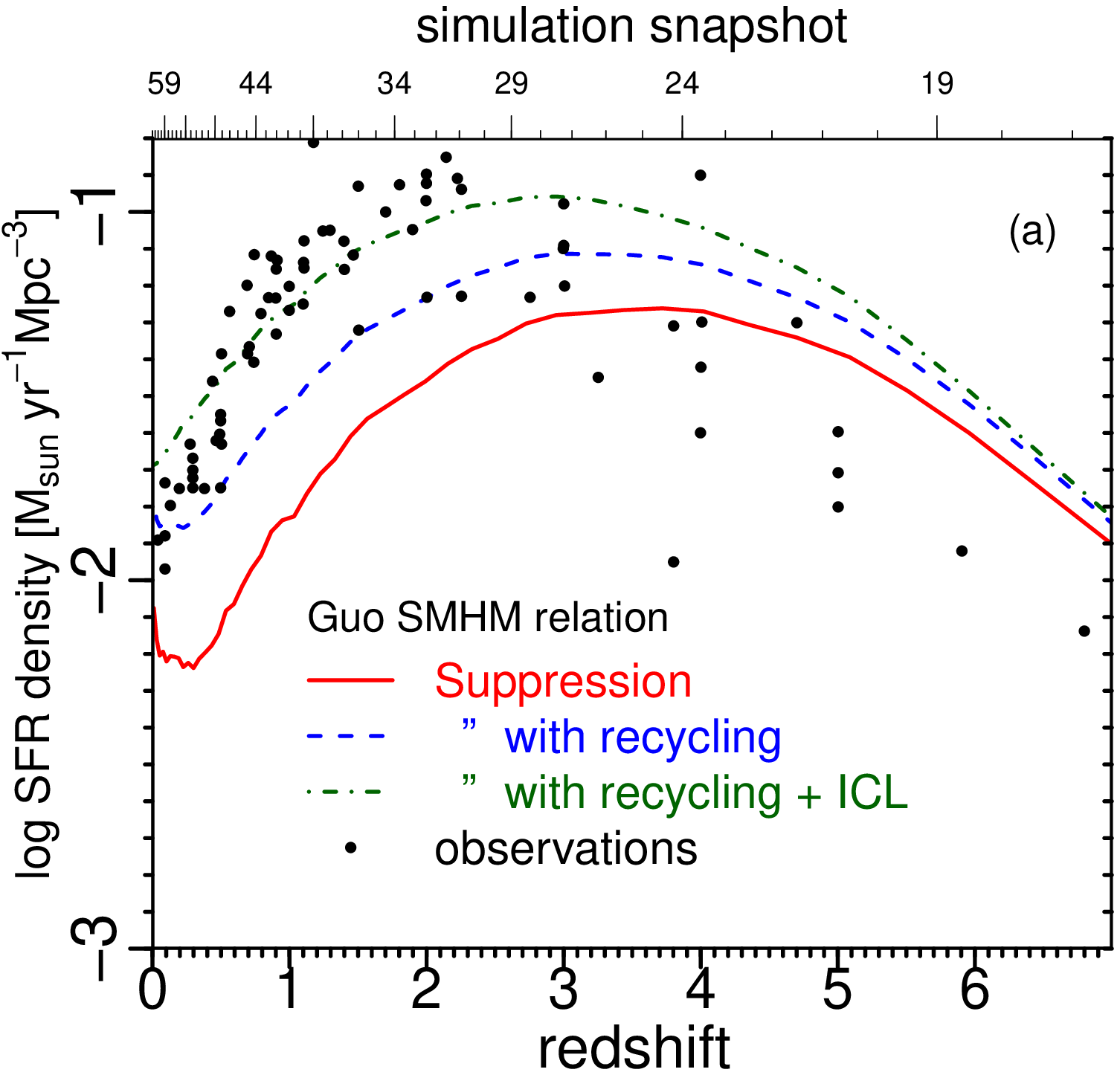}\hfil
\includegraphics[width=0.4\linewidth]{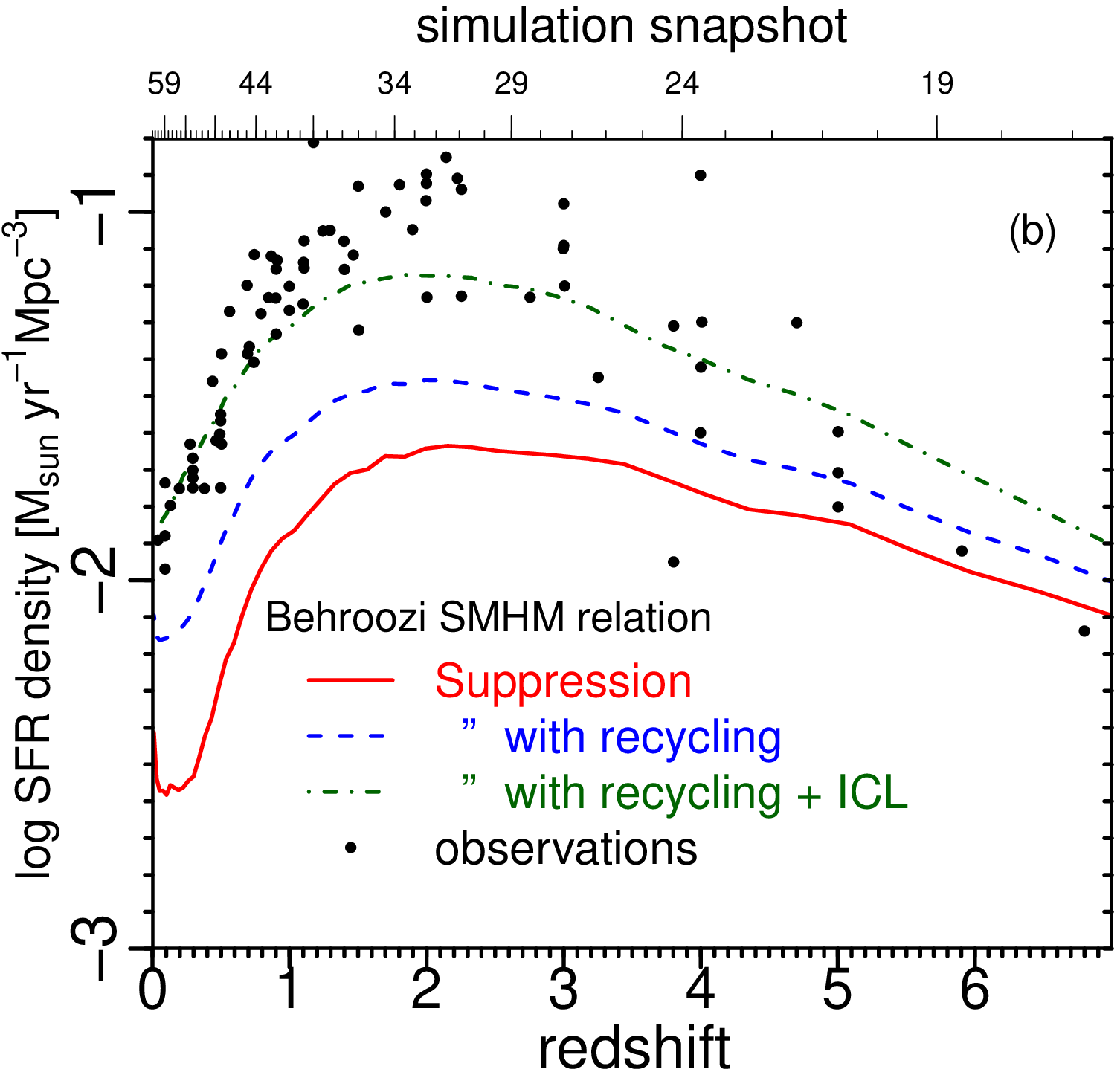}
\caption{Cosmic star formation rate density (SFRD) as a function of redshift.
(a) Reference SMHM model from \cite{guo2010}.
(b) Redshift-dependent SMHM model from \cite{behroozi2013}.
The black dots show observed SFRDs compiled by \cite{behroozi2013}.
The solid red line is the SFRD for our model with
suppression of past star formation to create a
monotonically increasing stellar mass in halo merger trees.  The
dashed blue line includes the additional star formation from instantaneous recycling
of mass lost from shorter-lived more massive stars to maintain the stellar
mass.  The dot-dashed green line results from assuming that rather than
suppressing star formation to enforce a monotonically increasing stellar
mass, all halos form the stars required by the SMHM relation but the excess
stellar mass is ejected from the halos to form intracluster light (ICL).
The models including ICL fit the observations best but require that most
stars reside in the ICL rather than in galaxies (see text for discussion).
}
\label{Fig:SFRD}
\end{figure*}

\section{Cosmic Star Formation Rates}

The cosmic star formation rate density (SFRD, the mass of stars formed
per unit time and per unit comoving volume) and its evolution with
redshift have played a key role in studies of galaxy formation
\citep[see][and references therein]{madau2014}.  We
have postponed consideration of the SFRD until now because, although
this relation is derived from observations and is often portrayed
as an ``observed'' relation, it is unsuitable for direct comparison
with our models.
In previous
studies (e.g., \citealp{behroozi2013}), predictions of the SFRD are compared with observations in
the theoretical domain, i.e., by combining observed galaxy counts
and colors with corrections for dust absorption but without corrections for missing
light at low luminosity and surface brightness.
We argue instead that it is better to project theoretical
models into the
observational domain by creating simulated data and making comparisons directly
with the observed quantities (``forward modeling'').  However, a complete study using that approach is beyond
the scope of this paper.

Nonetheless, a comparison between our model properties and SFRD
values from the literature is illuminating.  Figure~\ref{Fig:SFRD}(a)
shows the SFRD in our reference model using the SMHM relation from
\cite{guo2010}.  The solid red line is the SFRD that results from
our approach of enforcing a monotonically increasing stellar mass
in halo merger trees by retroactively suppressing star formation
(see \S\ref{Sec:Halo-StellMassRelation} and Appendix \ref{Sec:Appendix1}).
Observations from the compilation of \cite{behroozi2013} are also shown.
The model SFRD lies above the data points at high redshift
($z>4$) but below them at low redshifts ($z<2$).

Figure~\ref{Fig:SFRD}(b) shows the SFRD in our model using the
redshift-dependent SMHM function from \cite{behroozi2013}.
Since \cite{behroozi2013} forced their model to match
the observed SFRD, one might expect that this SMHM function
would also be consistent with the observations.
However, while our model SFRD (solid red line) is similar to the
observations at high redshift ($z>6$), it is still too low at
lower redshifts.

We have identified two significant effects that
can raise the SFRDs in our models.  First,
additional star formation is required to replace the stellar
mass that is lost as massive stars
reach the end of their lifetimes and die.
We compute the additional star formation from
the simple approximation given in Eqn.~(14) of \cite{behroozi2013}
for the fraction of mass lost from a single starburst as a function
of age; this function is integrated over time to determine the
ongoing star formation rate with recycled material.
The resulting enhanced SFRD, shown in Figure~\ref{Fig:SFRD}
as the dashed blue line, is increased by factors ranging from
1.5 at $z\sim3$ to 2.5 at $z\sim0$.
This recycling effect is
important and should be incorporated into future models, but it
does not fully explain the difference between the computed SFRDs and
the observations.

The second potentially important effect is the ejection of stars
from galaxies during merging.  This is a crucial feature of the
\cite{behroozi2013} calculation: the fraction of stars that escape
from galaxies is an adjustable parameter.  Those stars are then
subsequently replaced by additional star formation.  This parameter
enables fitting both the observed SFRD and the galaxy masses in
massive cluster-scale halos.  Most of the intergalactic stars end
up in clusters of galaxies and hence contribute to the diffuse
intracluster light (ICL).

To include the possibility of star formation enhancement through ICL,
we adopt a more conservative procedure.  Our standard approach with star formation
suppression reduces the star formation rates in merger trees to ensure that
the stellar mass of galaxies never decreases.  An alternative model is
to allow all galaxies to form the stars required by the SMHM relation, 
and then to eject the excess mass (if any) to the ICL.
The dot-dashed green lines in Figure~\ref{Fig:SFRD} show the enhanced
SFRDs that result from this procedure.  With this addition, the model SFRD for the \cite{behroozi2013} SMHM relation
has been increased by another factor of 2 and
agrees reasonably well with the observations at all redshifts.

Does this agreement imply that there \textit{must} be a large fraction of stellar mass ejected to
the ICL in order to match the observed SFRD?  Probably not.
In the \cite{behroozi2013} computations, 30\% of the stellar mass in the universe
lies outside galaxies, with the ICL mass fraction being higher in clusters
(P.~Behroozi, private communication).
Such a large fraction of extragalactic stars may be in conflict with observations
of the ICL, which suggest ICL stellar mass fractions closer to 10\%
\citep[e.g.,][]{mihos2005,zibetti2005,krick2007,gonzalez2007,presotto2014,montes2014}.
A model that relies on hiding much of the stellar mass outside
of galaxies in order to enhance the SFRD may conflict as strongly with the observations as one that
does not generate much intracluster light but has a lower SFRD.

Instead of treating the inferred SFRD and ICL as known
quantities, a better approach would be to apply the forward modeling methodology
discussed in this paper: from simple theoretical models of galaxy
masses and star formation rates, create simulated observations,
including the SFRD and the ICL.  Then the comparison between the models
and the observations should be carried out entirely in the observational
domain.  It seems quite plausible that a consistent model can be
created that fits the observed galaxy masses, observed SFRD indicators,
and measured ICL properties; such a model will not necessarily
demand a very large stellar mass in the ICL.

\section{Discussion}\label{Sec:Conclusion}

\subsection{Summary of Results}

This paper demonstrates that a credible simulated universe, consistent with deep \textit{HST\/} images, 
can be constructed using only a
few basic assumptions and a small number of parameters. The
simplicity of our model, in contrast with the complexity of most other 
models of galaxy evolution, enables robust
comparisons with the observed universe. 
A key element of this comparison is to project our models of evolving galaxy populations
into the observational domain and to extract measurements from the simulations
using the same tools as used to analyze the real images (``forward modeling''). Since our 
simulated \textit{HST\/} images include all
relevant cosmological, instrumental, and selection effects, they can be compared directly 
with real \textit{HST\/} images. In order to simulate the clumpiness and
internal structure of galaxies, we cut out images of SDSS galaxies 
as templates for our model galaxies (instead of the commonly used
smooth S\'ersic profiles), with their fluxes and sizes rescaled to
reflect the properties of model galaxies in our simulations. 

The key assumption of our theoretical modeling is that most of the
information needed for a first order match between the simulated
and observed universes is already encoded in the gravitational dynamics
of dark matter halos, as derived from cosmological $N$-body simulations. In this 
semi-empirical approach, we express the stellar mass and size
of a model galaxy as a function of the mass and size of its host
halo. This effectively avoids any detailed modeling of complex baryonic
physical processes, in contrast to semi-analytical and 
hydrodynamical simulations, with their larger number of assumptions and parameters. Our 
semi-empirical models are based on statistical matching between the
distribution of halo mass and size measured in the $N$-body
simulation to the distribution of stellar mass and size for real galaxies at $z=0$, 
thus guaranteeing a good match between the present-day simulated and observed galaxy populations.  
For the evolution of galaxy populations, we adopt several simple models for the 
redshift dependence of the stellar mass-halo mass (SMHM) relation. 
We also show that the evolution of the luminosities
and spectra of galaxies can be modeled, to good approximation, using
the star formation histories implied by the growth of the stellar
mass of each galaxy along the merger tree of its host halo.  The
galaxy spectra are determined by convolving the star formation
histories with stellar population synthesis models.

The analysis of simulated \textit{HST\/} images is a powerful tool for comparing
the predictions from different models of galaxy evolution.
By comparing directly in the observational domain, we can decide
which choices of parameters make an observable difference and which
do not. Our reference model assumes a non-evolving 
SMHM relation and  
non-evolving solar metallicity and dust content. Despite these simple assumptions,
it provides a good match to real \textit{HST\/} images, particularly
when comparing the luminosity and size distributions. Including plausible 
redshift-dependence in the SMHM relation does not radically alter the simulated \textit{HST\/} images. 
In contrast, a much less plausible, linear SMHM relation, with the stellar
mass proportional to the halo mass, leads to a far greater 
abundance of compact, high-luminosity galaxies in low-mass halos
at high redshift. From this and our other models, we conclude that the stellar
mass efficiency per unit halo mass should indeed peak at stellar
masses around $M_{*}$ and decrease for smaller and higher masses.  Metallicity
and, especially, dust have a strong effect on the observed galaxy
properties, with their luminosities increasing by up to
0.7~mag when removing dust or metals
from the galaxies.  From the analysis of galaxy sizes, we conclude 
that galaxies must be smaller in the past.
A simple linear
scaling between galaxy and halo sizes provides a good
match between the size distributions of simulated and real \textit{HST\/}
images.
If there were no evolution in the sizes of galaxies, the deep \textit{HST\/} images
would appear almost empty.

We also find that the measured values of size and luminosity of
galaxies in the simulated images are strongly biased. {\tt SExtractor} underestimates the luminosities and
sizes of simulated galaxies especially around the magnitude detection limit, as the extended low 
surface brightness components of galaxies are lost in the
image background noise. These biases vary among the models, depending primarily on the size-luminosity relation 
of galaxies. For example, our model with a linear SMHM relation predicts more light for small, faint galaxies, which makes them easier to detect compared with our reference model.

The detection efficiency of galaxies, measured by
comparing the number and luminosities of detected galaxies in the simulated images to those
for all model galaxies, also reveals some interesting
results.  For the reference model, the number detection
efficiency declines slowly with magnitude as objects
get fainter to $\sim80$\% just above the detection limit and it then drops more rapidly. 
This behavior is similar at low and high redshifts,
since it depends only on whether galaxies are bright enough to pass
the magnitude cut. The fraction of light recovered from galaxies,
on the other hand, falls considerably faster with magnitude; it is
$\sim$60-70\% right above the detection limit and then
drops sharply. The reason for this is that {\tt SExtractor}
not only fails to detect faint galaxies but also underestimates the fluxes
of the ones it does detect. Note that although the model with a linear SMHM
relation behaves similarly as a function of magnitude; it has
a much higher detection efficiency at small sizes, since it assigns
more luminosity to the small, high-redshift galaxies, making them
easier to detect.
The fraction of missing light may be underestimated in our simulations because they truncate the
luminosity profiles of galaxies at two Petrosian radii.

The redshift dependence of the detection efficiencies is also interesting but requires some additional considerations. 
The numerator of the efficiency (i.e., the number of detected galaxies or the total light detected) is measured 
with reasonable accuracy. However, the denominator (the total number or luminosity of galaxies)
is not well determined because our images are missing galaxies with masses below the resolution 
of the milli-Millennium simulation. We address this issue by fitting the power-law tail of the 
luminosity function, and estimate the missing light by extrapolating 
the fitting function toward zero luminosity. Our measurements for the reference model show that the 
fitted power-law slope becomes steeper at higher redshift, changing from $\alpha\sim -1$ at $z<1$ to $\alpha=-1.75$ at $z=7-8$. 
Since the slopes are shallower than the divergence value $\alpha=-2$, the extrapolated 
missing light is always finite and the fraction of missing light is modest in our case.
Indeed, the light detection efficiency drops from $\sim$90\% at small redshifts to only $\sim$50\% at $z=7-8$.
These values depend sensitively on the slope of the luminosity function, which in turn depends
on the dark matter halo mass distribution. Since the power-law tail of this mass distribution has a 
slope of $\alpha=-2$ (confirmed both by theory and simulations), it would not be difficult to 
find models with suitable combinations of SMHM relations and mass-to-light ratios that predict a 
power-law slope of the luminosity function closer to $\alpha=-2$ or even steeper, resulting in very small detection efficiencies.  
We demonstrated this by our model with a linear SMHM relation, which presents a lower light efficiency than models with more plausible SMHM relations.
The total number of galaxies, on the other hand, diverges for slopes steeper than the limit $\alpha<-1$. 
Since our simulated universe presents steeper slopes, we conclude that the number counts
detection efficiency as a function of redshift cannot be accurately measured using the
milli-Millennium simulation at most redshift.

\subsection{Future Directions}

The main goal of this paper was to develop some first-generation simulations of deep \textit{HST\/} images 
using semi-empirical models of evolving galaxy populations. In this spirit, 
we have tried to strike a balance between models that are realistic enough for meaningful conclusions 
and models that are simple enough for computational efficiency and ease of interpretation. Indeed, we regard the
simplicity of our models,
with relatively few assumptions and parameters, as a definite virtue in a preliminary exploration such as this. 
Now that this initial analysis has been successfully concluded, it is appropriate to consider how our approach can be 
enhanced for greater physical realism and accuracy. In the remainder of this section, 
we outline several directions for future studies of this type.

A relatively straightforward enhancement of our models would be to base them 
on dark-matter simulations with larger comoving volumes and smaller particle masses, such
as the Millennium II \citep{boylan-kolchin2009} or Bolshoi \citep{kyplin2011} simulations. The larger volume would permit a more accurate 
simulation of both the small and large-scale distribution of galaxies in deep images than is 
possible with the milli-Millennium simulation. In particular, it should be possible to lay down 
light cones through the simulation volume to compute galaxy distributions rather than relying on
sampling the halo distribution as we do in this paper.
Such simulations, in addition to providing direct estimates of correlation functions and other clustering 
statistics on large scales, will provide estimates of the variance in counts and other properties of galaxies in smaller fields.
The higher mass resolution of larger dark matter simulations will enable the modeling of galaxy populations 
to lower masses and luminosities. This is especially important for the simulation of \textit{JWST\/} images, which will reach much 
fainter magnitudes than the \textit{HST\/} images considered here. 
The extension of model galaxy populations to lower masses will also improve estimates of detection efficiencies 
for the counts and light of galaxies, which depend on extrapolations of luminosity functions below the detection thresholds. 

It would be also worthwhile to make simulations with a larger suite of SMHM relations. 
The resulting simulated \textit{HST\/} images could then be compared with observed images and a 
goodness of fit assigned to each of these SMHM relations (by e.g. maximum likelihood). 
The requirement that the models satisfy constraints on the SFRD and ICL (\S4) could be added to the procedure at this stage.
In this way, the ``best" SMHM relation and confidence regions around it could be determined. 
As a byproduct, such statistical tests would indicate how much useful information \textit{HST\/} images 
really contain about galaxy formation and evolution. In particular, we would like to know whether 
the  evolution of the SMHM relation can be pinned down uniquely or whether substantially different 
SMHM relations lead to similarly good fits in comparison with \textit{HST\/} observations. 
This issue has important
theoretical implications; different SMHM relations presumably 
reflect differences in the physical processes in galaxy formation and evolution. This issue also has practical implications, because 
simulations like those presented here could be used to guide future observing strategies, with different telescopes, cameras, 
and filters, to determine the SMHM relation most efficiently and robustly and to resolve
ambiguities in the models.

Another improvement would be to allow the metal and dust content of 
model galaxies to evolve with redshift. Our results, based on non-evolving metal and dust content, 
show that differences in these quantities are readily apparent in simulated \textit{HST\/} images. 
A physically plausible scheme for evolving the metal and dust content should be based on the star formation and thus the metal 
enrichment history of each galaxy in the model. Unfortunately, the stellar 
metallicity and ISM dust optical depth also depend on the evolution of 
the ISM of each galaxy, including inflows and outflows, and these cannot 
be specified without introducing additional assumptions and parameters. 
This is why we have neglected the evolution of the metal and dust content in the present exploratory study. 
Nevertheless, such effects could and probably should be included in future studies, 
provided one is willing to pay the price of extra assumptions and complexity.

\acknowledgments

We are grateful for support of this project by the STScI Director's Discretionary Research Fund.
We thank Henry Ferguson and Michael Peth for help on running {\tt SExtractor}, Ching-Wa Yip for
help on running {\tt GALAXEV} and Victor Paul for helping with Data-Scope maintenance. 
We thank Massimo Stiavelli and Jason Tumlinson for helpful discussions in the early phases of
this project and Peter Behroozi for helpful input in the later phases.

\appendix

\section{Enforcing a Self-Consistent, Monotonically Increasing Stellar Mass in the Merger Tree}\label{Sec:Appendix1}

We have developed a self-consistent and recursive method that ensures a
monotonic increase in the stellar mass content of dark matter halos
across their merger trees. This retroactive suppression mechanism
functions in the following way: when a halo is measured to decrease
its current $\Ms$ value, we suppress accordingly the value
of $\Ms$ in its progenitor halos located in the closest
previous time steps, which consequently causes scatter away from the
one-to-one SMHM relation. First, we start with a root halo picked
from the collection $\lbrace h_{i}\rbrace_{i=1}^{i=N}$ of all $N$
halos found in the last ($k$th) time step in the simulation (most
likely to be at $z=0$), and compute $\Ms=\Ms(M_{\rm
halo})$ for all halos in its merger tree using the one-to-one SMHM
relation. Then we pick its progenitors located in the previous $(k-1)$th
time step and verify that $\Sigma M_{\rm s,prog}\leq M_{\rm s,root}$.
If not, we reduce the stellar masses of these progenitors (proportionally
to their masses), so that $\Sigma M_{\rm s,prog}=M_{\rm s,desc}$
holds. We continue decreasing the $\Ms$ values in halos of
all the closest previous time steps until reaching the $j$th time step
($j<k$), where $\Sigma M_{\rm s,prog}\leq M_{\rm s,root}$ holds for
all progenitor halos located at the $(j-1)$th time step. This process
is repeated recursively and backward in time throughout the merger
tree of the root halo, with each one of its progenitors playing 
the role of the root halo, that is, each progenitor goes later under 
the same process that modifies $\Ms$ values along its own (smaller) merger tree.
All the previous steps are then repeated for the next root halo in $\lbrace h_{i}\rbrace_{i=1}^{i=N}$ 
until the whole dark matter simulation is traversed.
Evidently, the optimal case would be to have the $k$th time step
located (if available) at sometime far enough in the future, so
that all values of $\Ms$ at $z=0$ are modified and stabilized.

\section{Galaxy Image Cutouts from SDSS}\label{Sec:Appendix2}

This appendix describes the procedure used to select
SDSS galaxy image cutouts that represent model galaxies on our simulated \textit{HST\/} images.

As a repository of galaxy images, we use the 
SDSS Data Release 10 \citep{york2000,ahn2014}. This data
archive (a MS-SQL Server database available online via CasJobs\footnote{
\url{http://casjobs.sdss.org}}) is suitable for our needs, as it provides
imaging and spectroscopy of $\sim 10^6$ galaxies as well as many
other derived physical properties.

We use in particular the Main Galaxy Sample \citep[MGS,][]{strauss2002},
which spans a rich variety of galaxy types over a sky area of 7930
deg$^{2}$. These galaxies form a flux-limited sample, with an $r$-band
Petrosian apparent magnitude limit of $m_r \leq 17.77$. We define
our sample in the redshift range of $[z_{1},z_{2}]=[0.01,0.2]$,
with the peak of the redshift distribution located at $z \sim 0.1$.
We consider clean {\tt science primary} galaxies only on calibrated
images having the {\tt photometric} status flag. We discard suspicious
detections and objects with deblending and interpolation problems,
as well as objects whose spectral line measurements and properties
are labeled as unreliable\footnote{More details in 
\url{http://sdss3.org/dr11/algorithms}}.  Our main sample contains $\sim$400,000
galaxies.

The properties available for the MGS that we use in this paper are
the photometry in the $ugriz$ bands, in particular model
magnitudes for computing the colors \citep{stoughton2002}. A measure
of the galaxy size is given by the (redshift insensitive) $r$-band
Petrosian radius $R_{\rm P}$ \citep{petrosian1976} and its less
noisy proxy $R_{50}$ (the radius encircling 50\% of the Petrosian flux).
The galaxy stellar mass $\Ms$ is also available; it was obtained
from spectral analysis at MPA and
JHU\footnote{\url{http://www.mpa-garching.mpg.de/SDSS}}, as detailed by
\cite{kauffmann2003}.

We extract the galaxy cutouts from 50~TB of 5-band, 2048$\times$1489 pixel SDSS
image frames using parallel processing in
the Data-Scope,\footnote{\url{http://idies.jhu.edu/datascope}} a
computing facility designed specifically for data-intensive
applications \citep{szalay2012}. As a first step, we 
denoise the images using the anisotropic diffusion method \citep{perona1990},
which is particularly useful for the more noisy $u$ and $z$ bands.
This leaves the galaxy outline and internal structure intact,
since the smoothing is done in the direction locally parallel to
the edges or borders. Multi-band images are registered to the
same reference frame using cubic spline interpolation.

We extract individual galaxy cutouts from the original images,
considering only what is included inside an aperture of radius 2$R_{\rm
P}$ centered at each galaxy. To avoid the influence of neighboring
objects, we do not use galaxies where the sum of fluxes
from all neighbors that overlap the
aperture exceeds 1\% of the galaxy flux. This leaves us
with a total of $\sim$270,000 galaxies in the repository.

We select from the SDSS image database 
the galaxy that is the closest neighbor to the model
galaxy in the multi-dimensional space of galaxy properties.
For this paper, we use two sets of parameters to select matching
SDSS galaxies:
\begin{enumerate}
\item $u-r$ color and $\log \Ms$

\item $u-r$ color, $\log \Ms$ and $\log R_{50}$

\end{enumerate}

Other schemes using additional parameters were tried with very similar
results.
Each property is mean-subtracted and normalized by the standard
deviation before doing the matching. The $u-r$ color tracks the
flux difference before and after the $4000 \rm \AA$ break, and is a
good indicator of age. To compare photometry, the (evolved)
rest-frame spectrum of the model galaxy is 
redshifted to the median of the SDSS redshift distribution ($z=0.1$), 
and then convolved with the $u$ and $r$ filters.
Using the properties from a simulated universe based on our
reference model (\S\ref{Sec:Results}), we
find that the median of the rms error over all properties derived
from the second matching scheme ($\log {\rm rms}=-0.85$) is bigger
than that of the first scheme ($\log {\rm rms}=-1.18$), since there
is one more variable to match ($\log R_{50}$). However, the individual
median rms error for $\log R_{50}$ in the second matching scheme is
$\sim1$~dex smaller than that of the first. A good compromise is to
use the first matching scheme to get a smaller overall rms error, but
then to manually rescale the SDSS galaxy size to match the
model galaxy size as described in \S\ref{Sec:CreatingSimImage}.
This approach is used as part of our reference model as
well.

\end{document}